%% file: main.tex
\documentclass[twocolumn]{aastex63}
\usepackage[utf8]{inputenc}
\usepackage{natbib}
\setcitestyle{round}
\usepackage{amsmath,amssymb}
\usepackage{afterpage}
\usepackage{xcolor}
\usepackage[caption=false]{subfig}
\usepackage{soul}
\usepackage{enumitem}

\newcommand{\hi}{{\rm H\,{\small I}}}
\newcommand{\kms}{\ensuremath{\,{\rm km\,s^{-1}}}}
\newcommand{\kkms}{\ensuremath{\,{\rm K km\,s^{-1}}}}
\newcommand{\percc}{\ensuremath{\,{\rm cm^{-3}}}}
\newcommand{\persc}{\ensuremath{\,{\rm cm^{-2}}}}
\newcommand{\fwhm}{\ensuremath{{b_{\mathrm{FWHM}}}}}
\newcommand{\ts}{\ensuremath{T_S}}

\defcitealias{doi:10.1146/annurev-astro-081817-051810}{SZ18}

\received{February 20, 2020}
\revised{March 24, 2020}
\accepted{March 25, 2020}
\submitjournal{ApJ}

\shorttitle{Small-scale \hi{} structure}

\begin{document}

 \title{Small-scale structure traced by neutral hydrogen absorption in the direction of multiple-component radio continuum sources}

\author{Daniel R. Rybarczyk}
\affiliation{Department of Astronomy, University of Wisconsin--Madison, Madison, WI 53706, USA}

\author{Snezana Stanimirovi{\'c}}
\affiliation{Department of Astronomy, University of Wisconsin--Madison, Madison, WI 53706, USA}

\author{Ellen G. Zweibel}
\affiliation{Department of Astronomy, University of Wisconsin--Madison, Madison, WI 53706, USA}
\affiliation{Department of Physics, University of Wisconsin--Madison, Madison, WI 53706, USA}

\author{Claire E. Murray}
\affiliation{Department of Physics \& Astronomy, Johns Hopkins University, Baltimore, MD 21218, USA}

\author{John M. Dickey}
\affiliation{School of Natural Sciences, University of Tasmania, Hobart, TAS 7001, Australia}

\author{Brian Babler}
\affiliation{Department of Astronomy, University of Wisconsin--Madison, Madison, WI 53706, USA}

\author{Carl Heiles}
\affiliation{Department of Astronomy, University of California, Berkeley, CA 94720-3411, USA}

\begin{abstract}
We have studied the small scale distribution of atomic hydrogen (\hi{}) using 21-cm absorption spectra against multiple-component background radio continuum sources from the 21-SPONGE survey and the Millennium Arecibo Absorption Line Survey. We have found $>5\sigma$ optical depth variations at a level of $\sim0.03$--$0.5$ between 13 out of 14 adjacent sightlines separated by a few arcseconds to a few arcminutes, suggesting the presence of neutral structures on spatial scales from a few to thousands of AU (which we refer to as tiny scale atomic structure, TSAS). The optical depth variations are strongest in directions where the \hi{} column density and the fraction of \hi{} in the cold neutral medium (CNM) are highest, which tend to be at low Galactic latitudes. By measuring changes in the properties of Gaussian components fitted to the absorption spectra, we find that changes in both the peak optical depth and the linewidth of TSAS absorption features contribute to the observed optical depth variations, while changes in the central velocity do not appear to strongly impact the observed variations. Both thermal and turbulent motions contribute appreciably to the linewidths, but the turbulence does not appear strong enough to confine overpressured TSAS. In a majority of cases, the TSAS column densities are sufficiently high that these structures can radiatively cool fast enough to maintain thermal equilibrium with their surroundings, even if they are overpressured. We also find that a majority of TSAS is associated with the CNM.
For TSAS
 in the direction of the Taurus molecular cloud and the local Leo cold cloud, we estimate densities over an order of magnitude higher than typical CNM densities.
\end{abstract}

\section{Introduction} \label{sec:intro}
For over 40 years, high-resolution absorption measurements of the 21-cm line of atomic hydrogen (\hi{}) have revealed significant non-uniform structure of interstellar gas on angular scales $\lesssim1\arcmin{}$. This has traditionally been interpreted as evidence for small, overdense, overpressured \hi{} structures---the so-called
 tiny scale atomic structure (TSAS; \citealt{1997ApJ...481..193H,2000riss.conf....7H}; \citealt{doi:10.1146/annurev-astro-081817-051810}, hereafter SZ18). 
Under a simple geometric assumption, these structures have densities and pressures several orders of magnitude higher than either the cold neutral medium (CNM) or warm neutral medium (WNM), the two stable thermal phases that underlie most models of the Galactic neutral interstellar medium (ISM). 

While the origin of TSAS is still an open question, such structures have been proposed  to enhance molecular abundances (\citealt{2013MNRAS.429..939S}), increase heating of the neutral ISM (\citetalias{doi:10.1146/annurev-astro-081817-051810}), mark sites of turbulent dissipation (\citealt{2006EllenSINS}), or possibly represent ablated stellar or planetary material (\citealt{2017ApJ...836..135R}). Constraining observationally physical properties of TSAS is essential for understanding its origin and role in the ISM. 

One way TSAS is observed is by constructing high resolution optical depth images against extended radio continuum sources, typically using very long baseline interferometry (VLBI), where the pixel-to-pixel variations reveal TSAS (e.g., \citealt{1989ApJ...347..302D,2000ApJ...543..227D,2001AJ....121.2706F,2005AJ....130..698B,2008MNRAS.388..165G,2009AJ....137.4526L,2012ApJ...749..144R}).
In addition to spatial variations, the temporal variations of \hi{} absorption measured against extragalactic radio continuum sources and Galactic pulsars (whose transverse velocities are $\sim10$--$100$ AU $\mathrm{yr^{-1}}$) have been used to probe TSAS on scales $\lesssim 1000$ AU (e.g., \citealt{2003MNRAS.341..941J,2005AJ....130..698B,2005ApJ...631..376M,2008ApJ...674..286W,2010ApJ...720..415S}).

\begin{figure}
    \centering
    \includegraphics[width=\columnwidth]{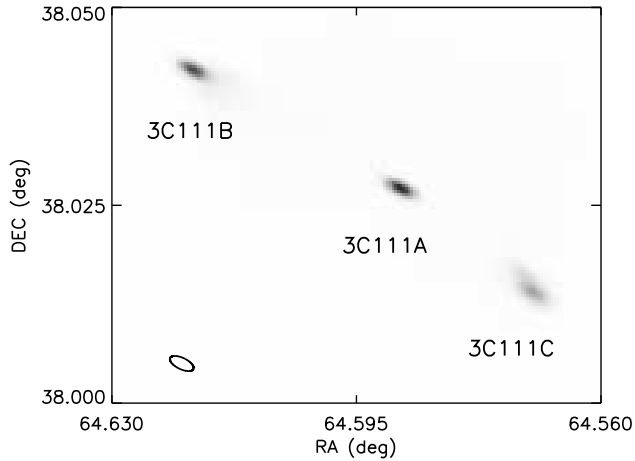}
    \caption{The triple-lobed radio galaxy 3C111 shown in 1420 MHz continuum (\citealt{2018ApJS..238...14M}). Absorption spectra were obtained toward each of its three components, labeled 3C111A, 3C111B, and 3C111C.}
    \label{fig:3c111cont}
\end{figure}

Neutral structures at spatial scales of hundreds to thousands of AU
can be detected by separately measuring absorption spectra toward each component of a multiple-component radio continuum source (see Figure \ref{fig:3c111cont}).
If the components are separated by a few arcseconds to a few arcminutes, the transverse linear separation of Galactic \hi{} structures between the two lines of sight, $L$, is typically $\lesssim 10^4$ AU. \cite{1979ApJ...233..558D}, \cite{1985A&A...146..223C}, and \cite{1986ApJ...303..702G} have all found significant \hi{} optical depth variations across multiple-component sources with angular separations $\lesssim6\arcmin{}$, suggesting non-uniform \hi{} structure on sub-parsec scales. 

Analogous studies have been carried out at optical wavelengths 
against stars in multiple star systems,
which typically probe spatial scales of a few thousand AU
(\citetalias{doi:10.1146/annurev-astro-081817-051810} and references therein). Several of these observations have suggested that AU-scale structure is especially common in the direction of supernova remnants (\citealt{2016ApJ...819...45D,2017MNRAS.467.1186K}) and perhaps stellar bow shocks (\citealt{2015AAS...22514123M}) and the interfaces of warm cloud collisions (\citealt{2012ApJ...752..119M}).

Under the assumption that the optical depth variations seen on small angular scales are caused by the presence of discrete \hi{} structures, the measured column density difference divided by the assumed line-of-sight length provides an estimate of the \hi{} number density.
Observationally inferred TSAS densities ($\sim10^4$--$10^6$ \percc{}) and thermal pressures ($\sim10^6$--$10^7$ K\percc{}) are several orders of magnitude higher than what is common in either the CNM or the WNM. 
This has led to several alternative interpretations for the origin of the \hi{} optical depth variations, including
the suggestion that the variations could be caused by the overlapping of \hi{} sheets and filaments along the line of sight (\citealt{1997ApJ...481..193H}), or 
the general turbulent cascade over a range of
\hi{} structures (\citealt{2000MNRAS.317..199D}). 

Simulations investigating the formation of TSAS have been rare because of the AU-scale spatial resolution required. \cite{2002ApJ...564L..97K} performed a two-dimensional numerical hydrodynamic simulation of the propagation of a strong shock into the ISM. They found that regions of a shock-compressed layer leading the shock front fragmented into small, cold cloudlets as a result of thermal instability. These cloudlets had temperatures of $\sim20$ K and densities of $n\sim2000$ \percc{}, similar to the properties attributed to TSAS. \cite{2007A&A...465..431H} numerically modeled a turbulent flow of interstellar atomic gas with spatial resolution of $2\times10^{-3}$ pc, finding that both diffuse and dense structures can form at small scales. The dense structures were produced by collisions between CNM fragments that form within the WNM. These simulations suggest that the formation and survival of TSAS may be driven by shocks. If the small-scale variations in \hi{} optical depth are indeed caused by small dense structures, then we ought to detect these variations preferentially in the direction of shocked regions of the ISM, specifically within a post-shock layer. This may explain TSAS's association with supernova remnants, stellar bow shocks, or warm cloud collisions as noticed at optical wavelengths (\citealt{2016ApJ...819...45D,2017MNRAS.467.1186K,2015AAS...22514123M,2012ApJ...752..119M}), and with the Local Bubble wall in the \cite{2010ApJ...720..415S} radio \hi{} absorption study against the pulsar B1929+10.

Here, we search for evidence of AU-scale atomic structure by measuring the optical depth variations in the direction of the 9 multiple-component background radio continuum sources included in the 21-SPONGE survey (21-cm Spectral Line Observations of Neutral Gas with the VLA; \citealt{2015ApJ...804...89M,2018ApJS..238...14M}) and 3 multiple-component background radio continuum sources included in the Millennium Arecibo Absorption-Line Survey (\citealt{2003ApJS..145..329H,2003ApJ...586.1067H}). This nearly doubles the sample size of multiple-component background sources with sensitive \hi{} absorption measurements, and represents the most sensitive study of its kind to date. As discussed by \cite{1985A&A...146..223C}, the \hi{} optical depth variations can be measured in two ways. The first is to measure the change in optical depth for each velocity channel, $\Delta \tau(v)$. This approach uses only measured quantities and is relatively simple, but it does not typically allow us to study the properties of individual structures. The second method of quantifying \hi{} optical depth variation is to fit Gaussian components to the optical depth spectra. Each Gaussian function can be considered to represent a unique structure with a peak optical depth, central velocity, linewidth, spin temperature, and column density. One can then compare the properties of the Gaussian components matched to adjacent lines of sight to determine how these structures change on small scales. This method gives more physically meaningful results than measuring the channel-by-channel variations in optical depth, but it assumes that \hi{} spectral lines are well represented with Gaussian functions, and also depends on the reliability of the Gaussian fitting. We focus here only on measuring the spatial variations of \hi{} optical depth profiles. Future follow-up observations will be necessary to measure temporal variations,
which can probe even smaller spatial scales.
While \cite{1997ApJ...481..193H} used ``TSAS'' for structures on scales of less than a few hundred AU
by focusing on pulsar HI spectra and VLBI imaging only, a variety of different observational techniques have expanded the TSAS parameter space in more recent years (see \citetalias{doi:10.1146/annurev-astro-081817-051810}). Therefore, in this paper we refer as TSAS all neutral structures on spatial scales from a few to thousands of AU.

The 21-SPONGE and Millennium survey data used in this study are presented Section \ref{sec:data}. In Sections \ref{sec:spectra} and \ref{sec:gaussians}, we discuss the tiny scale changes in observed \hi{} gas properties using the channel-by-channel method and the method of Gaussian fitting, respectively. We then discuss the implications of the variations seen using both methods in Section \ref{sec:discussion}.

\begin{figure*}[]
    \centering
    \includegraphics[width=\textwidth]{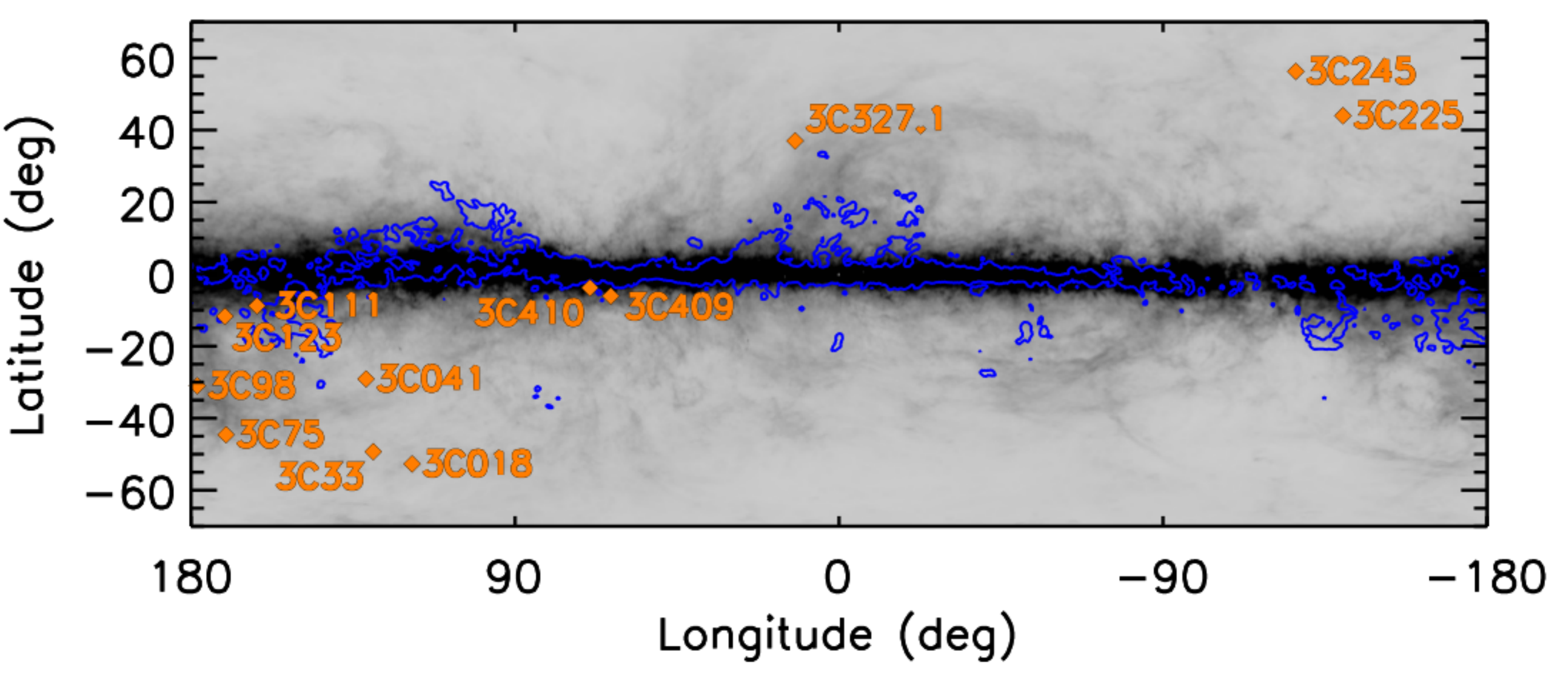}
    \caption{An all sky image of the \hi{} column density from the \hi{}4PI survey (\citealt{2016A&A...594A.116H}) with the multiple-component sources from the 21-SPONGE and Millennium surveys shown as orange diamonds and labeled in orange text. Blue contours of CO integrated intensity (4\kkms{}) from the \cite{2014A&A...571A..13P} survey of Galactic CO emission are overlaid.}
    \label{fig:allskymap}
\end{figure*}

\begin{table*}[]
\centering
\begin{tabular}{cccccc}
Component 1     &  Component 2  & $\ell$ & $b$ &  $\Delta\theta$& $\langle EW \rangle$\\
     &    & deg & deg &   & \kms{} \\
\hline
\hline
 3C018A  &   3C018B  &   118.6  &  $-52.7$  &             $49\arcsec{}$  &      $2.37 \pm 0.02$\\
\hline
 3C041A  &   3C041B  &   131.4  &  $-29.1$  &             $23\arcsec{}$  &      $0.41 \pm 0.03$\\
\hline
 3C111A  &   3C111B  &   161.7  &    $-8.8$  &  $2\arcmin{}$  &      $10.17 \pm 0.03$\\
\hline
 3C111A  &   3C111C  &   161.7  &  $-8.8$  &    $1.4\arcmin{}$  &     $10.79 \pm 0.04$\\
\hline
 3C111B  &   3C111C  &  161.7  &   $-8.8$  &    $3.4\arcmin{}$  &     $10.76 \pm 0.04$\\
\hline
 3C123A  &   3C123B  &    170.6  &  $-11.7$  &            $22\arcsec{}$  &     $8.93 \pm 0.01$\\
\hline
 3C225A  &   3C225B  &   220.0  &   44.0  &              $5\arcsec{}$  &     $1.57 \pm 0.02$\\
\hline
 3C245A  &   3C245B  &  233.1  &     56.3  &             $5\arcsec{}$  &     $0.09 \pm 0.03$\\
\hline
3C327.1A &  3C327.1B &     12.2  &   37.0  &            $12\arcsec{}$  &      $2.24 \pm 0.04$\\
\hline
 3C409A  &   3C409B  &   63.4  &    $-6.1$  &              $6\arcsec{}$  &     $8.60 \pm 0.02$\\
\hline
 3C410A  &   3C410B  &    69.2  &     $-3.8$  &            $6\arcsec{}$  &     $17.45 \pm 0.04$\\
\hline
3C33-$\mathrm{1^M}$ & 3C33-$\mathrm{2^M}$ & 129.4 & $-49.3$  & $4.2\arcmin{}$ & $0.45 \pm 0.06$\\
\hline
3C75-$\mathrm{1^M}$ & 3C75-$\mathrm{2^M}$ & 170.3 & $-44.9$  & $3.4\arcmin{}$   & $2.47\pm0.10$\\
\hline
3C98-$\mathrm{1^M}$ & 3C98-$\mathrm{2^M}$ & 179.8 & $-31.0$  & $4\arcmin{}$   & $2.96\pm 0.05$\\
\hline
\end{tabular}
\caption{The multiple-component sources from the 21-SPONGE survey
         (\citealt{2015ApJ...804...89M,2018ApJS..238...14M}) and the Millennium survey (\citealt{2003ApJS..145..329H,2003ApJ...586.1067H}). 
         \textbf{Col 1--2:} The first and second components of each component pair. Sources from the Millennium survey are denoted with an ``M'' superscript; all other sources are from the 21-SPONGE survey. 
         \textbf{Col 3:} Galactic longitude of the background source.
         \textbf{Col 4:} Galactic latitude of the background source.
         \textbf{Col 5:} The angular separation of the two components.
         \textbf{Col 9:} The average optical depth equivalent width of the two adjacent lines of sight: $\langle EW \rangle =\int \frac{\tau_1(v)+\tau_2(v)}{2} dv$ (see Section \ref{sec:spectra} for discussion).}
\label{tab:sources}
\end{table*}

\section{Observations and Data} \label{sec:data}

\subsection{Observations} \label{subsec:observations}
The 21-SPONGE survey measured Galactic \hi{} absorption spectra against background radio continuum sources with exceptional optical depth sensitivity ($\sigma_{\tau} < 10^{-3}$ at a velocity resolution of 0.42 \kms{}) using the Karl G. Jansky Very Large Array (VLA). Matching \hi{} emission spectra were measured at $\sim 4^\prime$ angular resolution using the Arecibo Observatory. In total, 21-SPONGE observed emission and absorption spectra along 57 unique lines of sight to study the physical properties of neutral hydrogen in the Milky Way (\citealt{2015ApJ...804...89M,2018ApJS..238...14M}). 
The 21-SPONGE background sources included 8 double-lobed galaxies (3C018, 3C041, 3C123, 3C225, 3C245, 3C327.1, 3C409, and 3C410) and one triple-lobed galaxy (3C111; Figure \ref{fig:3c111cont}). Absorption spectra were measured toward each lobe separately. We measure the optical depth variations across each pair of components; there are 11 component pairs in total (one pair for each double-lobed source, and three pairs for the triple-lobed source). For each component pair, Table \ref{tab:sources} lists the name of each component (Columns 1 and 2), the Galactic coordinates of the source (Columns 3 and 4), the angular separation of the the component pair (Column 5), and the average equivalent width of the \hi{} optical depth toward the two components, defined as $\int\langle\tau\rangle dv = \int\frac{\tau_1(v)+\tau_2(v)}{2}dv$ (Column 6).

\cite{2003ApJS..145..329H,2003ApJ...586.1067H} previously obtained high sensitivity emission-absorption measurements with the Arecibo Observatory along 79 lines of sight, including lines of sight toward the double-lobed galaxies 3C33, 3C75, and 3C98. At 0.42 \kms{} binning, they reach an optical depth sensitivity of $\sigma_{\tau}\approx0.01$. Table \ref{tab:sources} also lists the properties of the component pairs for these lines of sight. While the Millennium survey's measurements are less sensitive than those of 21-SPONGE, we include these three sources to increase our sample size. 

Four of the multiple-component sources are at low Galactic latitude (3C111, 3C123, 3C409, and 3C410), and 8 sources are at $|b|>29^{\circ}$ (3C018, 3C041, 3C225, 3C245, 3C327.1, 3C33, 3C75, and 3C98). The angular separations between components range from $5\arcsec{}$ to $4.2\arcmin{}$. The positions of these multiple-component background sources in Galactic coordinates are shown in Figure \ref{fig:allskymap} against an all-sky map of \hi{} column density from the \hi{}4PI survey (\citealt{2016A&A...594A.116H}) with contours of Galactic CO emission from the \cite{2014A&A...571A..13P} survey.

\subsection{Gaussian fits} \label{subsec:gaussianfits}
Both \cite{2003ApJS..145..329H} and \cite{2018ApJS..238...14M} fit Gaussian features to the \hi{} absorption and emission spectra for each line of sight. The reliability of the Gaussian fitting influences our interpretation (see Section \ref{sec:gaussians}).

\cite{2003ApJS..145..329H} used a least-squares fit of Gaussian components to the observed absorption and emission spectra to derive the properties of the Gaussian features. \cite{2018ApJS..238...14M} used the Autonomous Gaussian Decomposition algorithm (AGD; \citealt{2015AJ....149..138L}), a derivative spectroscopy program that employs machine learning to estimate the appropriate number of Gaussian features and their properties. The AGD was trained on spectra with the same properties as the \cite{2003ApJS..145..329H} data. \cite{2017ApJ...837...55M} tested the AGD fitting on synthetic spectra (\citealt{2014ApJ...786...64K}). They found that the completeness---defined as the ratio of the number of AGD-fitted components to the true number of components in the synthetic spectra---was 29\% for low Galactic latitudes ($|b| < 20^{\circ}$), 38\% for mid latitudes ($20^{\circ} < |b| < 50^{\circ}$), and 83\% for high latitudes ($|b| > 50^{\circ}$). The velocity crowding of structures at lower latitudes poses a challenge to the Gaussian fitting. 

The number of fitted Gaussian components for the sightlines considered here ranges from 0 in the case of 3C245B (the only sightline in this sample with no fitted optical depth features, since no features reach the signal-to-noise cutoff of 3) to 13 in the case of 3C410A, which is at the lowest Galactic latitude. The Gaussian fitting was used to estimate the central velocity, $v_0$, the full-width at half maximum (FWHM), \fwhm{}, and the peak optical depth, $\tau_0$, for Gaussian components seen in \hi{} absorption spectra. By using the corresponding \hi{} emission, \cite{2003ApJS..145..329H} and \cite{2018ApJS..238...14M} ran radiative transfer calculations to estimate the spin temperature, \ts{}, for each absorption structure. The column density, $N(\hi{})$ was then calculated using $N(\hi{})=1.064 \times C_0 \times \tau_0 \times \fwhm{} \times T_S$, where $C_0=1.823\times10^{18}$ \persc{}/K\kms{}. Some features were fitted with $T_{S} < 10$ K. This was deemed physically unreasonable, so the spin temperature and the \hi{} column density are not known for these features. Finally, the total column density along the line of sight and the CNM fraction were estimated.

\section{Analysis of Spectra} \label{sec:spectra}

Figure \ref{fig:allspectra} shows the \hi{} optical depth spectra for each pair of 
components, $\tau_1(v)$ and $\tau_2(v)$ (top panel), as well as their difference, $\Delta \tau(v) \equiv \tau_1(v) - \tau_2(v)$ (bottom panel) with corresponding uncertainties ($\pm3\sigma$). For 13 out of 14 pairs, we find channels for which $\Delta \tau(v) > 5\sigma_{\Delta \tau}$. Such optical depth variation is the trademark signature of TSAS.

Previously, \cite{1979ApJ...233..558D} found significant changes in \hi{} optical depth in the direction of only 3 out of 9 multiple-component sources, but \cite{1985A&A...146..223C} and \cite{1986ApJ...303..702G} detected significant variations across all of the multiple-component sources they observed, numbering 7 and 3, respectively. The two strongest detections made by \cite{1979ApJ...233..558D} had a sensitivity of $\sigma_{\Delta \tau}\approx 0.005$, the most sensitive of all their measurements. All non-detections had noise levels of $\sigma_{\Delta \tau}\gtrsim 0.01$. \cite{1985A&A...146..223C} reached noise levels of $\sigma_{\Delta \tau}\approx 0.005$--0.05 and \cite{1986ApJ...303..702G} reached noise levels of $\sigma_{\Delta \tau}\gtrsim 0.05$. Our results (Figure \ref{fig:allspectra}) and the results of these previous \hi{} absorption studies against multiple-component sources suggest that optical depth variations of at least $\sim0.05$ are common for cold \hi{} in the Milky Way on angular scales smaller than a few arcminutes. 3C245 is the only multiple-component source in any of the multiple-component studies with a noise level $\lesssim0.01$ that does not show $>5\sigma$ variation in \hi{} optical depth.

Some multiple-component sources---like those detected by \cite{1979ApJ...233..558D} and \cite{1986ApJ...303..702G}---are highly variable ($\Delta \tau \gg 0.01$) over angular scales of $\lesssim1\arcmin{}{}$ and have shown repeated variations in several studies. Out of 12 background sources in this study, we see the most variation in the direction of 3C111, which is one of the sources where \citealt{1986ApJ...303..702G} also saw significant changes in the \hi{} optical depth. 3C111 is located behind the Taurus molecular cloud. Since it is triple-lobed, optical depth variations can be measured across three component pairs (AB, AC, and BC). We find optical depth variations as high as 0.38, 0.35, and 0.47 for the three component pairs. We also find large spatial variations of the \hi{} optical depth profiles in the direction of 3C409 ($\max\{\Delta\tau(v)\}=0.28$) and 3C410  ($\max\{\Delta\tau(v)\}=0.52$). All three sources  show variations at a level of over 40$\sigma$, and all are located at low Galactic latitudes (Table \ref{tab:sources}).

We discuss the changes in the integrated \hi{} optical depth spectra across all of our multiple-component background sources in Section \ref{subsec:integratedproperties} and then present a channel-by-channel analysis of the \hi{} optical depth variation in Section \ref{subsec:channelvariations}. 

\begin{figure*}[h!]
\gridline{\fig{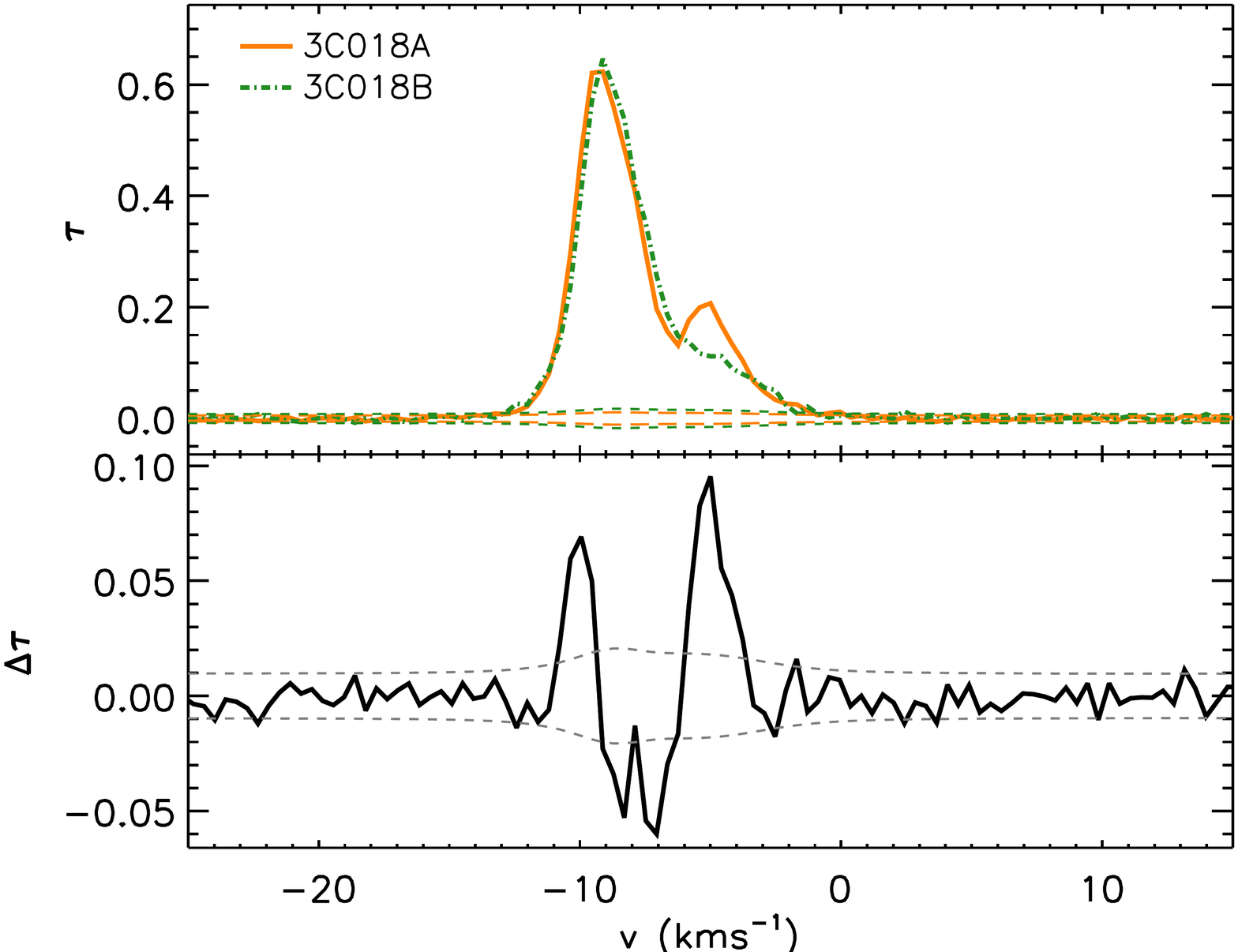}{0.5\textwidth}{3C018 (21-SPONGE)}
          \fig{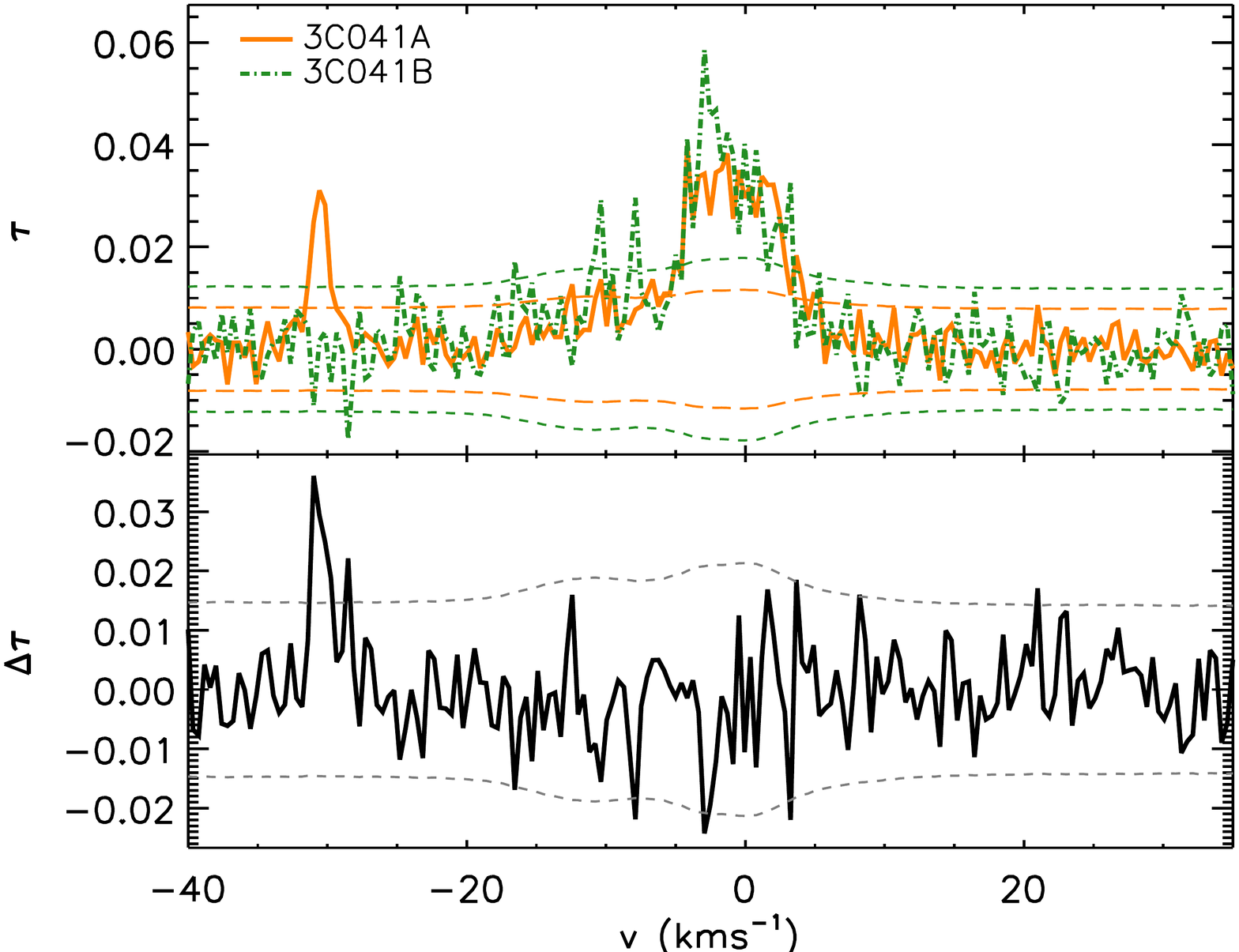}{0.5\textwidth}{3C041 (21-SPONGE)}}
\gridline{\fig{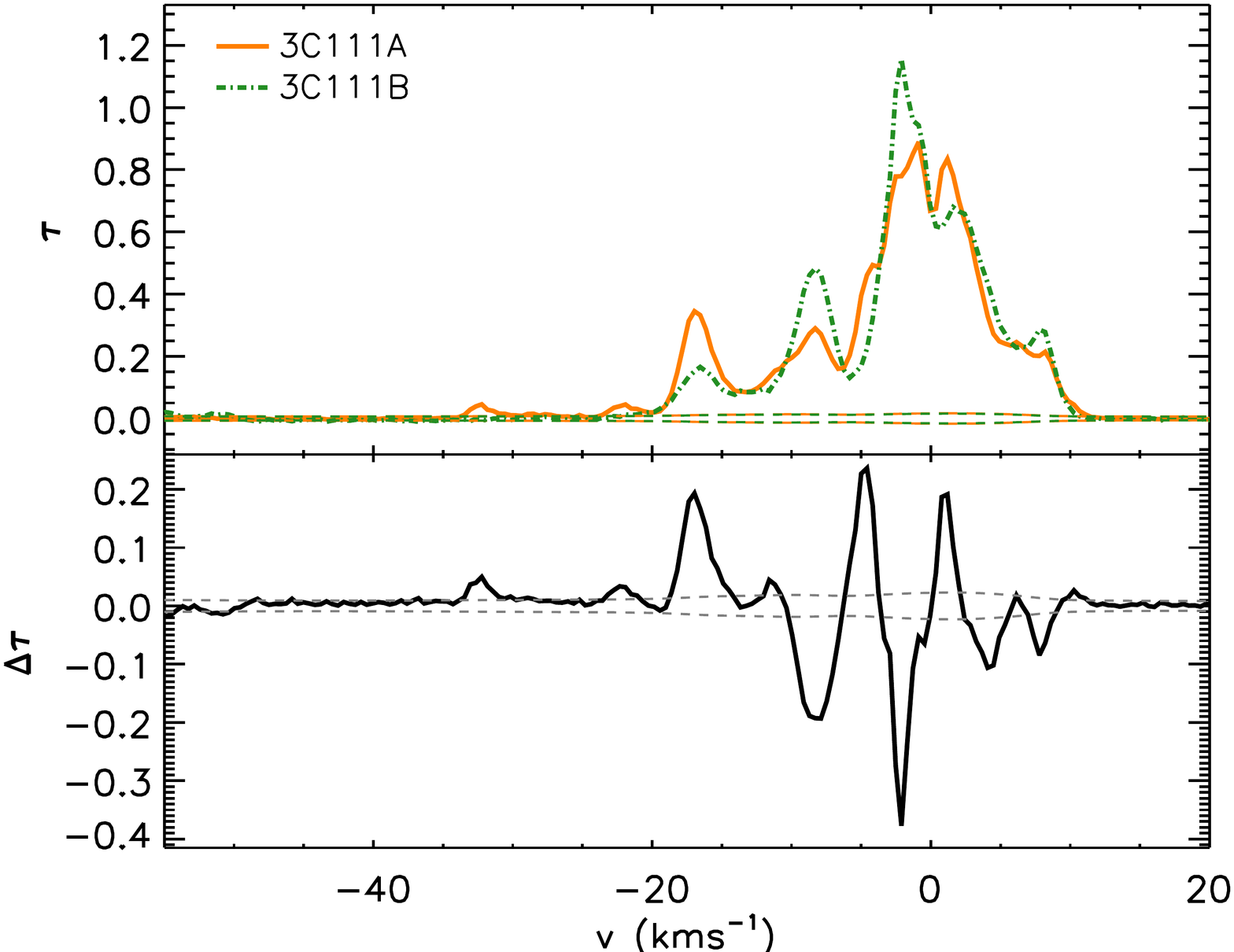}{0.5\textwidth}{3C111-AB (21-SPONGE)}
          \fig{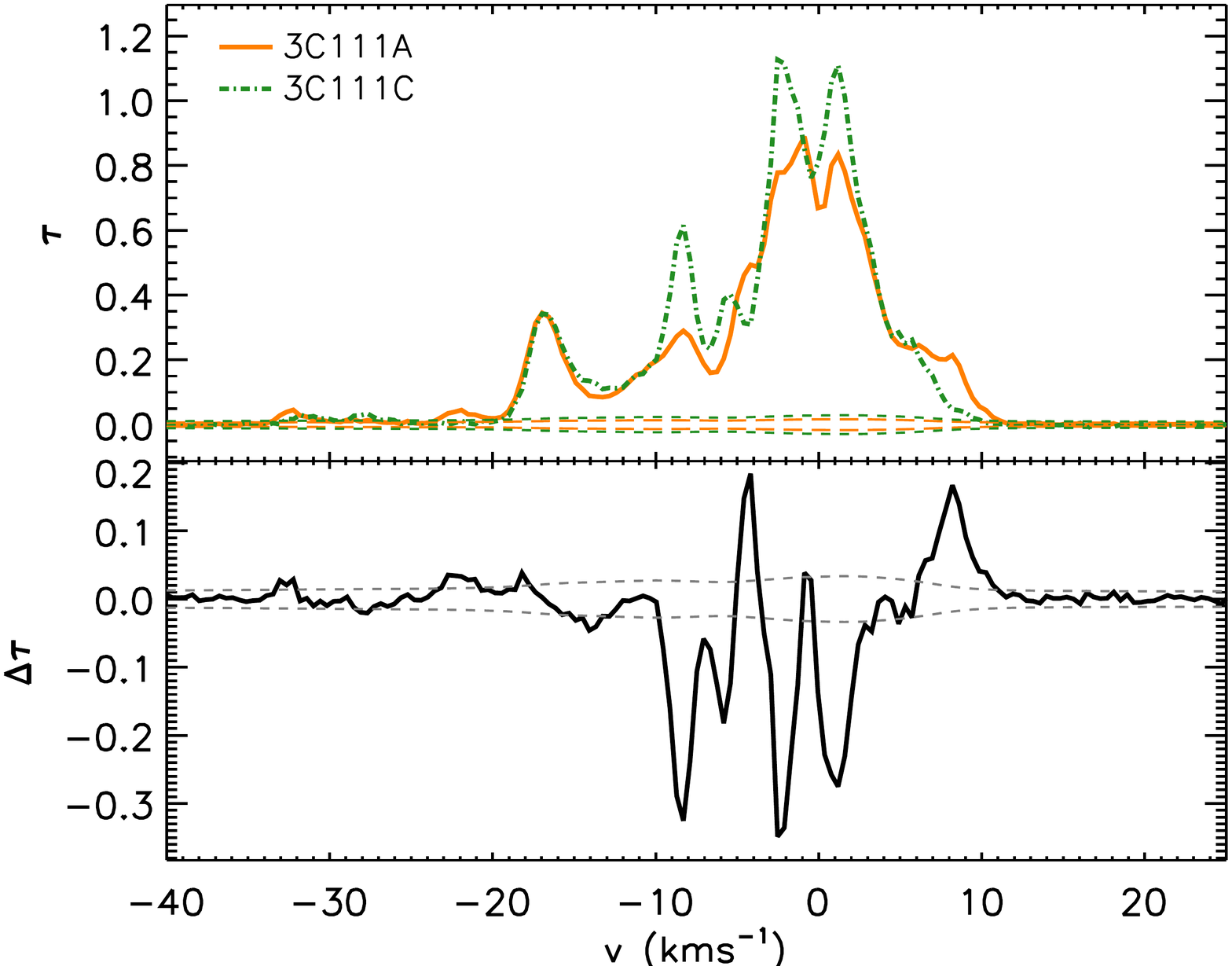}{0.5\textwidth}{3C111-AC (21-SPONGE)}}
\gridline{\fig{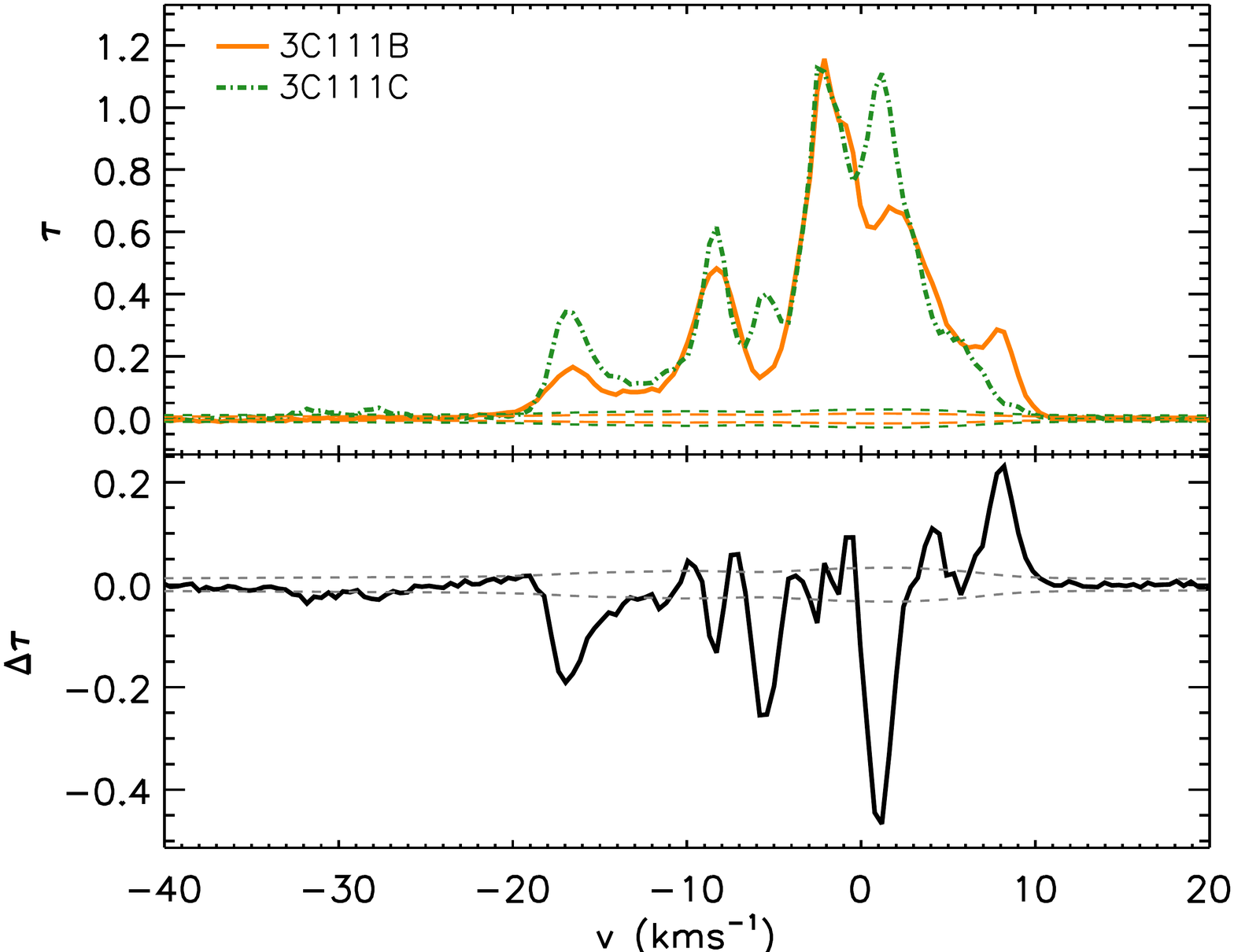}{0.5\textwidth}{3C111-BC (21-SPONGE)}
          \fig{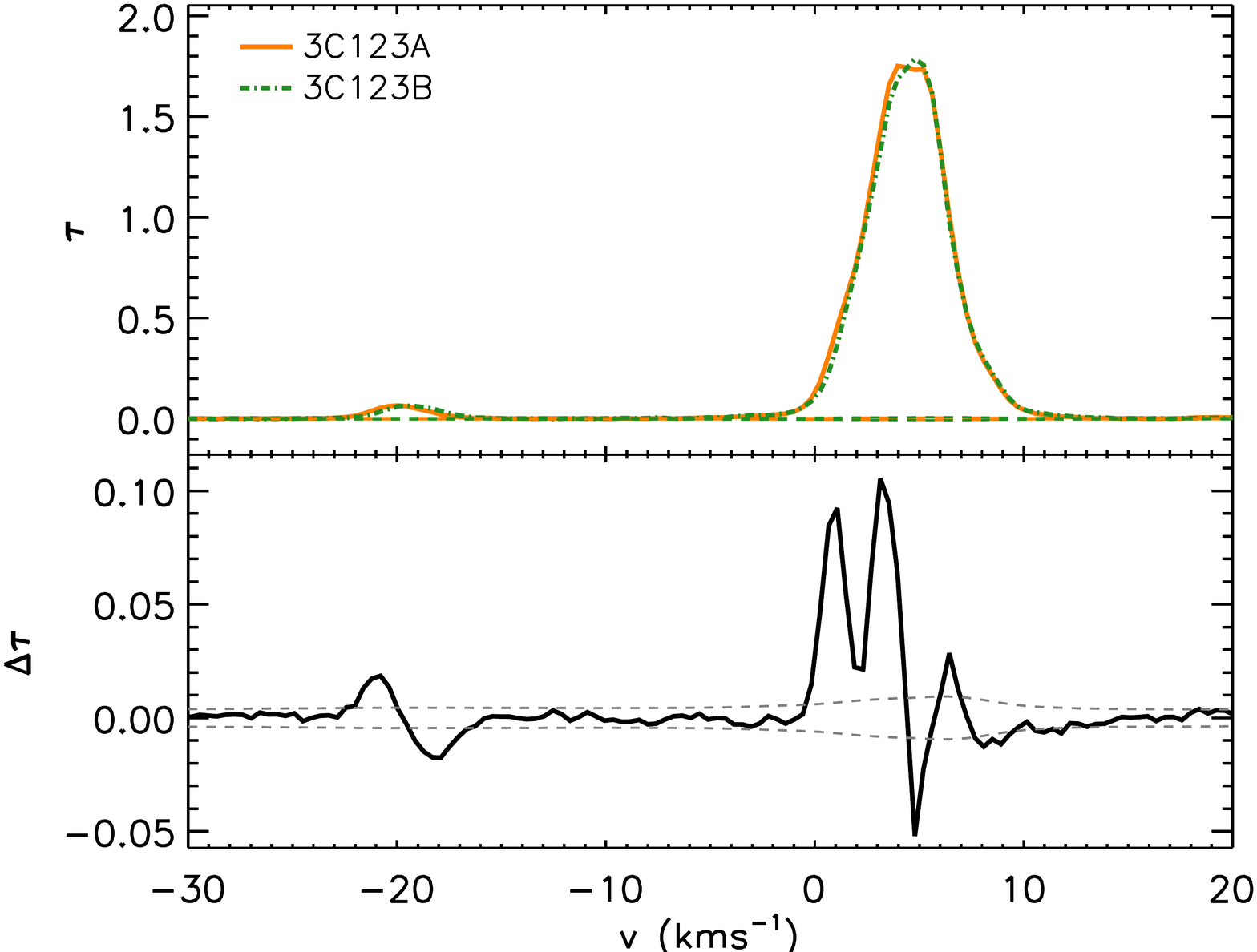}{0.5\textwidth}{3C123 (21-SPONGE)}}
\caption{}
\end{figure*}
\begin{figure*}\ContinuedFloat
\gridline{\fig{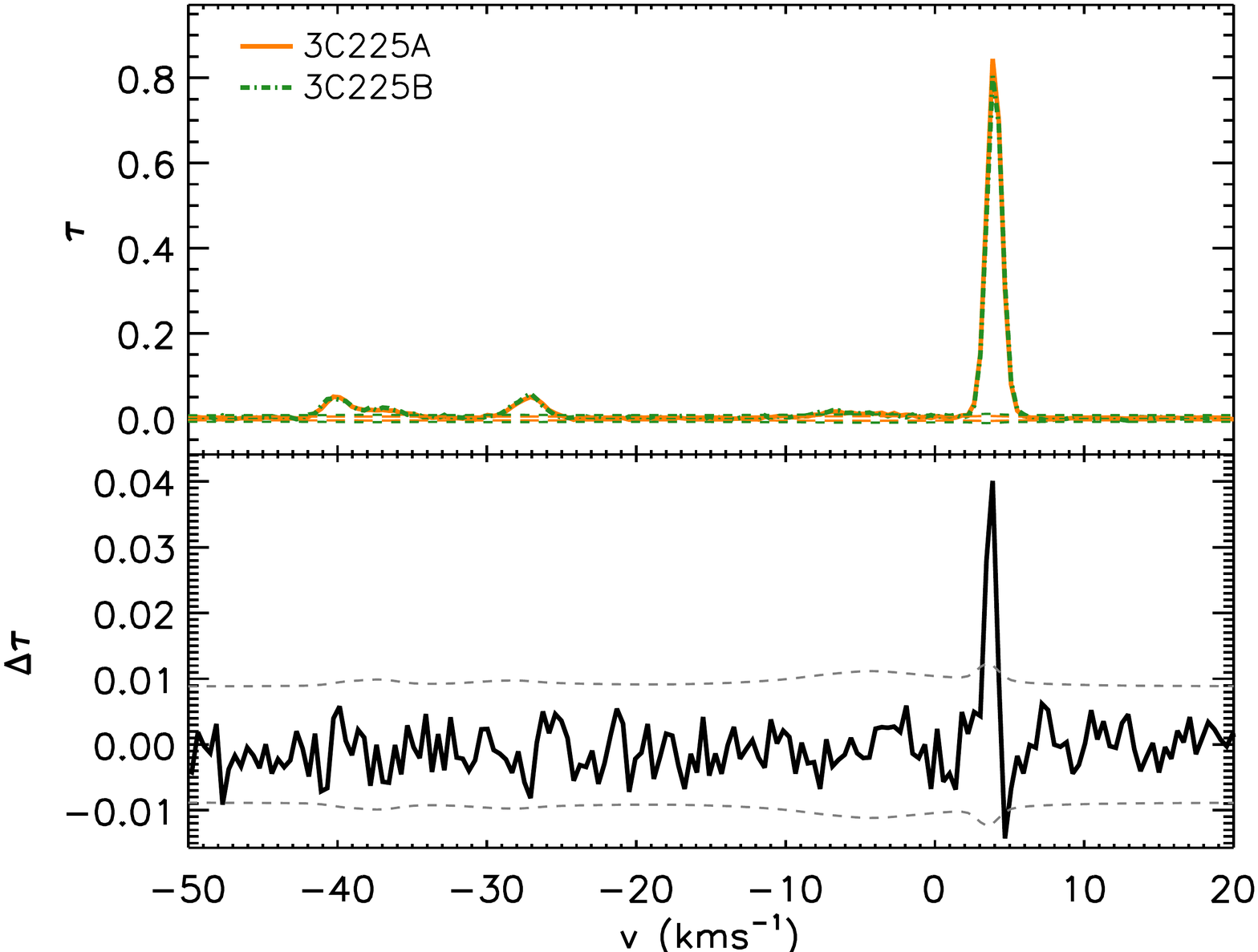}{0.5\textwidth}{3C225 (21-SPONGE)}
          \fig{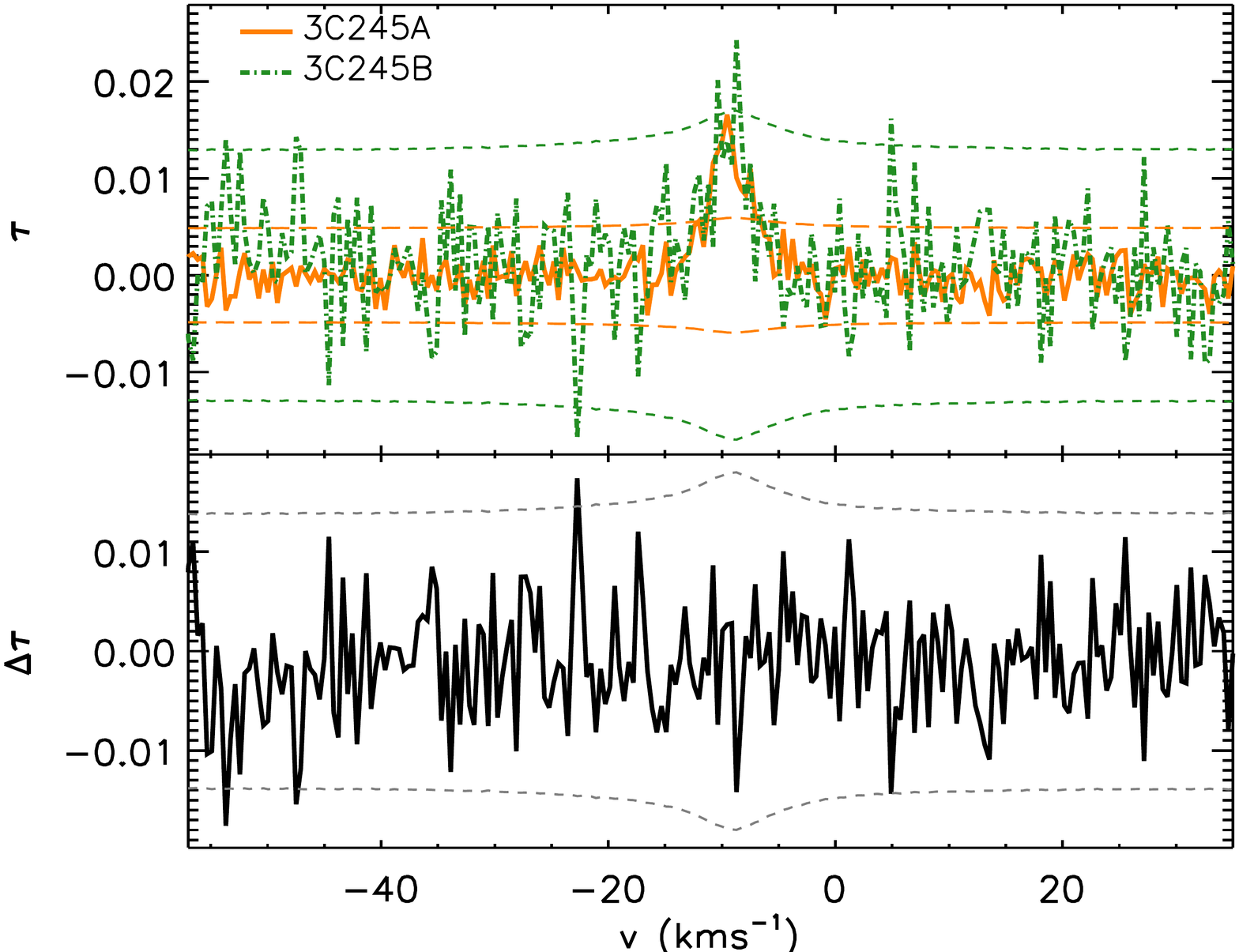}{0.5\textwidth}{3C245 (21-SPONGE)}}
\gridline{\fig{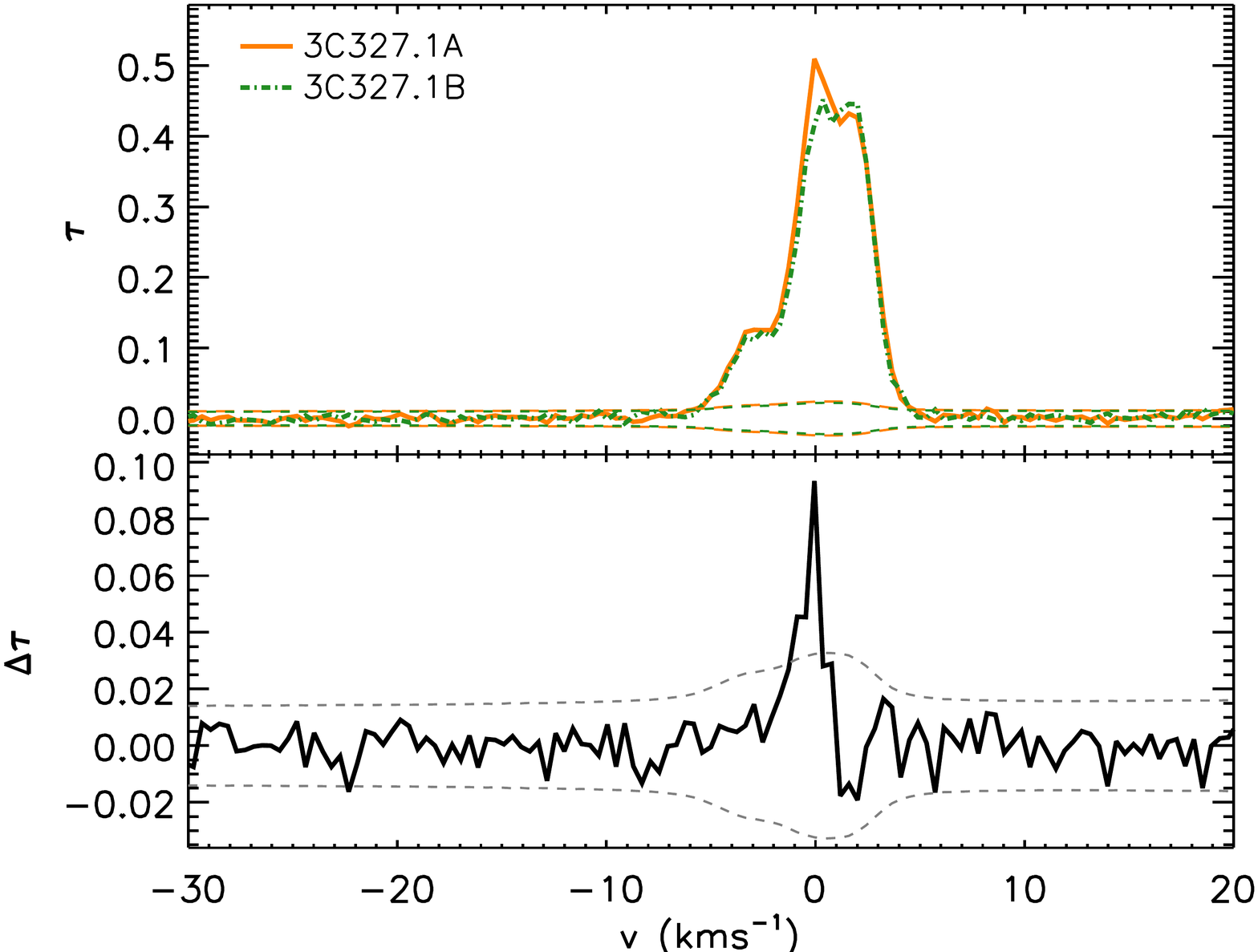}{0.5\textwidth}{3C327.1 (21-SPONGE)}
          \fig{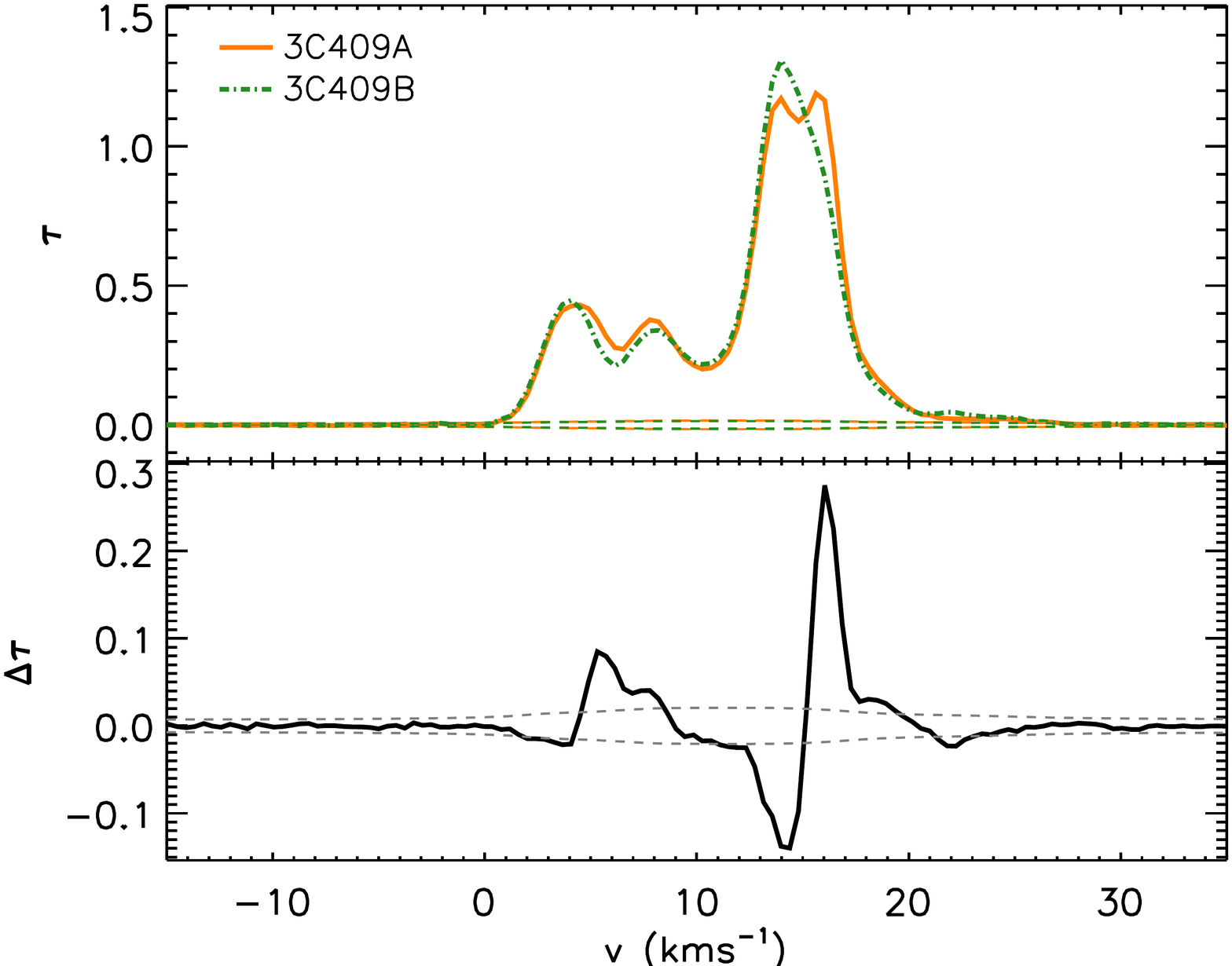}{0.5\textwidth}{3C409 (21-SPONGE)}}
\gridline{\fig{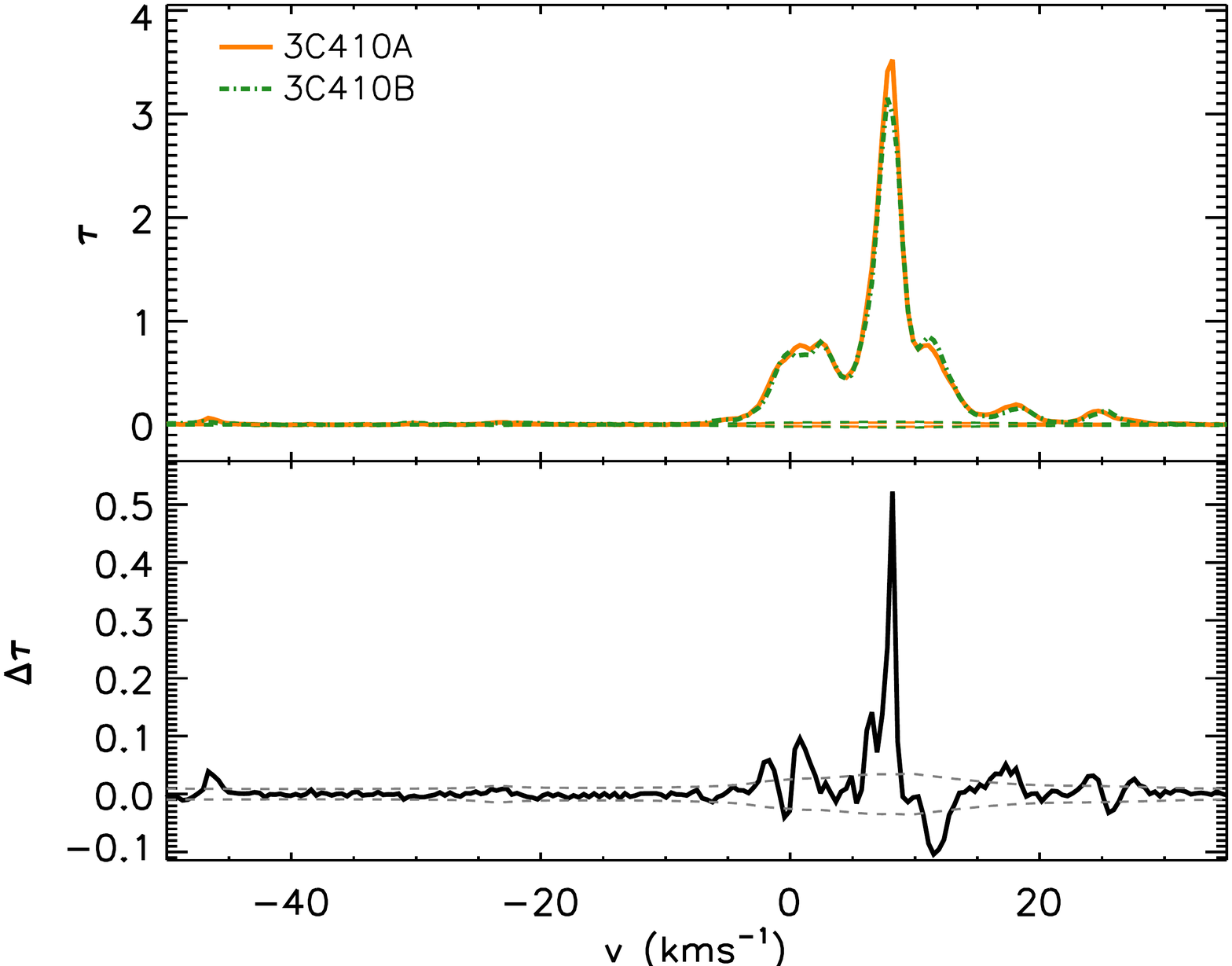}{0.5\textwidth}{3C410 (21-SPONGE)}
          \fig{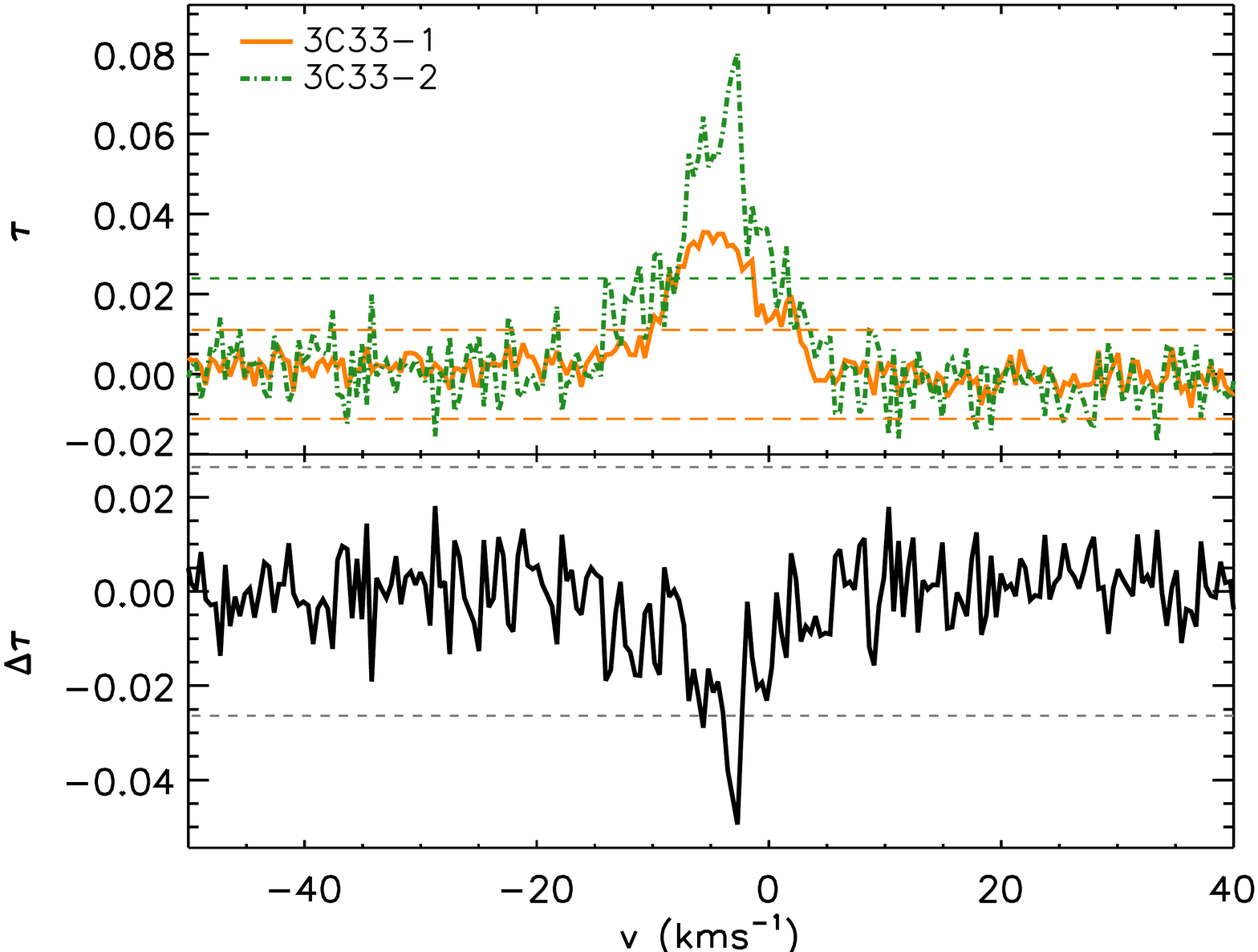}{0.5\textwidth}{3C33 (Millennium)}}
\caption{}
\end{figure*}
\begin{figure*}\ContinuedFloat
\gridline{\fig{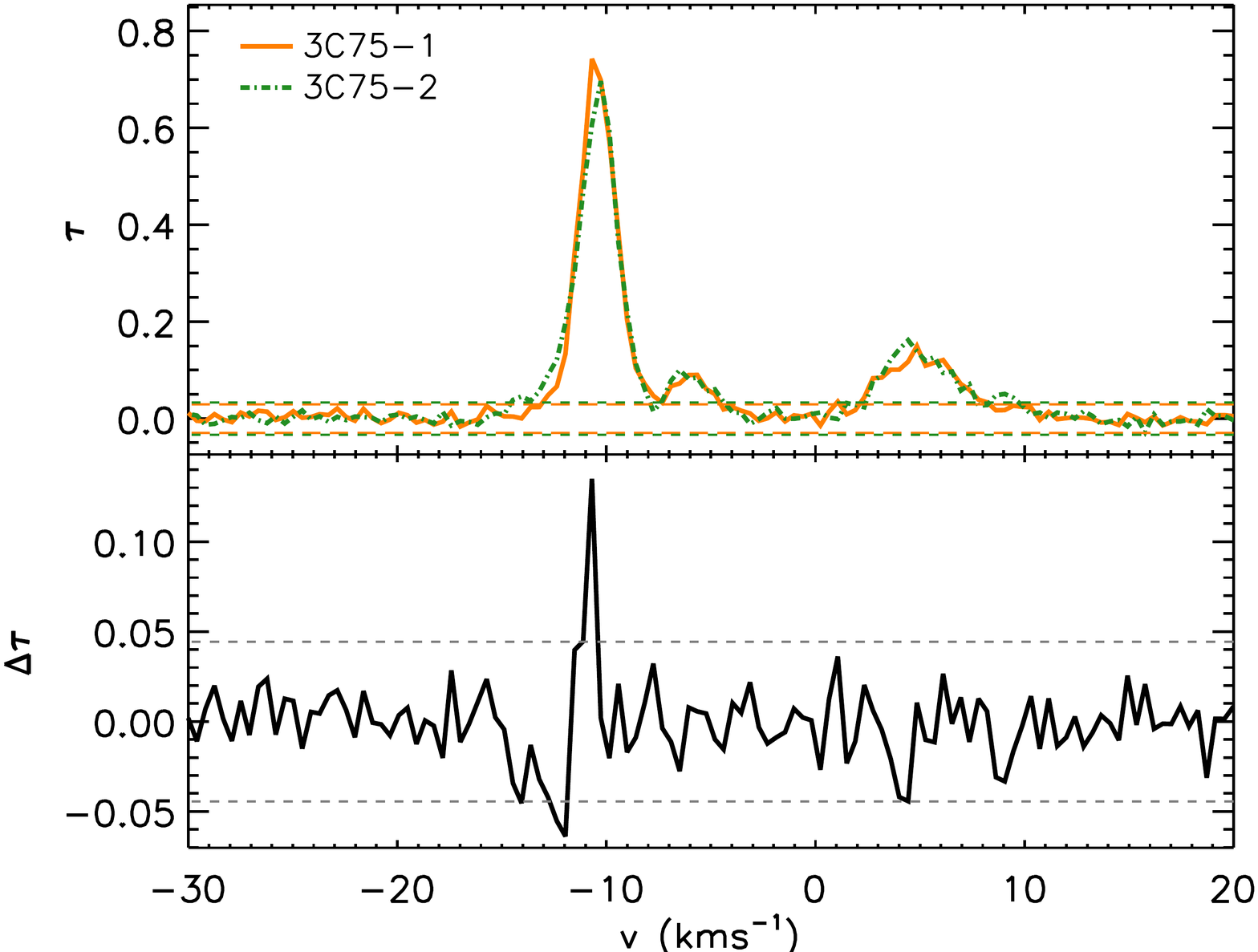}{0.5\textwidth}{3C75 (Millennium)}
          \fig{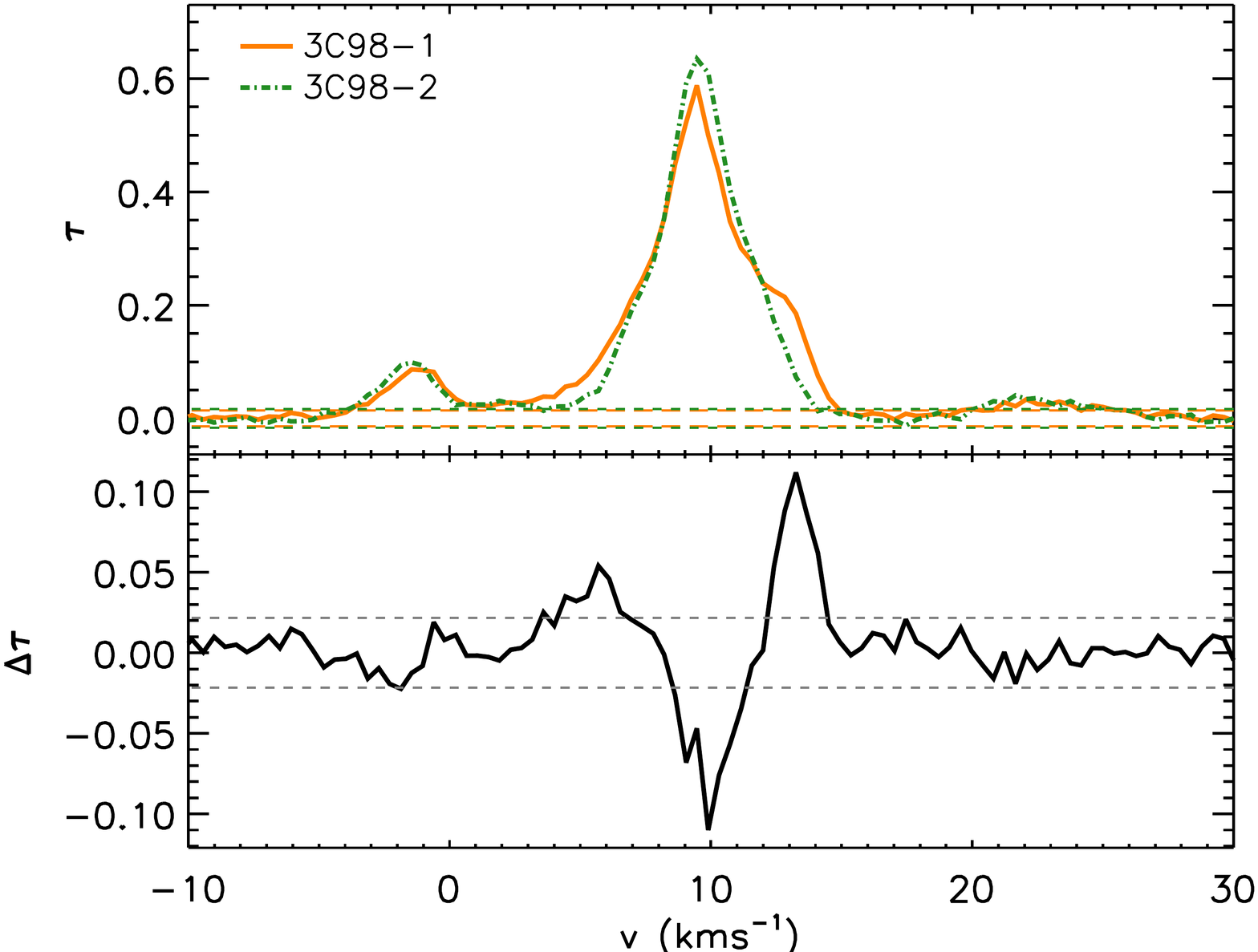}{0.5\textwidth}{3C98 (Millennium)}}
\caption{The optical depth, $\tau(v)$, for both components (top panels), and the change in optical depth, $\Delta \tau(v)$, (bottom panels) for the eleven component pairs from 21-SPONGE and the three component pairs from the Millennium Arecibo Survey multiple-component sources considered in this work. 3$\sigma$ uncertainties are shown as dashed lines for $\tau(v)$ (both components) and for $\Delta \tau(v)$.}
\label{fig:allspectra}
\end{figure*}

\subsection{Integrated properties} \label{subsec:integratedproperties}
For each source, we calculate the equivalent width of the change in optical depth, 
\begin{equation}
    \Delta EW = \int \Delta \tau(v) dv.
    \label{eq:ew}
\end{equation}
We also calculate a modified version of the equivalent width,
\begin{equation}
    \Delta EW_{abs,3\sigma} = \int \Big(\left| \Delta \tau(v) \right| - 3\sigma_{\Delta\tau}\Big)\,
                \delta(\Delta\tau \geq 3\sigma_{\Delta\tau}) \, dv,
        \label{eq:EWabs}
\end{equation}
where
$$
\delta(\Delta\tau \geq 3\sigma_{\Delta\tau})= \left\{
        \begin{array}{ll}
            0 & \quad \mathrm{if} \left|\Delta\tau(v)\right| < 3\sigma_{\Delta \tau}\\
            1 & \quad \mathrm{if} \left|\Delta\tau(v)\right| \geq 3\sigma_{\Delta \tau}
        \end{array}
    \right.
$$
for each velocity channel. This modified equivalent width has the advantage of counting both positive and negative changes in \hi{} optical depth, whereas positive and negative values can cancel out using Equation \ref{eq:ew}. For the 21-SPONGE spectra, Equations \ref{eq:ew} and \ref{eq:EWabs} are evaluated over the entire velocity range. For the three Millennium survey sources, they are evaluated from $-60$ \kms{} to 60 \kms{}, comparable to the 21-SPONGE velocity range.

\begin{figure}
    \centering
    \includegraphics[width=\columnwidth]{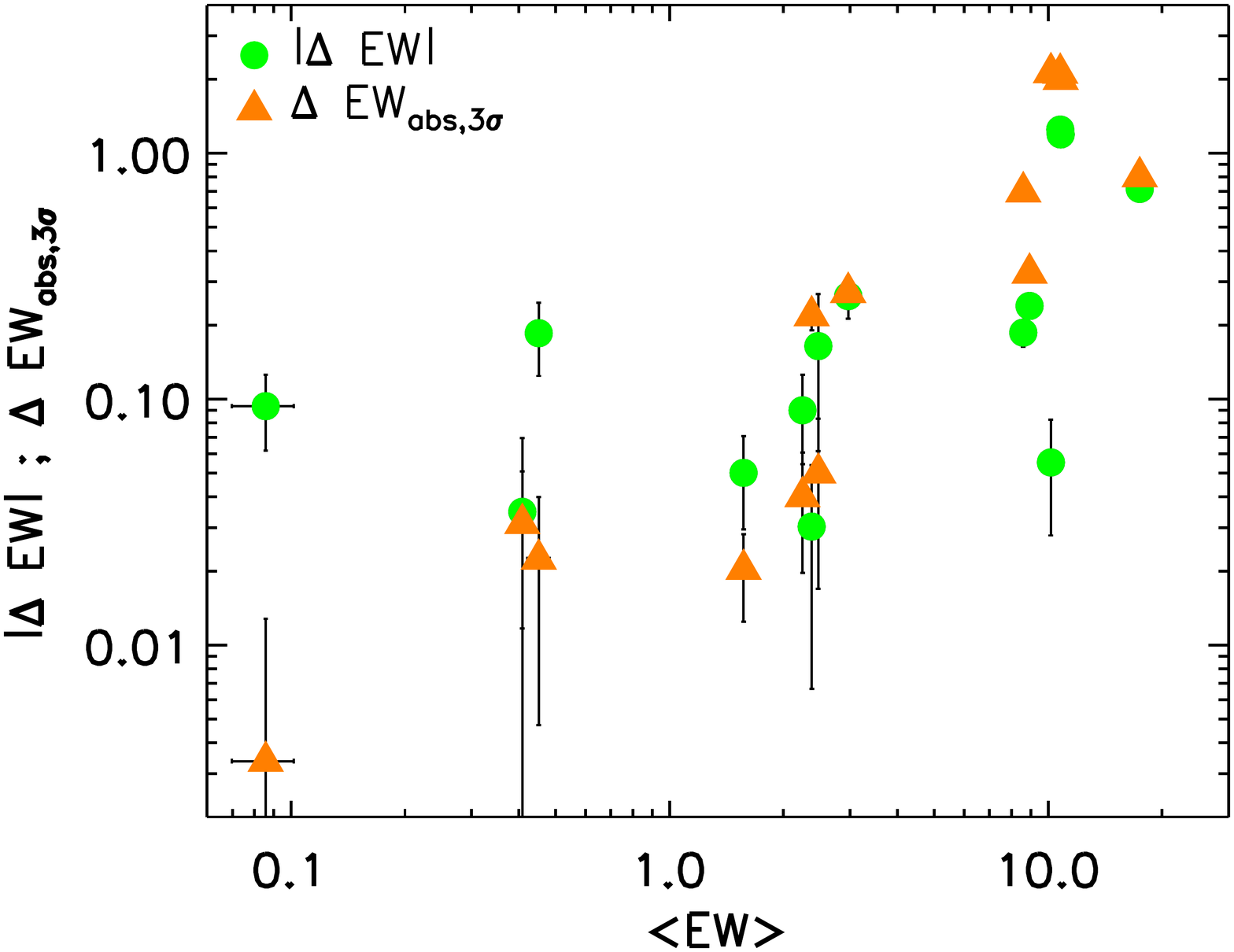}
    \caption{The absolute value of the change in $EW$ (green; Equation \ref{eq:ew}) and $EW_{abs,3\sigma}$ (orange; Equation \ref{eq:EWabs}) versus the average equivalent width, $\langle EW \rangle=\int\frac{\tau_1(v)+\tau_2(v)}{2}dv$, for each component pair. This is analogous to Figure 4 of \cite{1994ApJ...436..144F} and Figure 4 of \cite{2010ApJ...720..415S}.}
    \label{fig:EW_v_avgEW}
\end{figure}

Both $\left|\Delta EW\right|$ and $\Delta EW_{abs,3\sigma}$ tend to be higher for sources with higher average equivalent width, $\langle EW \rangle=\int\frac{\tau_1(v)+\tau_2(v)}{2}dv$ (Figure \ref{fig:EW_v_avgEW}). 
This suggests that directions with significant amount of cold \hi{} (often located at low Galactic latitudes) are more likely to host TSAS.
This is consistent with what \cite{1994ApJ...436..144F} and \cite{2010ApJ...720..415S} found in their multi-epoch studies of \hi{} absorption against pulsars. 

However, the fractional changes in the optical depth equivalent width in our sample, $\left|\Delta EW \right|/\langle EW \rangle$, are somewhat larger than those found by \cite{2010ApJ...720..415S} at much smaller scales ($6-50$ AU). Here, $\left|\Delta EW \right|/\langle EW \rangle$ ranges from $<1\%$ to $>100$\% with a median of $6.6\%$. The fractional variation of our modified equivalent width, $\Delta EW_{abs,3\sigma}/\langle EW \rangle$, ranges from $\sim2\%$ to $\sim20\%$, with a median of $7.6\%$, which is still larger than what was found by \cite{2010ApJ...720..415S}, $\sim4\%$. This may suggest that the fractional variation of EW is larger when probing larger angular/linear separations.
While we do not know the distance to the absorbing \hi{}, based on Table \ref{tab:knownsources}, most of our multiple-component sources likely probe linear separations of $\sim10^3$ to $\sim10^5$ AU. This estimate comes from assuming scale height for the CNM of 100 pc (e.g., \citealt{1975ApJ...198..281B,1978A&A....70...43C}) and calculating distance as $D=100$ pc$/ \sin |b|$.
The sources that show the highest fractional variation are 3C245 (110\%) and 3C33 (41\%), but the fractional variation of $\Delta EW_{abs,3\sigma}$ for these sources is typical (4\% and 5\%, respectively). 3C111, one of our low latitude sources (which is in the direction of the Taurus molecular cloud) shows the highest fractional variation of $\Delta EW_{abs,3\sigma}$, about 20\% between any two components.

In summary, directions with a significant amount of cold \hi{} show more variation in $EW$ than directions with a lower cold \hi{} fraction. As most of our sources likely probe 
linear scales larger than what is commonly probed with pulsars or VLBI imaging, our results may suggest that larger fractional variation in $EW$ is found on larger spatial scales.

\subsection{Channel-by-channel properties} \label{subsec:channelvariations}

\begin{figure}
    \centering
    \includegraphics[width=\columnwidth]{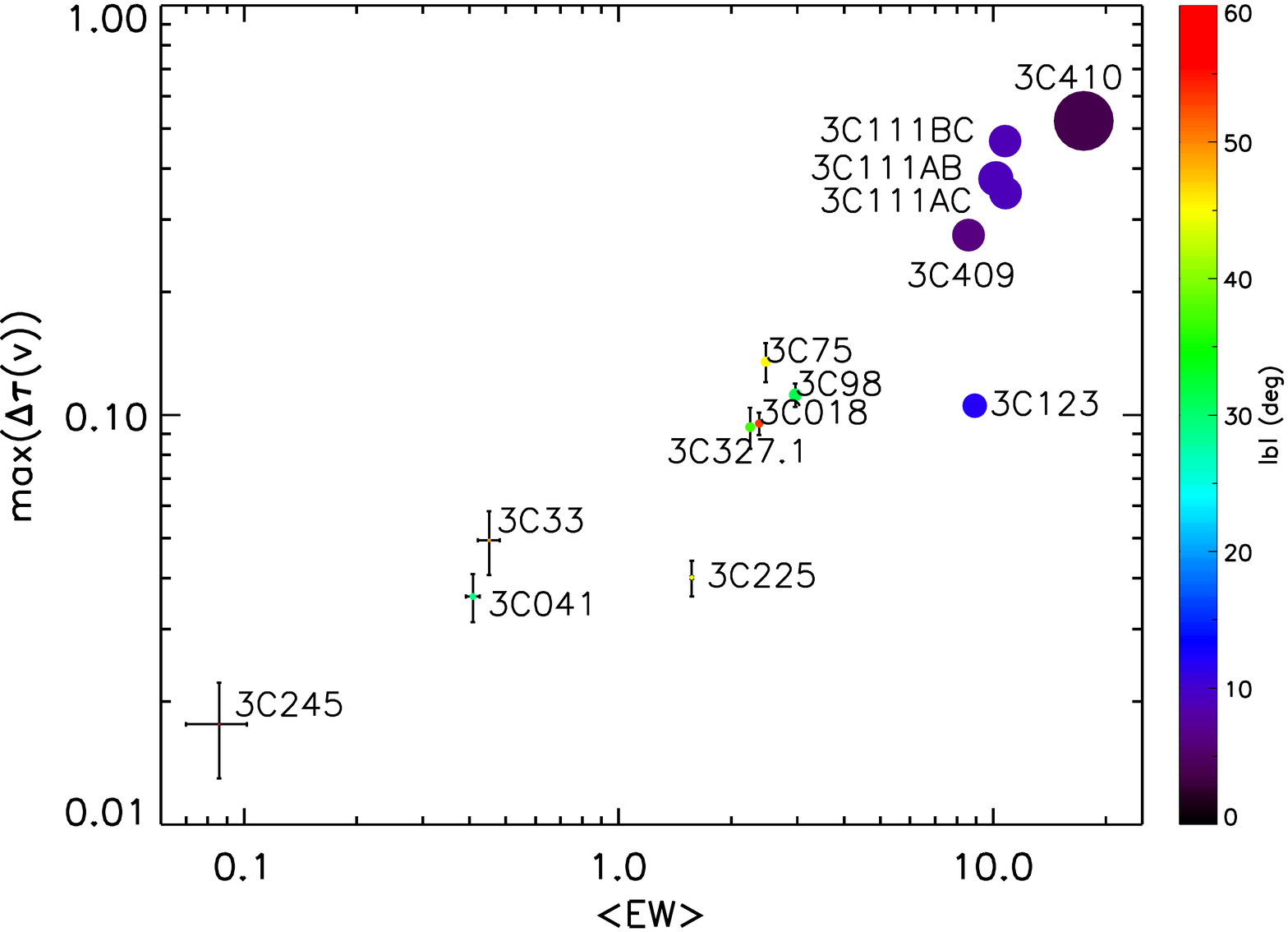}
    \caption{The maximum change in \hi{} optical depth versus the average equivalent width for all fourteen component pairs. Points are colored according to Galactic latitude and sized according to total \hi{} column density, $N(\hi{})_{\mathrm{tot}}$. $N(\hi{})_{\mathrm{tot}}$ ranges from $2.1\times10^{20}$ \persc{} to $4.8\times10^{21}$ \persc{}.}
    \label{fig:maxdtau_v_ew_nhi}
\end{figure}

A more fine-grained analysis of optical depth variation can be done by considering $\Delta \tau(v)$ for each velocity channel. In Figure \ref{fig:maxdtau_v_ew_nhi}, we show the maximum change in \hi{} optical depth, $|\max\{\Delta \tau(v)\}|$, versus the average \hi{} optical depth equivalent width , $\langle EW \rangle$, for all component pairs. The color in this figure shows the Galactic latitude and the sizes of the points indicate the total \hi{} column density, $N(\hi{})_{\mathrm{tot}}$, from \cite{2018ApJS..238...14M} and \cite{2003ApJ...586.1067H}.
The maximum variations in \hi{} optical depth for these sources show a significant correlation with $\langle EW \rangle$, in agreement with Figure \ref{fig:EW_v_avgEW}. This figure is especially striking as it covers a range in maximum optical depth variations from $\sim0.03$ to $\sim0.5$, a factor of $\sim15$, demonstrating that our sources are probing diverse  interstellar environments. 
Sources with the highest optical depth variations are those at low Galactic latitudes (3C111, 3C123, 3C409, and 3C410 are all at $|b|<12^{\circ}$). They also have the highest \hi{} column densities (Figure \ref{fig:maxdtau_v_ew_nhi}) and tend to have the highest CNM fractions, defined as the ratio of the CNM column density to the total column density, $f_{\mathrm{CNM}}=N(\hi{})_{\mathrm{CNM}}/N(\hi{})_{\mathrm{tot}}$.
This again indicates that regions at low latitudes with the highest amount of cold \hi{} tend to show the most \hi{} optical depth variation.
A similar trend was found by \cite{1986ApJ...303..702G} for the multiple-component sources 3C111 and 3C348. 

We note that \cite{2018ApJS..238...14M} and \cite{2003ApJ...586.1067H} also estimated the fraction of thermally unstable WNM along each line of sight, although often these estimates have large uncertainties. We do not find a correlation between the magnitude of the optical depth variation and the fraction of thermally unstable WNM.

\begin{figure}
    \centering
    \includegraphics[width=\columnwidth]{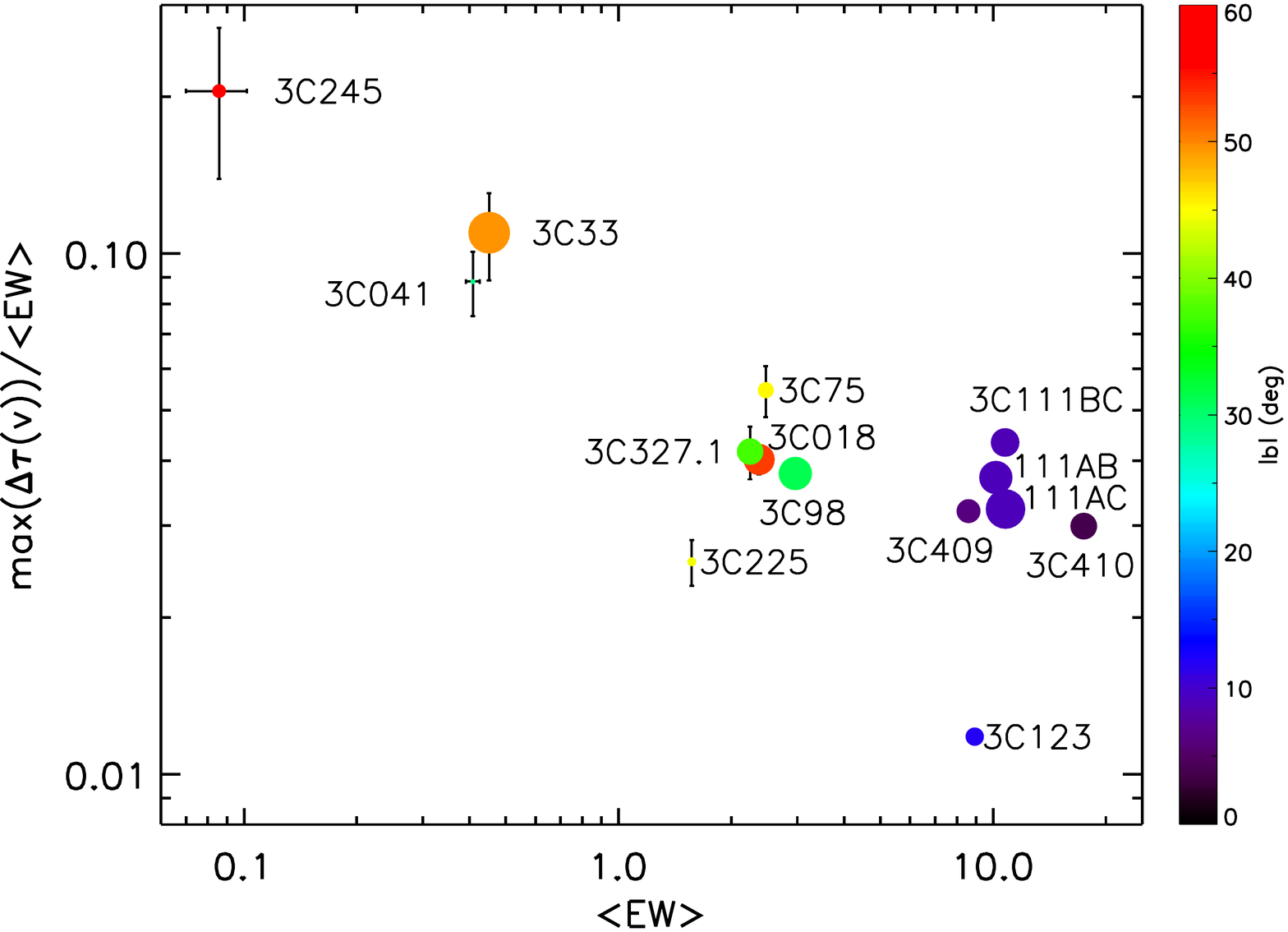}
    \caption{The maximum change in optical depth as a fraction of the of the average equivalent width versus the average equivalent width for all fourteen component pairs. Points are colored according to Galactic latitude and sized according to their CNM fractions, $f_{\mathrm{CNM}}=N(\hi{})_{\mathrm{CNM}}/N(\hi{})_{\mathrm{tot}}$. $f_{\mathrm{CNM}}$ ranges from 0.0 to 0.67.}
    \label{fig:maxdtauratio_v_ew_cnm}
\end{figure}

The maximum change in optical depth as a fraction of the average equivalent width, $\max\{\Delta \tau(v)\}/\langle EW \rangle$, is the greatest for sources with lowest optical depth and at the highest Galactic latitudes (Figure \ref{fig:maxdtauratio_v_ew_cnm}). The point sizes in Figure \ref{fig:maxdtauratio_v_ew_cnm} indicate the CNM fraction, discussed above.  
The sources that show the highest fractional variations are 3C245, 3C33, and 3C041, with fractional changes of 0.21, 0.11, and 0.09, respectively. All three sources  probe \hi{} environments with a very low peak \hi{} optical depth ($<0.1$) and relatively low total \hi{} column density ($\lesssim5\times10^{20}$ \persc{}); two of the three sources (3C245 and 3C041) have a very low CNM fraction ($f_{\mathrm{CNM}}\lesssim0.1$; \citealt{2018ApJS..238...14M}); and all three have an optical depth feature that appears only in one of two lines-of-sight. 
In the optical depth spectrum of 3C041, the TSAS feature at $-30$ \kms{} appears only toward the A component and is isolated from the main line in velocity, so clearly stands out in $\Delta\tau/\tau$.
3C245, although with a lower signal-to-noise, has a similar TSAS feature at $-20$ \kms{} which appears only in one line of sight. 
In the case of 3C33, which also has low signal-to-noise, the main line at 0 \kms{} significantly changes between two components due to appearance of a TSAS feature. 
For all other background sources, $\max\{\Delta \tau(v)\}/\langle EW \rangle\sim3$--5\%. The low latitude source 3C123 has the lowest fractional change, $\sim1\%$.

In agreement with Figure \ref{fig:EW_v_avgEW}, then, we see that TSAS is more abundant where the \hi{} optical depth, the \hi{} column density, and the CNM fraction are higher. However, the fraction of the \hi{} optical depth that is occupied by TSAS is the largest and the most prominent where the \hi{} optical depth (and the total \hi{} column density) are the lowest.
\cite{1993PhDT........56B}, \cite{2007ApJ...656...73L}, and \cite{2013MNRAS.434..163H} found a similar trend in the temporal variations of narrow quasar absorption lines in external galaxies.
While investigating several possible reasons for the variability of absorption lines, including bulk motions of the absorbers and the changing levels of ionization, \cite{2007ApJ...656...73L} concluded that the most likely explanation is the change in the covering factor of TSAS (their Figure 2). 
Essentially, for the lines of sight with a large $EW$, TSAS occupies only a small fraction of the absorption profile and is harder to distinguish than in the directions with a lower optical depth and $EW$, where TSAS stands out and is much easier to notice.

\section{Gaussian Analysis of Features} \label{sec:gaussians}

\cite{2003ApJS..145..329H} and \cite{2018ApJS..238...14M} fit Gaussian functions to the \hi{} absorption and emission spectra, deriving the peak optical depth ($\tau_0$), the central velocity ($v_0$), the FWHM (\fwhm{}), the spin temperature (\ts{}), and the \hi{} column density ($N(\hi{})$) of the \hi{} structure associated with each feature. Here we investigate the variations in these fitted quantities across the multiple-component sources to isolate the root of the variations seen in global measures like $EW$.

We note that Gaussian fitting is not unique, and is especially challenging for lines of sight with multiple \hi{} structures that are close in velocity space (e.g., the low latitude sources; see Figure \ref{fig:allspectra} and discussion in Section \ref{subsec:gaussianfits}). However, because Gaussian decomposition is useful to investigate important physical characteristics of TSAS---including its thermal and turbulent properties, discussed below---we proceed with the assumption that the Gaussian fits from \cite{2003ApJS..145..329H} and \cite{2018ApJS..238...14M} adequately describe the true properties of the \hi{} structures along the line of sight. The Gaussian fitting also implicitly assumes that the linewidths of absorbing \hi{} structures are larger than the velocity resolution, $0.42$ \kms{}. If linewidths are narrower than the velocity resolution, then the fitted Gaussian components may span multiple discrete structures.

\subsection{Matching Components from Different Lines of Sight} \label{subsec:matching}

Because the lines of sight to the different lobes of multiple-component galaxies are so close, we expect to see many of the same features in their optical depth spectra (i.e., we expect the same Galactic \hi{} structures to absorb radiation along both lines of sight). For components fitted to adjacent lines of sight to be considered a unique feature  (the same absorbing structure), we demand that 
\begin{equation} \label{eq:delta}
    \delta_{v 1,2} \equiv \frac{\left|v_{0,1}-v_{0,2}\right|}{\fwhm{}_{,1}/2.355} 
    \leq 1,
\end{equation}
where the subscripts 1 and 2 indicate the two separate lines of sight. We also require that $\delta_{v 2,1} \leq 1$, essentially requiring that the central velocities are within one standard deviation of the linewidths. If multiple features in the adjacent line of sight satisfy these criteria for a given feature, we select the feature that minimizes $\langle \delta \rangle = (\delta_{v1,2}+\delta_{v2,1})/2$ as the match.

We identify 54 matches in total. We shall henceforth refer to these as the matched features. For 8 of the matched features, we do not have information about $N(\hi{}$) or \ts{} (these were the features fitted with the unrealistic $T_S<10$ K; see Section \ref{subsec:gaussianfits}).

For 56 features, we do not find a matching feature in the neighboring line of sight (i.e., these features appear along only one line of sight). We shall refer to these as the unmatched features. The unmatched features are particularly interesting candidates for TSAS, since the sizes of the absorbing structures appear to be smaller than the linear separation of the \hi{} between the two lines of sight (i.e., $\lesssim 10^5$ AU). 
Unfortunately, though, for 28 of these features, we do not have $N(\hi{})$ or \ts{}.
While some of these components could be artifacts of Gaussian fitting, for components where there is a large disparity in angular resolution between \hi{} emission and absorption \ts{} is impossible to constrain (e.g., \citealt{2015ApJ...804...89M}).

Table \ref{tab:allgausstable} lists the properties of the fitted Gaussian components for all matched and unmatched features from \cite{2018ApJS..238...14M} and \cite{2003ApJS..145..329H}. In the following scatter plots, the data corresponding to unmatched features are shown as black triangles. The data corresponding to matched features are shown as circles with colors indicating the value of $\langle \delta \rangle$ (a lower $\langle \delta \rangle$ suggests a better match). Likewise, in the stacked histograms, the relative contributions of the unmatched and matched features are shown in green and orange, respectively.

\subsection{Variations of Gaussian Components} 

\subsubsection{Peak Optical Depth Variations}

\begin{figure}
    \centering
    \includegraphics[width=\columnwidth]{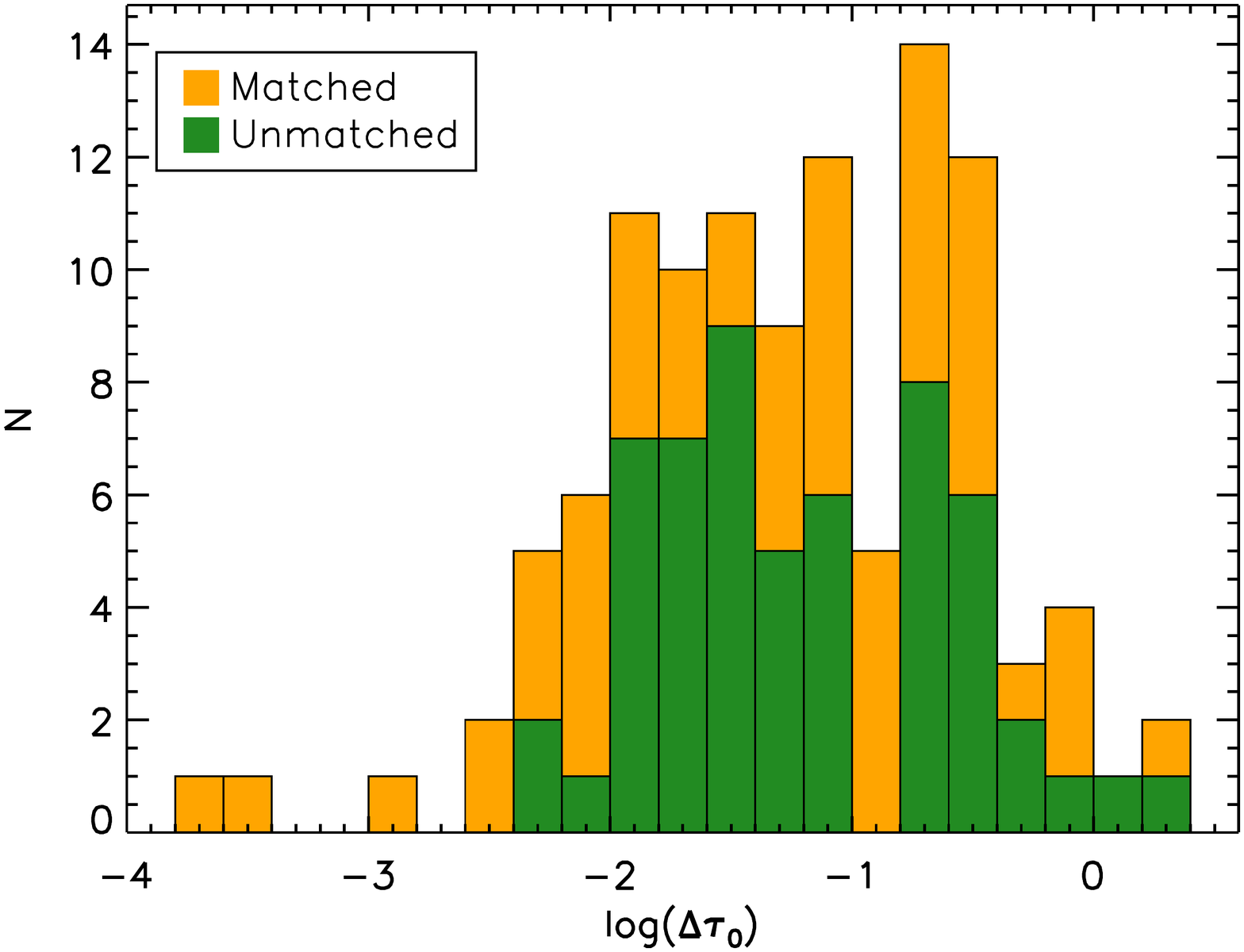}
    \caption{The distribution of optical depth variations, in $\log_{10}$, for all Gaussian features. For matched features, shown in orange, $\Delta \tau_{0}=|\tau_{0,1}-\tau_{0,2}|$. For unmatched features, shown in green, $\Delta \tau_{0}=\tau_{0}$.}
    \label{fig:dtau_hist_stacked}
\end{figure}

The distribution of $\Delta \tau_{0}$ (where $\Delta \tau_{0}=|\tau_{0,1}-\tau_{0,2}|$ for matched features and $\Delta \tau_{0}= \tau_{0}$ for unmatched features) is shown in Figure \ref{fig:dtau_hist_stacked}. A majority of the variations are small ($\Delta \tau < 0.1$), but there are a number of features with very large variations, including 8 with $\Delta \tau > 0.5$. Table \ref{tab:gauss_summary} compares the optical depth variations seen in this study to previous multiple-component studies (\citealt{1985A&A...146..223C,1986ApJ...303..702G}). We see a larger average change in the peak optical depth than these previous studies, but the velocity resolution of the absorption spectra used here is $\sim0.42$ \kms{}, which is a factor of 2--3 times higher than the velocity resolution of the previous studies. Moreover, these previous studies had non-uniform velocity resolution across different sources, whereas the data in this survey all have the same velocity resolution. If we smooth 21-SPONGE spectra to 2 \kms{} resolution, which is closer to the typical resolutions of \cite{1985A&A...146..223C} and \cite{1986ApJ...303..702G}, we find $\langle\Delta\tau\rangle\approx0.10$ (although fewer components are fitted). Similarly, although the average optical depth probed in this study is close to that of previous studies (Table \ref{tab:gauss_summary}), if we restrict our sample to features with optical depths $>0.1$, similar to their sensitivities, we find $\langle\tau\rangle\approx0.47$. We also see a larger spread in optical depth variations than previous studies, which is likely because we are sampling a larger, more diverse sample of interstellar environments, and because we are sensitive to weaker optical depth features.

\begin{figure}
    \centering
    \includegraphics[width=\columnwidth]{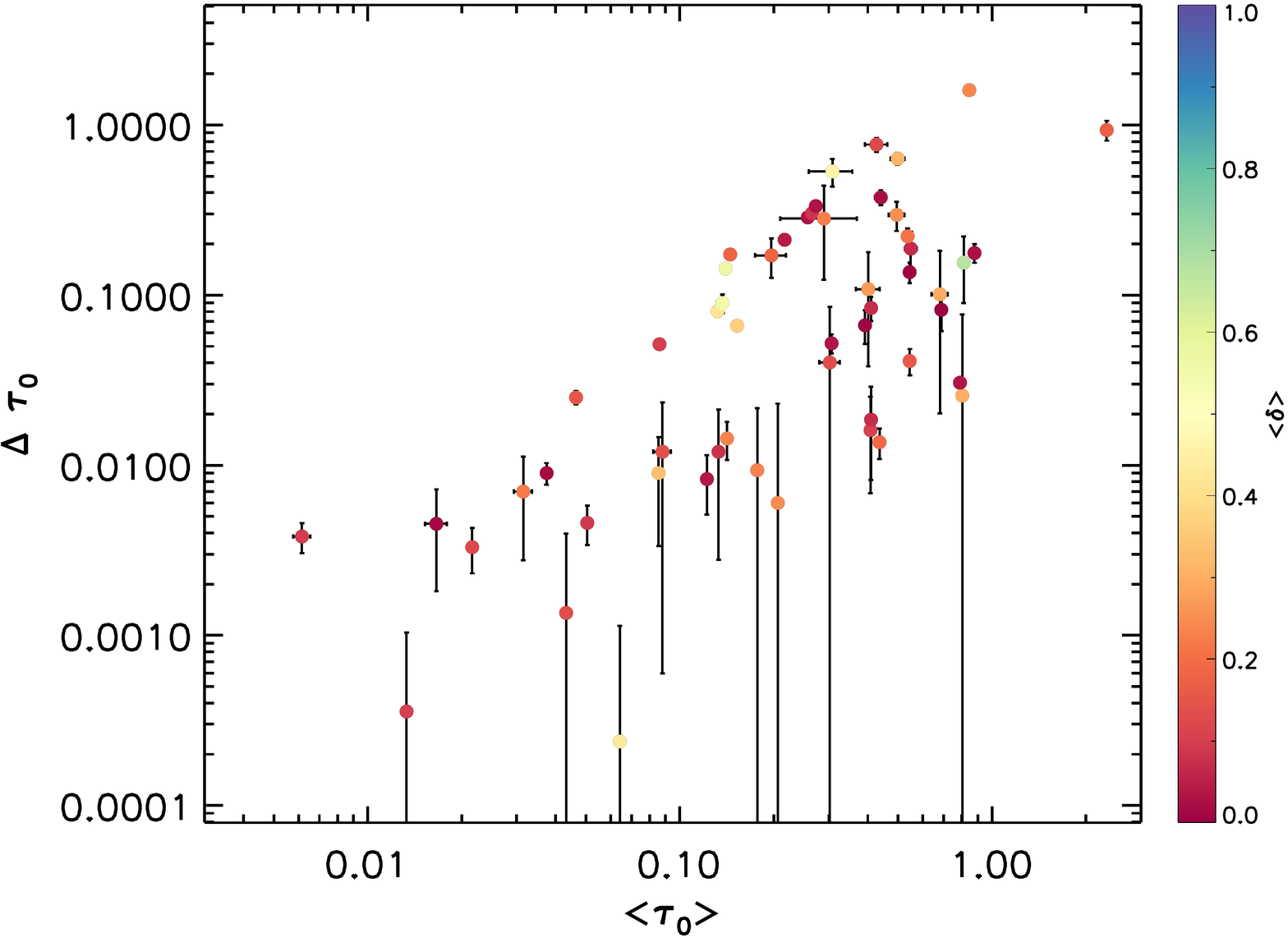}
    \caption{The change in peak optical depth, $\Delta \tau_0$, versus the average peak optical depth, $\langle \tau_0 \rangle$,  for all matched features. Colors indicate the value of $\langle \delta \rangle$ (see Section \ref{subsec:matching}). Unmatched features are not shown since all fall on a 1-to-1 line.}
    \label{fig:g_dtau_v_tau}
\end{figure}

Figure \ref{fig:g_dtau_v_tau} shows the change in optical depth, $\Delta \tau_0$, versus the average optical depth, $\langle \tau_{0}\rangle$, for all matched features. As in Figure \ref{fig:maxdtau_v_ew_nhi}, we find that the change in the peak optical depth of Gaussian features scales with the optical depth. TSAS is more prominent along lines of sight with higher \hi{} optical depth (i.e., lines of sight with more cold atomic gas).

\subsubsection{Central Velocity Variations}
\begin{figure}
    \centering
    \includegraphics[width=\columnwidth]{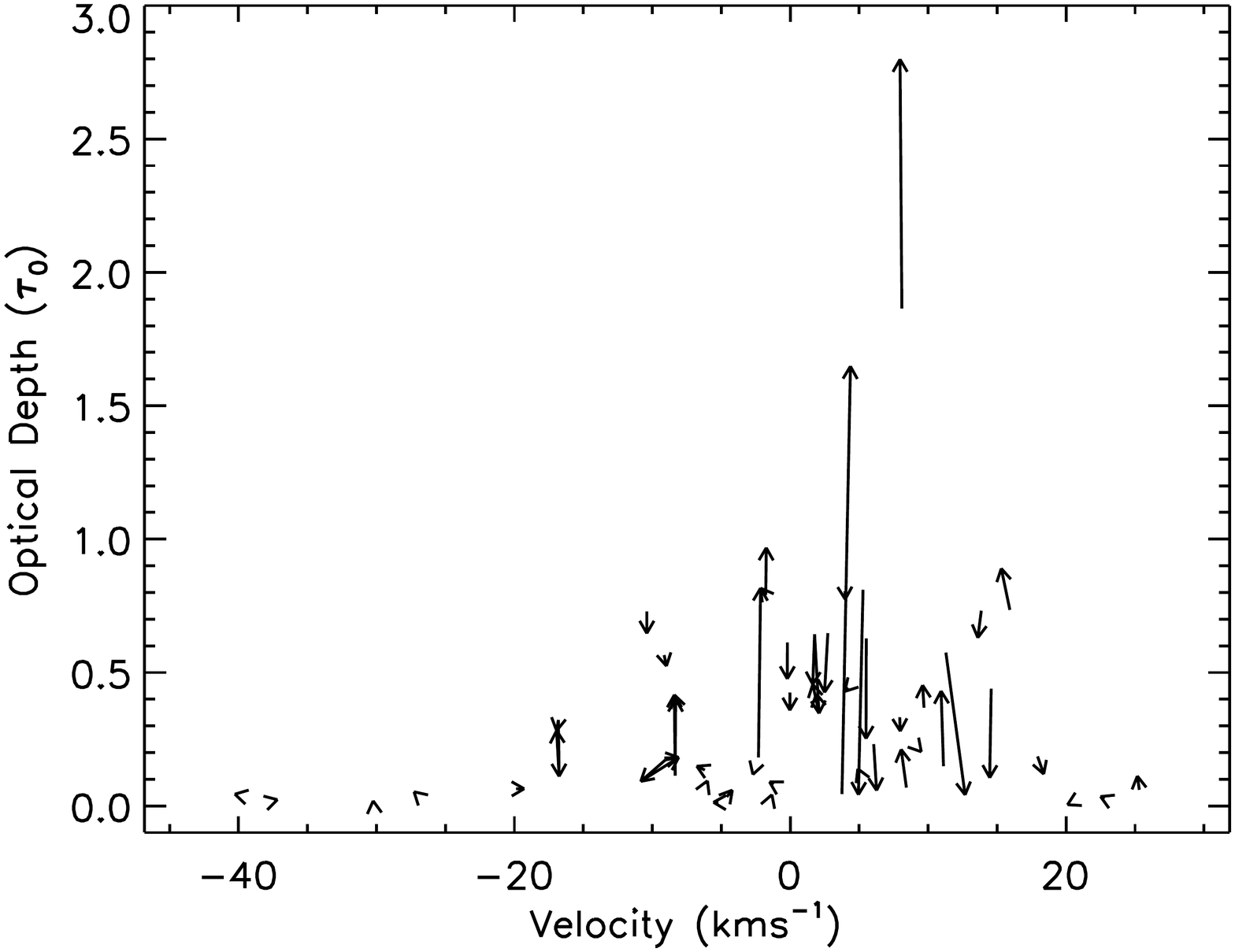}
    \caption{Analogous to Figure 6 of \cite{2007ApJ...656...73L}, the magnitude of the change in peak optical depth versus the central velocity for each matched feature. The direction of the arrows is arbitrary, pointing from Component 1 to Component 2, as listed in Table \ref{tab:sources}.}
    \label{fig:tau_v_v}
\end{figure}

The observed optical depth variations seen in Figure \ref{fig:allspectra} do not appear to be a product of changes in the central velocities of \hi{} structures. Figure \ref{fig:tau_v_v}, analogous to Figure 6 of \cite{2007ApJ...656...73L}, shows the magnitude of the change in peak optical depth, $\Delta \tau_0$, versus the central velocity, $v_0$, for each matched feature. It indicates that changes in $v_0$ tend to be small, while changes in $\tau_0$ can be dramatic. This is the same conclusion drawn by \cite{1986ApJ...303..702G}.
Table \ref{tab:gauss_summary} shows that the small changes we see in central velocities of fitted components are similar to those seen in previous multiple-component absorption studies. Moreover, if we downgrade the velocity resolution of the absorption spectra to 1--2 \kms{} and match components, we find that the changes in central velocity of matched components still tend to be $<0.4$ \kms{}. Neither the change in peak optical depth nor the change in optical depth equivalent width (Section \ref{subsec:dew_g}, below) are correlated with the change in central velocity.

\subsubsection{Velocity Width Variations} \label{subsec:linewidth_variability}
\begin{figure}
    \centering
    \includegraphics[width=\columnwidth]{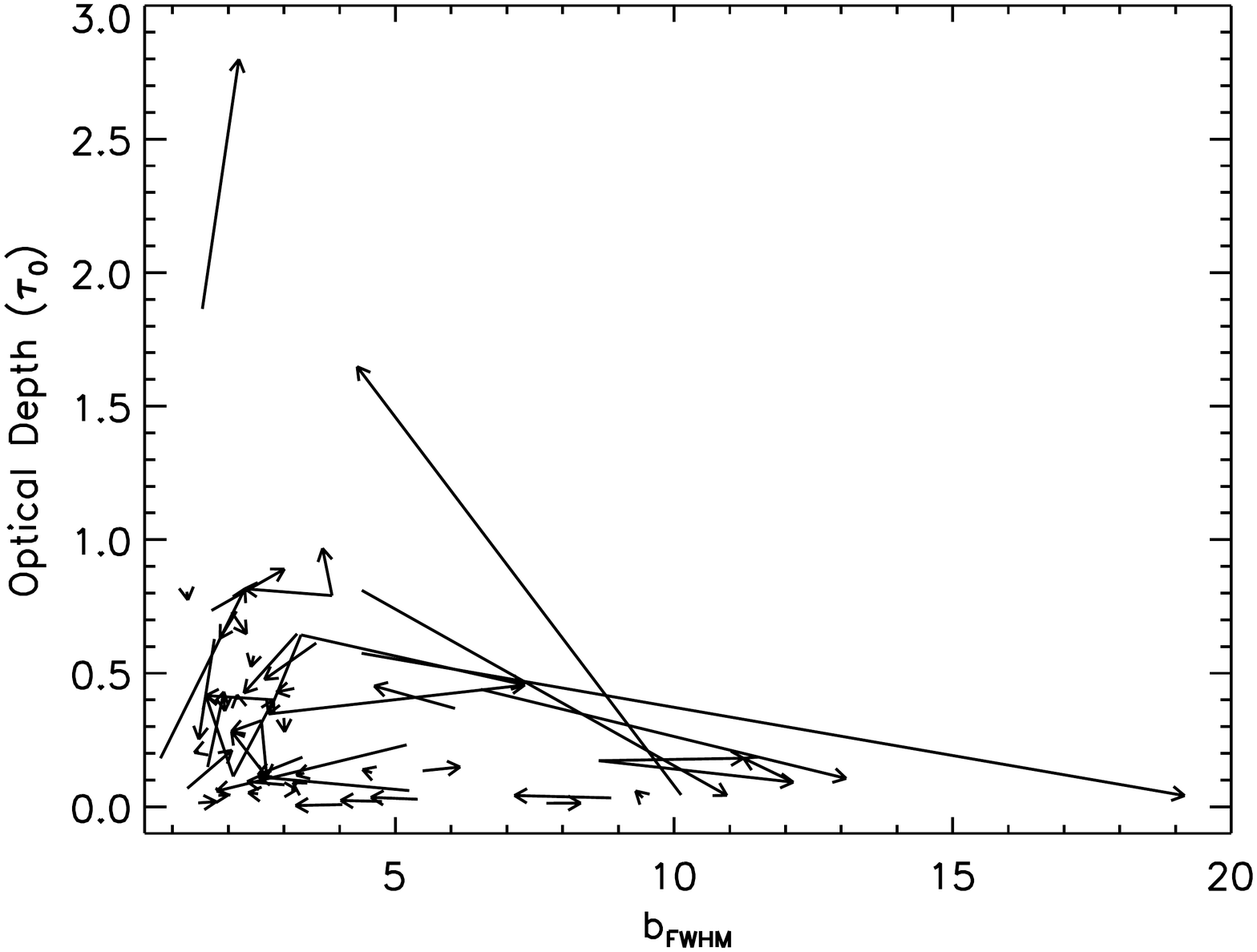}
    \caption{Analogous to Figure 5 of \cite{2007ApJ...656...73L}, the magnitude of the change in peak optical depth versus the linewidth for each matched feature. The direction of the arrows is arbitrary, pointing from Component 1 to Component 2, as listed in Table \ref{tab:sources}.}
    \label{fig:tau_v_fwhm}
\end{figure}

While changes in the central velocities of matched features tend to be small, changes in their linewidths can be considerable. Figure \ref{fig:tau_v_fwhm} shows that both the change in peak optical depth ($\Delta \tau_0$) and the change in the velocity width ($\Delta \fwhm{}$) can be large for a given feature---both  contribute to the observed optical depth variation. This helps explain the variation of the optical depth equivalent width, since both $\tau_0$ and \fwhm{} contribute to the equivalent width (Section \ref{subsec:integratedproperties} and Section \ref{subsec:dew_g}, below). 

\begin{figure}
    \centering
    \includegraphics[width=\columnwidth]{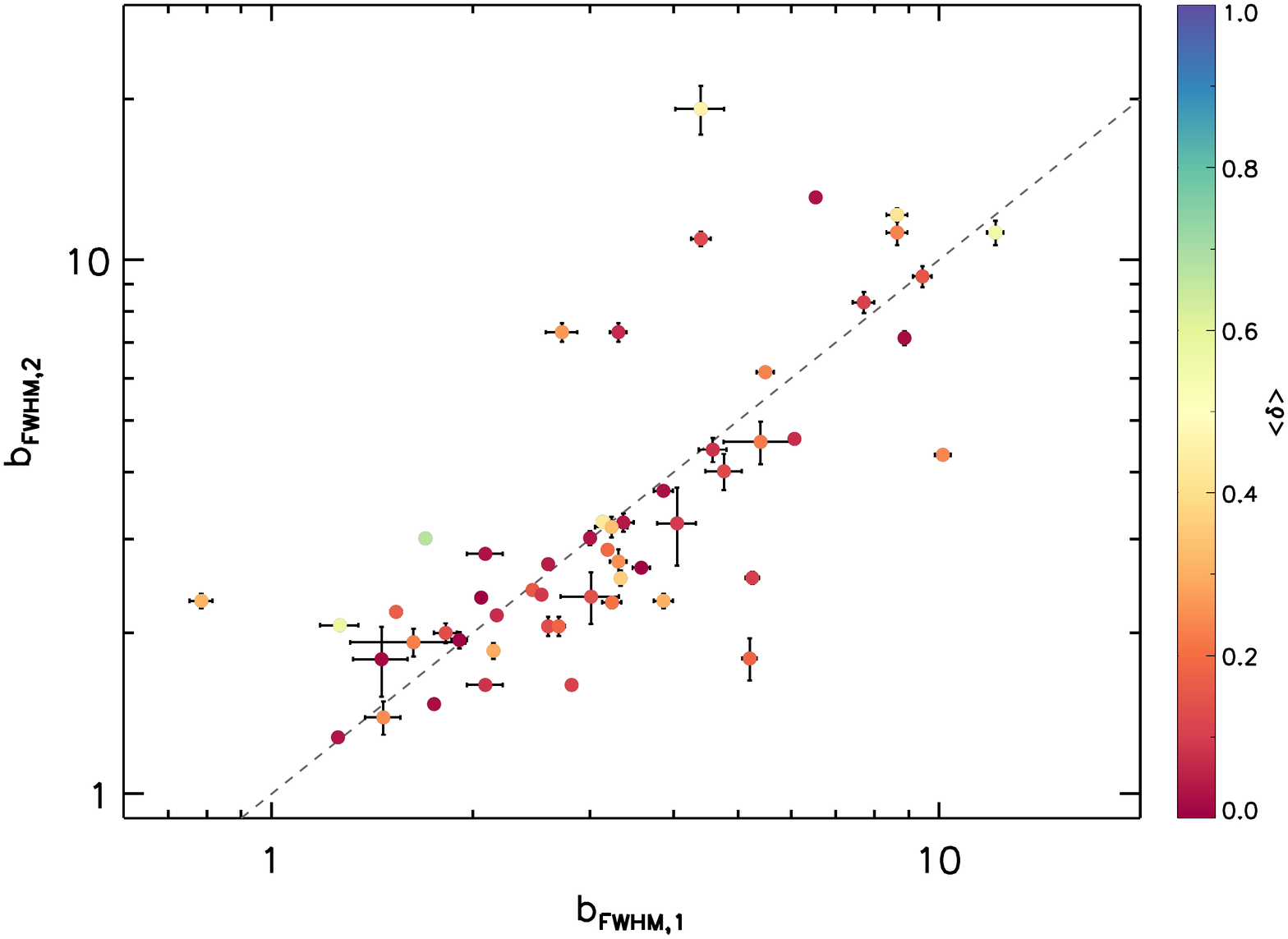}
    \caption{A comparison of the FWHMs of matched features, $\fwhm{}_{,2}$ versus $\fwhm{}_{,1}$. Colors indicate the average value of $\delta$ of the matched features (see Section \ref{subsec:matching}).}
    \label{fig:fwhm1_v_fwhm2}
\end{figure}

While there are features with drastic linewidth changes (five features with $\Delta\fwhm{}\sim4$--7 \kms{} and one outlier with $\Delta\fwhm{}\sim15$ \kms{}),
in most cases we find a smaller but clearly noticeable linewidth change of 1--2 \kms{}. This can be seen in Figure \ref{fig:fwhm1_v_fwhm2} which compares the two FWHMs of each matched feature, $\fwhm{}_{,1}$ and $\fwhm{}_{,2}$. The dashed line represents a 1-to-1 match. We see that extreme outliers tend to be more poorly matched, even though they satisfy our matching criterion (see Section \ref{subsec:matching}). A majority of components lie just below (or slightly above) the 1-to-1 line, offset by $\lesssim1$--2 \kms{}. As the velocity linewidth comes from both thermal motions and turbulent motions, we investigate both further (Sections \ref{subsec:dts_g} and \ref{subsec:turbulentproperties}, respectively). 

\subsubsection{Equivalent Width Variations} \label{subsec:dew_g}

\begin{figure} 
    \centering
    \includegraphics[width=\columnwidth]{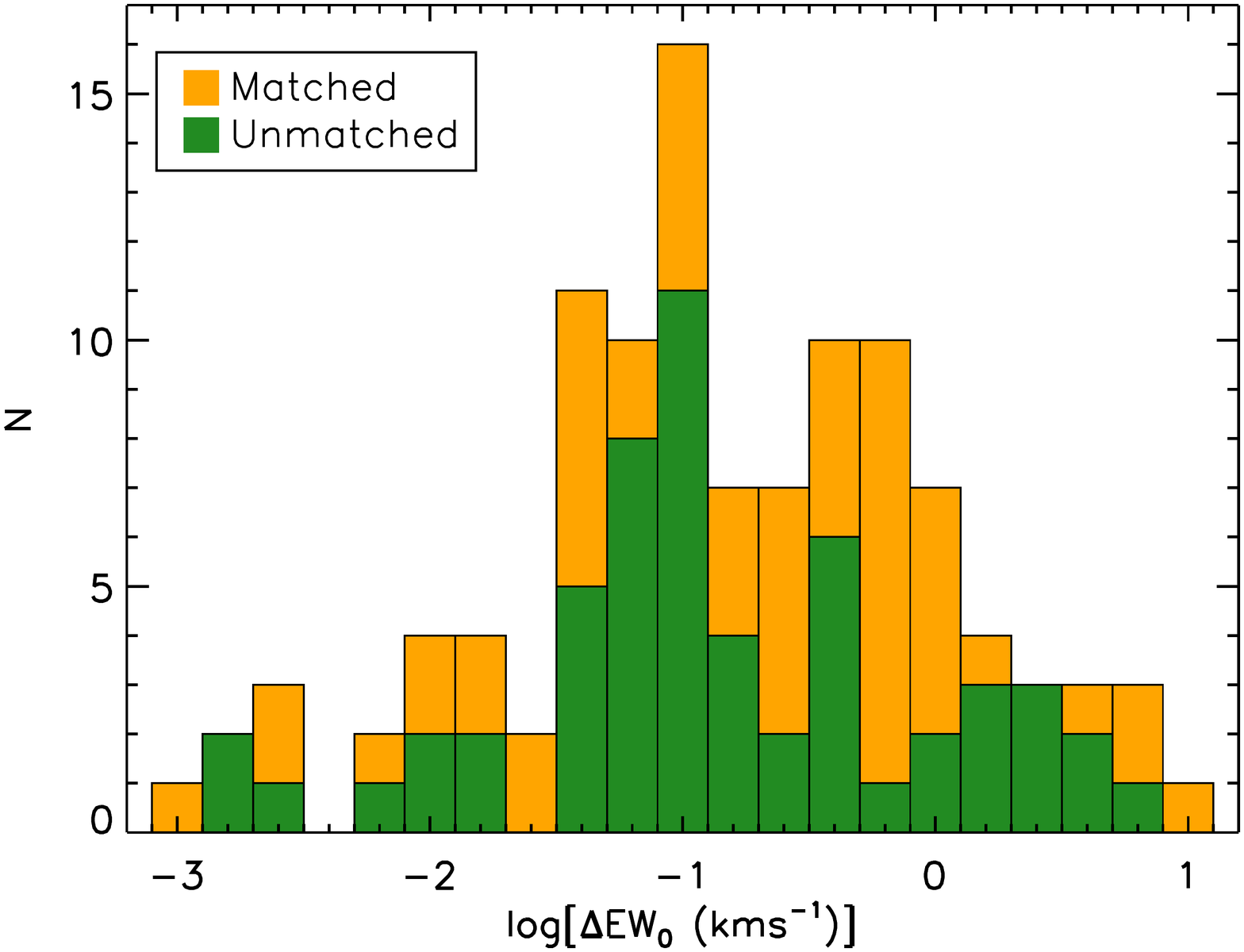}
    \caption{The distribution of optical depth equivalent width variations (Equation \ref{eq:ew0}), in $\log_{10}$, for all Gaussian features. For matched features, shown in orange, $\Delta EW_0=|EW_{0,1}-EW_{0,2}|$. For unmatched features, shown in green, $\Delta EW_0=EW_0$.}
    \label{fig:dew_hist_stacked}
\end{figure}

\begin{figure} 
    \centering
    \includegraphics[width=\columnwidth]{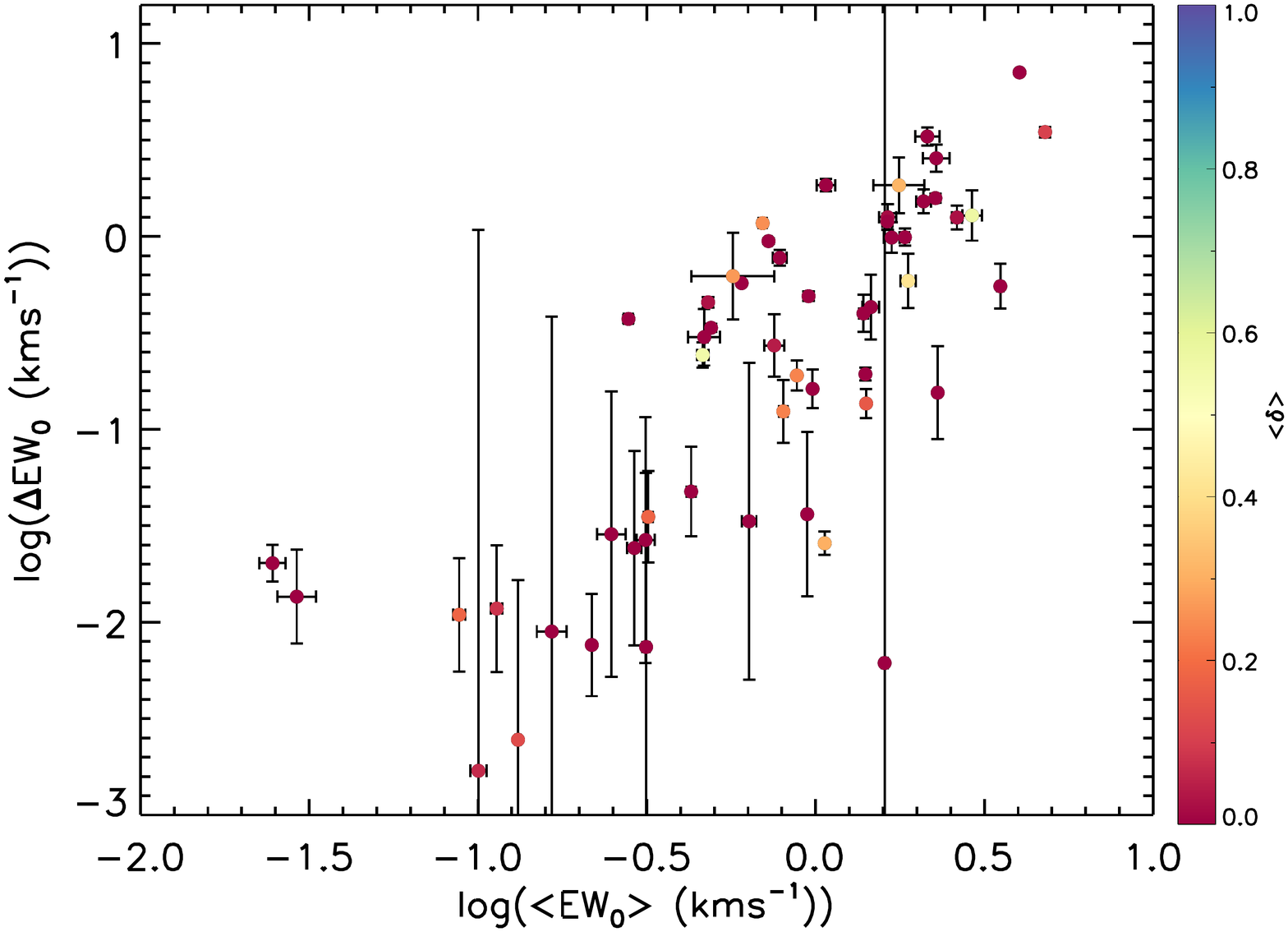}
    \caption{The change in \hi{} optical depth equivalent width, $\Delta EW_0$, as a function of the average optical depth equivalent width, $\langle EW_0\rangle$, for matched features. Points are colored based on the value of $\langle\delta\rangle$ (see Section \ref{subsec:matching}). Unmatched features are not shown since all lie on the 1-to-1 line.}
    \label{fig:dew0_v_ew0}
\end{figure}

The equivalent width of a given Gaussian feature is 
\begin{equation} \label{eq:ew0}
    EW_0 = 1.064 \times \tau_{0} \times \fwhm{}.
\end{equation} 
In Figure \ref{fig:dew_hist_stacked}, we show the distribution of the change in equivalent width ($\Delta EW_0= |EW_{0,1} - EW_{0,2}|$ for matched features and $\Delta EW_0=EW_0$ for unmatched features). 
This level of variation is consistent with the equivalent width variation seen in previous multiple-component absorption experiments (Table \ref{tab:gauss_summary}), although we see a larger spread in the variations. This is particularly noteworthy because the optical depth equivalent width does not depend on the velocity resolution, unlike, e.g., the peak optical depth.
Figure \ref{fig:dew0_v_ew0} indicates that greater equivalent width variations tend to be associated with higher average equivalent widths, $\langle EW_0 \rangle$, for matched features\footnote{We do not plot here unmatched features as they all lie on the 1-to-1 line}.
This is consistent with what we see in Figures \ref{fig:EW_v_avgEW} and \ref{fig:maxdtau_v_ew_nhi}, and agrees with the trend seen in narrow quasar absorption lines by \cite{2013MNRAS.434..163H}. The \hi{} optical depth variation is strongest toward lines of sight sampling more cold atomic gas (Section \ref{subsec:channelvariations}). As discussed above, this variation is likely a product of changes in both the peak optical depths and the linewidths.

\subsubsection{Spin Temperature Properties} \label{subsec:dts_g}

The spin temperature, which is a good estimate of the kinetic temperature for the CNM, was determined for most features in the 21-SPONGE survey (\citealt{2018ApJS..238...14M}) and the Millennium Arecibo 21-cm survey (\citealt{2003ApJS..145..329H}). 
However, as described in detail in \cite{2018ApJS..238...14M}, the 21-SPONGE survey estimated \ts{} using both the VLA \hi{} absorption spectra as well as \hi{} emission spectra from the Arecibo Observatory. The emission data therefore have a resolution of $\sim4\arcmin{}$, which is larger than the typical separation between multiple-component pairs in this study, meaning that our \ts{} estimates for the different components of each source were not from independent \hi{} emission data. Thus, we unfortunately do not have a clear probe of spin temperature variations across multiple components. Higher angular resolution \hi{} emission observations are needed to investigate the question of whether \ts{} changes significantly between multiple components.

If we neglect turbulent fluctuations and assume only thermal motions, though, we can place some constraints on how large temperature fluctuations are needed to explain the observed change in linewidth (Section \ref{subsec:linewidth_variability})\footnote{We discuss the contribution of turbulence to the linewidths in Section \ref{subsec:turbulentproperties}}. For example, for a Gaussian feature that is 3 \kms{} wide to change linewidth by 1--2 \kms{}, an increase in kinetic temperature of $\sim130$--260 K would be required.
For a Gaussian feature that is 8 \kms{} wide to change linewidth change by 1--2 \kms{}, we would need an increase in kinetic temperature of $\sim350$--700 K. For a typical CNM temperature of 50 K, this means that a significant temperature increase would be needed to explain the observed changes in linewidth. We cannot dismiss even this simplified scenario, though, as \cite{2007A&A...465..431H} showed that small, dense structures suggestive of TSAS form with a steep temperature gradient. For example, a structure $\lesssim10^5$ AU in size with a central temperature of $\sim200$ K may rise to $\sim4000$ K in the outskirts of the structure. The 3D magnetohydrodynamic (MHD) simulation of \cite{2012ApJ...759...35I}, in which \hi{} accretion flows lead to the formation of a highly inhomogeneous molecular cloud comprising dense cold clumps embedded in a warmer diffuse medium, found similarly strong temperature gradients on scales $<1$ pc (their Figure 4).

\begin{figure}
    \centering
    \includegraphics[width=\columnwidth]{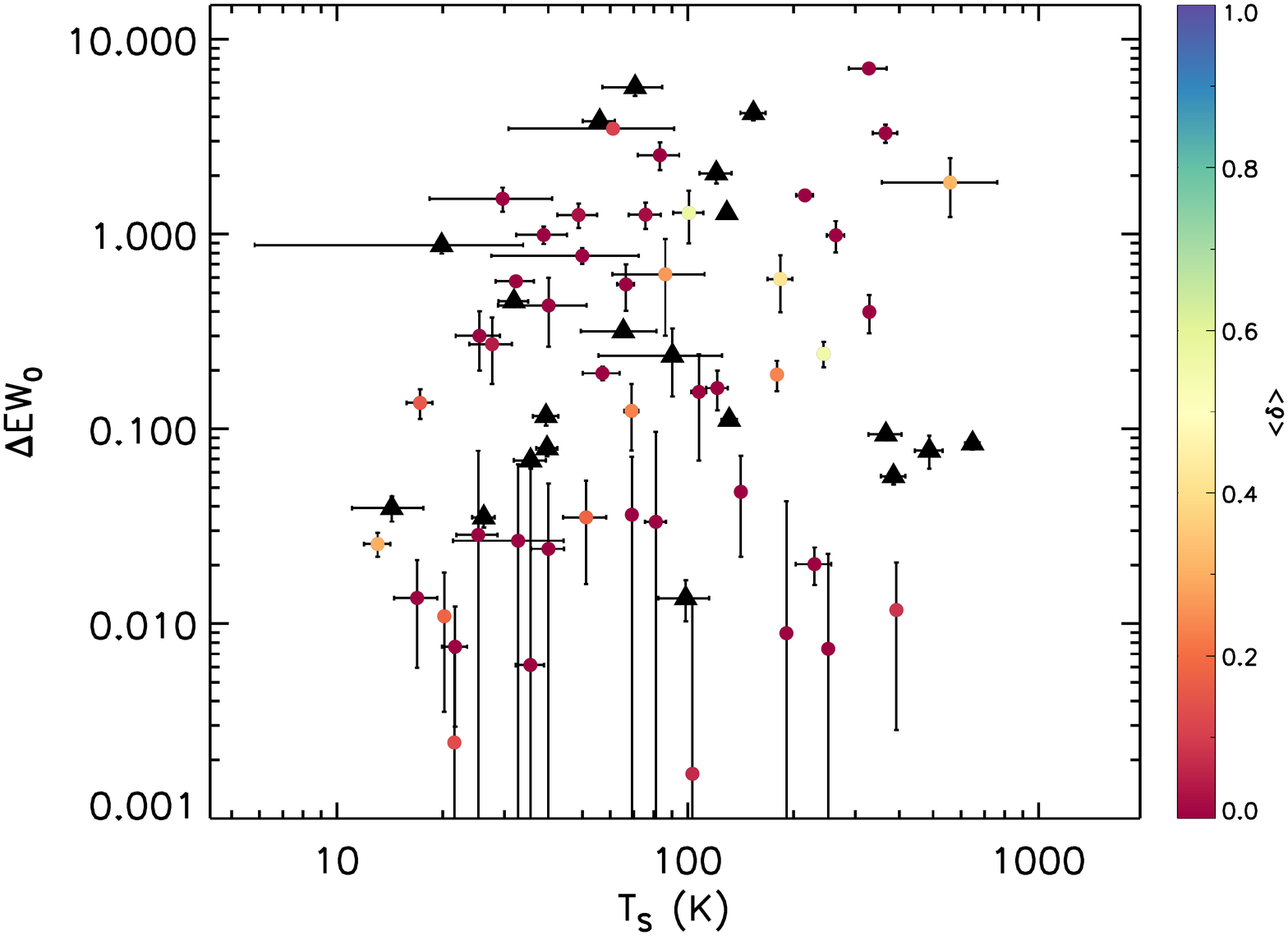}
    \caption{The change in optical depth equivalent width (Equation \ref{eq:ew0}) versus spin temperature, \ts{}. For matched features, $\Delta EW_0=|EW_{0,1}-EW_{0,2}|$ and the x-axis shows $\langle T_S \rangle$. Matched features are shown as circles with colors indicating the value of $\langle\delta\rangle$ (see Section \ref{subsec:matching}). For unmatched features, $\Delta EW_0=EW_0$ and the x-axis is \ts{}. Unmatched features are shown as black triangles.}
    \label{fig:dew_v_ts}
\end{figure}

We can also investigate whether Gaussian components with optical depth variations show unusual physical properties. 
For example, in a pulsar study of TSAS \cite{2010ApJ...720..415S} found that 
the spin temperature in the direction of B1929+10---the only source in their sample that showed repeated variations of \hi{} absorption spectra over time---was significantly higher (150--200 K) than what is typically found for the CNM (e.g., 20--70 K). The same line of sight has also a very low CNM fraction and is believed to be sampling the Local Bubble wall. 
In Figure \ref{fig:dew_v_ts} we show the difference in optical depth equivalent width (Equation \ref{eq:ew0}) as a function of spin temperature for both matched and unmatched features. For matched features, the x-axis shows the average spin temperature of the matched features, $\langle\ts{}\rangle$. For unmatched features, the x-axis is \ts{}.
While we do not see a correlation, it is clear from this plot that most of the variation is associated with the CNM ($T_S\lesssim 260$ K; \citealt{2003ApJ...587..278W}).
{}

\subsubsection{Turbulent properties} \label{subsec:turbulentproperties}
Turbulence has been suggested as a possible driver of TSAS formation, since compressible turbulence in the ISM produces density fluctuations and may even be sufficient to confine overpressured structures (\citetalias{doi:10.1146/annurev-astro-081817-051810}). Here, we are able to measure the turbulent velocity of the Gaussian features, $\sigma_{v_0,t}$, which allows us to compare the thermal pressure and turbulent pressure of TSAS structures and to test the influence of turbulence on the observed optical depth variations. The 1D turbulent velocity for each component is given by 
\begin{align}
    \sigma_{v_0,t}^2 & = \sigma_{v_0}^2 -k_BT_S/m_\mathrm{H} \nonumber \\
    & = \Bigg(\frac{b_{\mathrm{FWHM}}}{2.355}\Bigg)^2-k_BT_S/m_\mathrm{H},
    \label{eq:pturb}
\end{align}
where $m_\mathrm{H}$ is the mass of the hydrogen atom, and the spin temperature, \ts{}, is a proxy for the kinetic temperature, $T_{kin}$, which is reliable for the CNM (this is equivalent to Equation 16 of \citealt{2003ApJ...586.1067H}). 
Because $\sigma_{v_0,t}$ depends on the temperature, we cannot reliably measure the change in turbulent velocity between matched features (see Section \ref{subsec:dts_g}). Nevertheless, we can estimate a characteristic turbulent velocity, $\langle\sigma_{v_0,t}\rangle$, the average 1D turbulent velocity between the two matched components, evaluated with Equation \ref{eq:pturb} using $\fwhm{}_{,1}$ and $\ts{}_{,1}$ for the first component and $\fwhm{}_{,2}$ and $\ts{}_{,2}$ for the second component.

\begin{figure}
    \centering
    \includegraphics[width=\columnwidth]{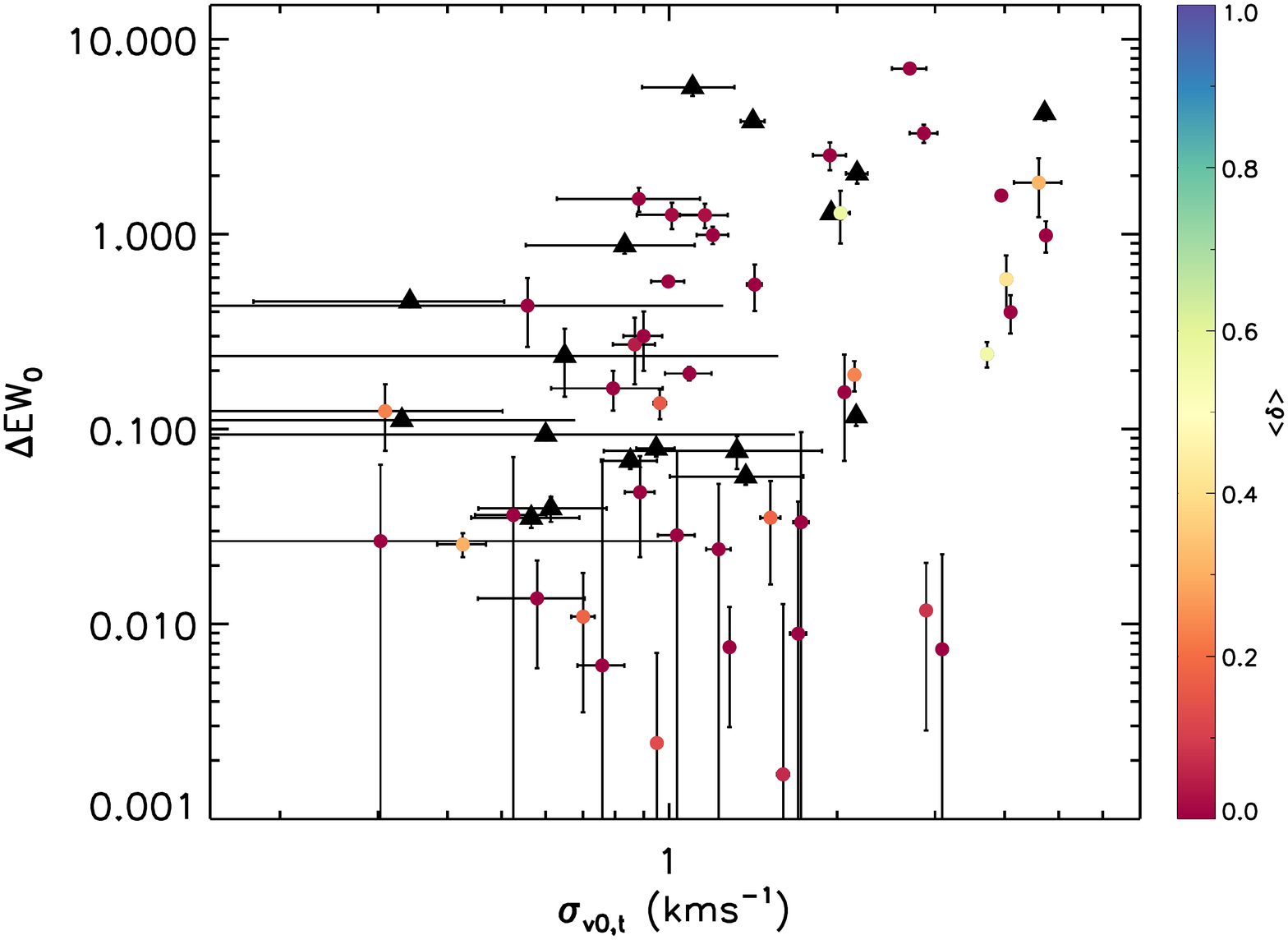}
    \caption{The change in optical depth equivalent width versus the 1D turbulent velocity for all Gaussian features with known \ts{}. For matched features, $\Delta EW_0=|EW{0,1}-EW_{0,2}|$ and the x-axis shows the average 1D turbulent velocity of the two features. Matched features are shown as circles with colors indicating the value of $\langle\delta\rangle$ (see Section \ref{subsec:matching}). For unmatched features, $\Delta EW_0=EW_0$ and the x-axis is the 1D turbulent velocity of the single feature. Unmatched features are shown as black triangles.}
    \label{fig:dew_v_vturb}
\end{figure}

Figure \ref{fig:dew_v_vturb} shows the change in optical depth equivalent width (Equation \ref{eq:ew0}) versus the 1D turbulent velocity for each feature. 
There is a marginally increasing trend ($\log\Delta EW_0\propto (0.98\pm0.28)\log\sigma_{v_0,t}$), with the largest variations tending to be associated with the highest turbulent velocities. 
We see that the 1D turbulent velocities range from $\sim0.3$ to $\sim4.7$ \kms{}, and are therefore a non-negligible contribution to the velocity linewidths.
\cite{2005AJ....130..698B} argued that spin temperature variation is not the primary cause of TSAS after they found strong agreement between the linewidths of \hi{} features observed with both the Very Long Baseline Array (VLBA) and the Arecibo Observatory (\citealt{2003ApJ...586.1067H}). But this assumes that thermal broadening dominates turbulent broadening, which these results suggest is not a safe assumption. Furthermore, these turbulent velocities are similar to or larger than the typical sound speed in the CNM. This suggests the need of a driver to sustain supersonic turbulence.

\begin{figure}
    \centering
    \includegraphics[width=\columnwidth]{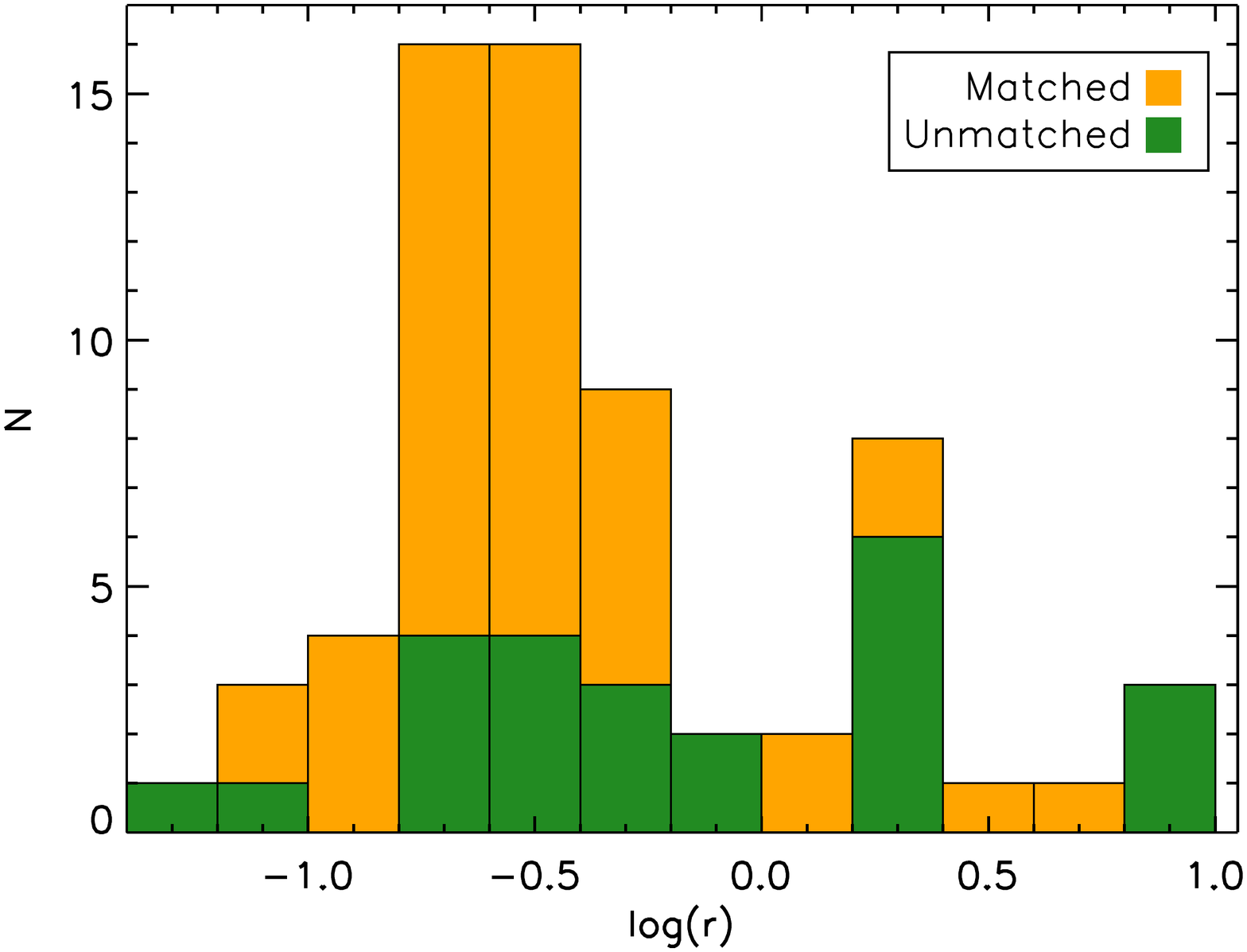}
    \caption{The distribution of $r$ (Equation \ref{eq:pratio}), in $\log_{10}$, for all Gaussian features. The contribution from the unmatched features is shown in green, while the contribution from the matched features is shown in orange.}
    \label{fig:r_hist_stacked_log}
\end{figure}

We also compare the turbulent pressure, $P_{turb}=\rho\sigma_{v_0,t}^2$, to the thermal pressure, $P_{th}=nk_BT$, using the ratio
\begin{equation} \label{eq:pratio}
    r = \frac{k_BT_{S}}{m_{\mathrm{H}}\sigma_{v_0,t}^2},
\end{equation}
where \ts{} is again a proxy for the kinetic temperature. For matched features, we use the average 1D turbulent velocity, $\langle \sigma_{v_0,t}\rangle$, and the average spin temperature, $\langle\ts{}\rangle$, when evaluating Equation \ref{eq:pratio}. The uncertainties on the spin temperatures and turbulent velocities for matched features due to variations in the spin temperature (Section \ref{subsec:dts_g}) lead to a typical uncertainty of $\sim20$--30\%. The distribution of $r$ is shown in Figure \ref{fig:r_hist_stacked_log}. For a majority of features, turbulent pressure is dominant (i.e., $r<1$).

\begin{figure}
    \centering
    \includegraphics[width=\columnwidth]{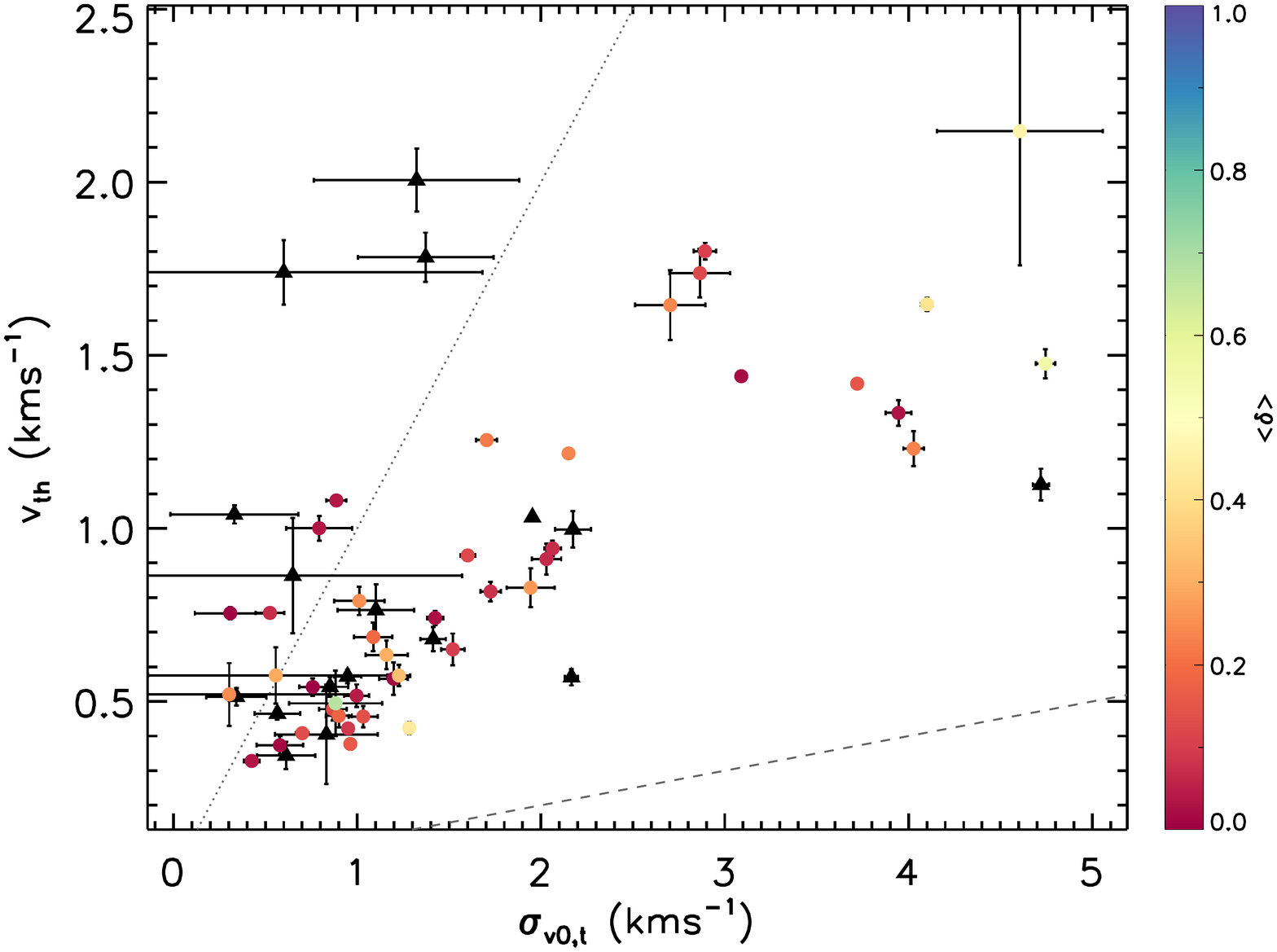}
    \caption{The estimated 1D rms velocity, $\sqrt{k_{B}\ts{}/m_{\mathrm{H}}}$, versus the 1D turbulent velocity (Equation \ref{eq:pturb}). Matched features are shown as circles whose colors indicate the value of $\langle\delta\rangle$ (see Section \ref{subsec:matching}). Unmatched features are shown as black triangles. The dotted line shows the 1-to-1 relationship. The dashed line shows the 1-to-10 relationship that \citetalias{doi:10.1146/annurev-astro-081817-051810} estimate would be necessary for turbulent confinement.}
    \label{fig:vth_v_vturb}
\end{figure}

Based on \cite{1992ApJ...399..551M}, \citetalias{doi:10.1146/annurev-astro-081817-051810} estimate that the turbulent velocity must exceed the rms velocity by a factor of at least 10 to confine a structure that is overpressured by a factor of 100. In Figure \ref{fig:vth_v_vturb}, we show the 1D rms velocity versus the 1D turbulent velocity. We show lines indicating a 1-to-1 ratio (dotted), and a 1-to-10 ratio (dashed). All points lie above the dashed line, indicating that the turbulent velocity is likely too small to lead to pressure confinement of the observed TSAS features, suggesting that our TSAS features are expanding or not highly overpressured.

\subsubsection{Column Density Variations} \label{subsec:columndensity}
The column density of each feature calculated by \cite{2018ApJS..238...14M} and \cite{2003ApJS..145..329H} is listed in Table \ref{tab:allgausstable}. Since the column density depends on the spin temperature, though, the column densities for the two components of each set of component pairs are not independently measured (see beginning of Section \ref{subsec:dts_g}). Nevertheless, we use these column density estimates as reasonable approximations because doing so allows us to investigate the column density and density of TSAS. 

\begin{figure}
    \centering
    \includegraphics[width=\columnwidth]{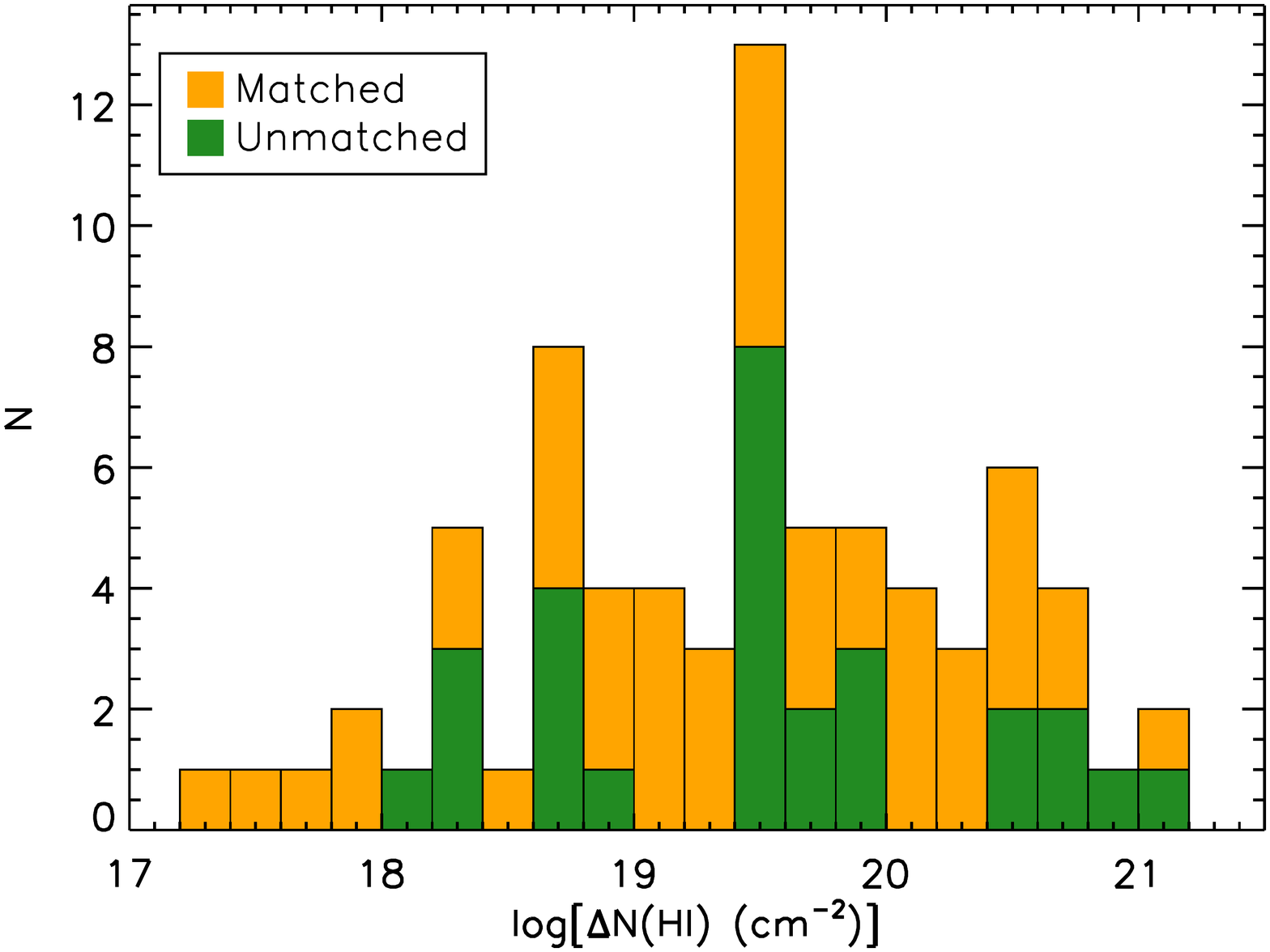}
    \caption{The distribution of column density variations, in $\log_{10}$, for all Gaussian features. For matched features, shown in orange, $\Delta N(\hi{})=|N(\hi{})_{1}-N(\hi{})_{2}|$. For unmatched features, shown in green, $\Delta N(\hi{})=N(\hi{})$.}
    \label{fig:dnhi_hist_stacked}
\end{figure}

In Figure \ref{fig:dnhi_hist_stacked}, we show the distribution of the change in \hi{} column density ($\Delta N(\hi{})=|N(\hi{})_1-N(\hi{})_2|$ for matched features and $\Delta N(\hi{})=N(\hi{})$ for unmatched features). The median change in column density is $\Delta N(\hi{})=3.5\times10^{19}$ \persc{}. This \hi{} column density variation is consistent with previous TSAS studies, which have primarily found $\Delta N(\hi{})$ between $10^{19}$ \persc{} and $10^{21}$ \persc{} (see Table 1 of \citetalias{doi:10.1146/annurev-astro-081817-051810} and references therein).

A cloud with a sufficiently high column density, $N_{\mathrm{min,c}}$, may radiatively cool fast enough to prevent expansion, meaning that it can reach thermal equilibrium while being overpressured.  \citetalias{doi:10.1146/annurev-astro-081817-051810} estimated the critical column density to be $N_{\mathrm{min,c}}=1.2\times10^{15}\,T^{3/2}\,e^{92/T}$ \persc{} by demanding that the cooling time ($3kT/2n\Lambda$, where $\Lambda$ is the radiative loss function) be less than the dynamical time ($R/\sqrt{kT/m}$, where $R$ is the size of the cloud). Taking $\Delta N(\hi{})$ to be the TSAS column density and \ts{} to be a proxy for the kinetic temperature, we find that 59 of the 79 features with known column densities have column densities exceeding $N_{\mathrm{min,c}}$, i.e., a majority of the structures could presumably be overpressured but in thermal equilibrium with the ambient ISM.

\subsubsection{Estimating density}\label{subsec:density}
Under a simple geometric assumption, the change in \hi{} density between adjacent lines of sight is $\Delta N(\hi{})/L$, where $L$ is the linear separation of \hi{} structures along the line of sight. The linear separation is unknown for most features in our sample because we do not know the distance, $D$, to the absorbing \hi{} structures ($L=D\Delta\theta$, with $\Delta\theta$ listed in Table \ref{tab:sources}). Unfortunately, kinematic distances are unreliable or have extremely large uncertainties ($\gtrsim 100\%$) for most features in this study (e.g., \citealt{2018ApJ...856...52W}). Another way of getting a rough distance estimate is by assuming a scale height for the CNM of 100 pc (e.g., \citealt{1975ApJ...198..281B,1978A&A....70...43C}) and taking $D=100$ pc$/ \sin |b|$.
For the range of angular separations probed here, this yields typical linear separations of $\sim10^3$ to $\sim10^5$ AU.

Three of our sources are in the direction of structures with well-known distances. We estimate TSAS densities only for these sources.
3C111 and 3C123 are in the direction of the Taurus molecular cloud (Figure \ref{fig:dameco}) and 3C225 is in the direction of the well-studied local Leo cold cloud, first identified by  \cite{1969ApL.....4...85V}. For features whose radial velocities indicate that they are associated with these known clouds ($\sim-4$--$10$ \kms{} for Taurus and 4 \kms{} for the Leo cloud; \citealt{1987ApJS...63..645U,1969ApL.....4...85V}), we have fairly tight constraints on distance (Taurus from \citealt{2019A&A...624A...6Y} and the Leo cloud from  \citealt{2011ApJ...735..129P}), so we can estimate the linear separation, $L$, and the density, $\Delta n\approx \Delta N(\hi{})/L$, of the TSAS. Table \ref{tab:knownsources} lists the estimated \hi{} densities for these features. Previous studies have found TSAS densities between a few thousand \percc{} (e.g., \citealt{2000ApJ...540..863D,2008MNRAS.388..165G}) and $\gtrsim10^6$ \percc{} (e.g., \citealt{1995ApJ...452..671M}). The results listed in Table \ref{tab:knownsources} are on the lower end of this range, but still exceed typical CNM densities by roughly an order of magnitude ($n_{\mathrm{CNM}}\lesssim100$ \percc{}; \citealt{2003ApJ...587..278W}).

\begin{deluxetable*}{ccccc}
\tablecolumns{4}
\tablecaption{Summary of the properties of the Gaussian features. Data from this work are separated into two columns---one for matched features, and another for unmatched features. This is analogous to Table 5 of \cite{1986ApJ...303..702G}. The uncertainties on all averaged quantities from this work are all very small. In this table, the standard deviations are shown. \label{tab:gauss_summary}}
\tablehead{Parameter & This Work & This Work & Greisen \& Liszt (1986) & Crovisier et el. (1985) \\
& Matched & Unmatched & & $L < 1$ pc / $L > 1$ pc }
\startdata
\input{summarytable.tex}
\enddata
\end{deluxetable*}

\subsection{Individual Sources} \label{subsec:sources}
Several of the multiple-component sources in this study have previously been studied for signs of TSAS (\citetalias{doi:10.1146/annurev-astro-081817-051810} and references therein). Some are also associated with known Galactic structures. Here, we compare our observations of these sources to previous studies. 

\subsubsection{3C111} \label{subsubsec:3C111}
\begin{figure}
    \centering
    \includegraphics[width=\columnwidth]{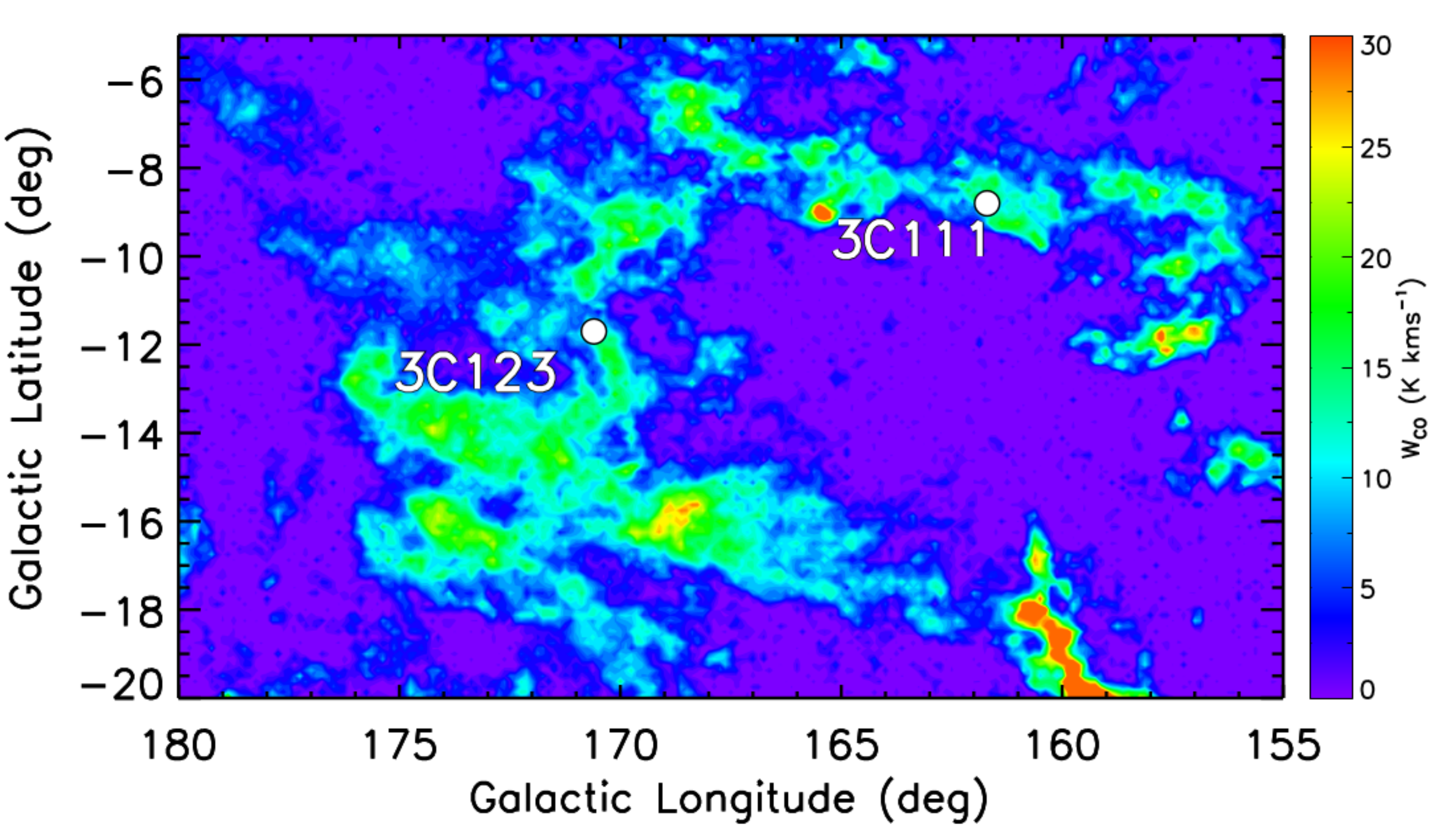}
    \caption{Integrated intensity of \textsuperscript{12}CO(1--0), $W_{CO}$, measured by \cite{2001ApJ...547..792D}, in the direction of the Taurus molecular cloud. The positions of two of our background sources, 3C111 and 3C123, are coincident with the molecular cloud.}
    \label{fig:dameco}
\end{figure}

3C111, the only triple-lobed background source in our sample, has been observed extensively in absorption studies. It is in the direction of the Taurus molecular cloud (Figure \ref{fig:dameco}).

\cite{1978AA....70..415D} measured \hi{} absorption across the B and C components of 3C111, noting significant variation ($\Delta \tau = 0.32$) at $v=8.3$ \kms{}. \cite{1986ApJ...303..702G} also detected a feature at $\sim8$ \kms{} present in the B component but absent in both the A and C components. With 21-SPONGE, we see a strong absorption feature at $\sim8$ \kms{} toward the B component ($\tau=0.212$) that is greatly diminished toward the (central) A component ($\tau = 0.069$), and vanishes at the C component. Unfortunately, we do not know \ts{} or $N(\hi{})$ for these features and are therefore unable to estimate the corresponding change in density. 

\cite{1986ApJ...303..702G} detected significant spatial variation in an optical depth feature at $-2$ \kms{} that appeared toward all three lobes, but increased to the east. We also see strong absorption toward all three lobes at $v\sim-2$ \kms{} with an eastward increase in optical depth, although the lowest optical depth is observed toward the central A component. Table \ref{tab:knownsources} suggests a density of a few hundred \percc{} for this TSAS feature, presumably associated with the Taurus molecular cloud complex.

\cite{2008MNRAS.388..165G} compared the \hi{} absorption from the A and B components of 3C111, identifying strong variations ($\Delta \tau \sim 0.3$) at $v\approx-19$ \kms{} and $v\approx-11$ \kms{}.
While we do not find components at $-19$ \kms{} and $-11$ \kms{}, we do identify variable features at $-16.8$ and $-8.3$ \kms{} that bear a striking resemblance to their features. If, as we suspect, their features are the same as ours, offset by a velocity of $\sim2$ \kms{}, then we confirm this variation ($\Delta \tau_0\approx 0.2$ for the $-16.8$ \kms{} feature and $\Delta \tau_0\approx 0.3$ for the $-8.3$ \kms{} feature).

\cite{1993ApJ...419L.101M} were the first to suggest TSAS in the direction of 3C111 on the basis of molecular absorption variations. They detected significant changes in the $\mathrm{H_{2}CO}$ optical depth toward 3C111 over a period of 2.05 yr, suggesting molecular structures with $n\sim 10^6$ \percc{} on scales of $\lesssim10$ AU. These results were later supported by \cite{1995ApJ...452..671M}, who extended the observational time span to 3.4 yr . \cite{1995ApJ...452..671M} also detected significant spatial variations in the \hi{}, $\mathrm{H_{2}CO}$, and OH absorption across 3C111. 
However, $\mathrm{H_{2}CO}$ absorption variation may be strongly dependent on the gas dynamics, chemistry, or radiation field, rather than the density and column density (\citealt{1996A&A...312..973T}). This would make it a poor tracer of TSAS compared to dust absorption or \hi{} absorption, which do not suggest TSAS densities as extreme as $10^6$ \percc{}.

Here, we see significant \hi{} optical depth variations across 3C111, with $\Delta EW_{abs,3\sigma}\approx 20\%$ between any two components---the \hi{} optical depth in the direction of 3C111 is highly variable, and some features appear to undergo large changes in density over spatial scales $\lesssim 0.1$ pc (Table \ref{tab:knownsources}). These changes in density are smaller than those suggested by the $\mathrm{H_{2}CO}$ absorption studies of \cite{1993ApJ...419L.101M} and \cite{1995ApJ...452..671M}, although we are probing spatial scales several orders of magnitude larger.

The \hi{} absorption features in the direction of 3C111 span a velocity range of over $30$ \kms{}. In Section \ref{subsec:density}, we estimate the densities of \hi{} structures whose radial velocities suggest that they are part of the Taurus molecular cloud complex, assuming that components with radial velocities outside this range are well-separated along the line of sight. But it is possible that many or all of the strong \hi{} absorption features are associated with Taurus---the large spread in velocity could be due to the contribution of multiple flows within the molecular cloud complex, driven by, e.g., Herbig-Haro outflows (although there are no known Herbig–Haro objects along the 3C111 sightlines; \citealt{2000yCat.5104....0R}). Optical studies have found that strong variations of Na\,{\footnotesize I} absorption line profiles over a wide velocity range can stem entirely from local gas. For example, \cite{1996ApJ...473L.127W} found strong variations in multiple Na\,{\footnotesize I} absorption components spanning $\sim40$ \kms{} in the direction of the binary stars HD32039 and HD32040 (their Figure 1), but these stars are at a distance of only 230 pc, meaning that the absorbing gas must all be at a distance of $<230$ pc. Here, it is unclear how many of the observed \hi{} absorption components are associated with the Taurus molecular cloud complex---we lack a reliable distance measurement to the absorbing \hi{} structures in this direction. If we were to assume that they are all associated with Tuarus, then the range of density variations would be $\sim5 \text{--}2000$ \percc{} for the features with known column densities.

\subsubsection{3C123} \label{subsubsec:3C123}
3C123 is also in the direction of the Taurus molecular cloud (Figure \ref{fig:dameco}) and has previously shown evidence for TSAS. 
\cite{2008MNRAS.388..165G} constructed an \hi{} optical depth image of 3C123. They identified strong absorption features at $v=-20.1$ \kms{} and $v=4.2$ \kms{}. They found that the $-20.1$ \kms{} feature showed almost constant absorption across the source, while there was a marginal change (2--3$\sigma$) in optical depth for the 4.2 \kms{} component. 

Here, we match features at $-19.9$ and $-19.3$ \kms{} whose Gaussian fitted optical depths are identical ($\tau_{0,1}=\tau_{0,2}=0.064$). Several Gaussian features are fit at velocities close to 4 \kms{}, and Figure \ref{fig:allspectra} confirms that there is indeed significant optical depth variation ($\gtrsim20\sigma$) at these velocities, probably associated with gas in the Taurus molecular cloud complex. The estimated densities in the direction of 3C123 (Table \ref{tab:knownsources}) are high but all have uncertainties $\gtrsim100\%$.

\subsubsection{3C225} \label{subsubsec:3C225}
\cite{1980A&A....88..329C} and \cite{1985A&A...146..223C} measured \hi{} absorption across 3C225 using the Arecibo Observatory and the Westerbork Synthesis Radio Telescope (WSRT), respectively. They both identified a strong absorption feature at $\sim4$ \kms{} that varied across the source. This feature is associated with one of the cold \hi{} clouds identified by \cite{1969ApL.....4...85V} (``cloud A''; \citealt{1985A&A...149..209C}). This cloud has a remarkably high pressure and density ($P/k\sim 6\times10^4$ K\percc{} and $n\sim3000$ \percc{}; \citealt{2012ApJ...752..119M}). Here, a 4.0 \kms{} feature is fitted to both components, with $\tau_{0,1}=0.805$ and $\tau_{0,2}=0.774$. Given its distance and change in column density, we find evidence for large density variations on a scale of $\sim100$ AU toward 3C225 ($\Delta n\sim 4200\pm3800$ \percc{}), but the uncertainty on this variation is rather large.

\begin{deluxetable*}{c|c|c|c|c|c|c|c} 
\tablecolumns{5}
\tablecaption{Properties of Gaussian features with known distances and known column densities. The distance to the Taurus molecular cloud, $D=145^{+12}_{-16}$, comes from Yan et al. (2019). The distance to the local Leo cloud comes from Peek et al. (2011). Feature with only one velocity indicated are the unmatched features; features with two velocities are the matched features. Some features seen toward 3C111 appear multiple times (for the different pairings, AB, AC, and BC). \label{tab:knownsources}} 
\tablehead{Component 1 & Component 2 & v1 & v2 & $D$ & $L$ & $\Delta N(\hi{})$ & $\Delta n$ \\
& & \kms{} & \kms{} & pc & pc & $10^{20}$\persc{} & \percc{}}
\startdata
\input{densities.tex}

\enddata
\end{deluxetable*}

\section{Discussion} \label{sec:discussion}

\subsection{Ubiquity of TSAS}

Using highly sensitive \hi{} absorption observations,
we find $>5\sigma$ \hi{} optical depth variations at a level of $\gtrsim0.05$ in the direction of 13 of the 14 component pairs. Our sources probe a wide range of environments---peak optical depths toward the 14 component pairs range by more than an order of magnitude (Figure \ref{fig:maxdtau_v_ew_nhi}). These results suggest that variation of \hi{} optical depth profiles  
on angular scales from $5\arcsec{}$ to $4\arcmin{}$ may be common. 
In the several cases where we have a good distance estimates, the \hi{} optical depth variations appear to be associated with TSAS. However, for the majority of observed \hi{} structures,
we do not know the distance and therefore cannot determine if these features are overdense and overpressured, as has historically been claimed of TSAS.
Our results are consistent with previous sensitive \hi{} absorption studies toward multi-component background sources (\citealt{1985A&A...146..223C,1986ApJ...303..702G}).   
While our sample size is limited, future large-scale \hi{} absorption observations (e.g. GASKAP; \citealt{2013PASA...30....3D}) will provide excellent samples to establish TSAS's ubiquity and other statistical properties.

The abundance of TSAS is
an important yet controversial topic when considering its role in the ISM. 
For example,
\citetalias{doi:10.1146/annurev-astro-081817-051810} argued on energetic grounds that TSAS clouds overpressured by a factor of $\sim100$ could occupy at most a volume filling factor of roughly $<(2$--$7)\times10^{-4}$. Larger volume filling factors would have a significant effect on the ISM thermal balance.

Optical studies of absorption variations in the direction of large samples of stars have found that only a few percent of sightlines show the absorption variations characteristic of TSAS at scales $\lesssim380$ AU (e.g.,  \citealt{2011MNRAS.414...59S,2015MNRAS.451.1396M}), while such variations are more common at scales $\gtrsim 480$ AU (e.g., \citealt{1996ApJ...473L.127W,2013A&A...550A.108V}). Multi-epoch observations of \hi{} absorption against pulsars have also suggested that TSAS is uncommon at scales $\lesssim500$ AU, with only a select few sightlines showing significant optical depth variation (e.g., \citealt{2003MNRAS.341..941J,2008ApJ...674..286W,2010ApJ...720..415S}). 

\cite{2005AJ....130..698B} argued that TSAS is ubiquitous, and that non-detections represented a spatial-scale selection effect. According to this interpretation, TSAS searches sampling only scales $\ll100$ AU---including multi-epoch absorption studies against pulsars and VLBI absorption studies---lack an adequate dynamic range of spatial scales to have a high probability of detecting TSAS. Moreover, as discussed in Section \ref{sec:intro} and Section \ref{sec:spectra}, previous multiple-component studies have found significant variations ($>3$--$5\sigma$) along most observed sightlines, especially when reaching an optical depth sensitivity of $\lesssim0.01$. Our results are consistent with these previous multiple-component studies.

\cite{2010ApJ...720..415S} discuss the survival of large versus small TSAS in the context of the \cite{2002ApJ...564L..97K} simulations, wherein TSAS represents the cold, dense cloudlets formed in thermally unstable regions of a post-shocked layer. These small cloudlets eventually evaporate into the hot ambient medium, but the timescale for this evaporation is highly dependent on the size of the cloudlets---clouds 30 AU in size evaporate in $10^2$--$10^3$ yr, while clouds $10^4$ AU in size evaporate in $10^5$--$10^7$ yr (\citealt{2010ApJ...720..415S}). Multiple-component absorption studies (including this work) typically probe the latter scale. The dramatically longer lifetime of TSAS clouds at this scale may explain why significant \hi{} optical depth variations have been found more commonly in these studies than in multi-epoch \hi{} absorption studies against pulsars.
Nevertheless, the sample size of TSAS observations---both at the larger and smaller scales---remains small, so it is unclear if the discrepancy in the observed abundance of TSAS indicates an observational bias or a true physical distinction.

\subsection{Properties of the Optical Depth Variations} \label{subsec:source_of_variability}

Both methods of measuring optical depth variations---the channel-by-channel variations and the variations in fitted Gaussian features---show that the highest variation is seen along lines of sight with the highest \hi{} optical depths. TSAS appears to be most significant where there is a greater quantity of cold \hi{}. Figure \ref{fig:maxdtau_v_ew_nhi} affirms this, indicating that the \hi{} optical depth variations are strongest for sources at lower latitudes and with higher \hi{} column densities (and also higher CNM fractions). As seen by \cite{1993PhDT........56B}, \cite{2007ApJ...656...73L}, and \cite{2013MNRAS.434..163H} in variable narrow quasar absorption lines, we find that the fractional variation is largest for sources with low optical depth (Figure \ref{fig:maxdtauratio_v_ew_cnm}), which may simply be a consequence of the low optical depth itself (see Section \ref{subsec:channelvariations} and \citealt{2007ApJ...656...73L}). 

We find that both variations in the peak optical depths and variations in the linewidths of fitted features contribute to the overall optical depth variation (Figure \ref{fig:tau_v_fwhm}). Moreover, contrary to the argument presented in \cite{2005AJ....130..698B}, it appears that the turbulent velocity contribution to the linewidths is non-negligible (Figure \ref{fig:vth_v_vturb}). To precisely determine the relative contributions of thermal motions and turbulent motions to the changes in linewidths, we await higher resolution emission measurements.

Table \ref{tab:knownsources} is consistent with the interpretation 
that the \hi{} optical depth variations are caused by dense structures along the line of sight. Under this classical interpretation, we find that the density variations toward structures with known distances are on the order of $\sim1000$ \percc{}, which is an order of magnitude higher than what is typical in the CNM. At least one of these structures---the local Leo cold cloud, in the direction of 3C225---is known to be highly overpressured ($P/k\approx 60,000$ K\percc{}; \citealt{2012ApJ...752..119M}). However, if the optical depth variations are the result of sheets or filaments, then the densities listed in Table \ref{tab:knownsources} may be overestimated by more than an order of magnitude.

If we assume that optical depth features represent discrete, tiny \hi{} structures,
then we find that these structures are likely expanding (Section \ref{subsec:turbulentproperties}), 
but that a majority of them may also be in thermal equilibrium with the surrounding ISM, whether or not they are overpressured (Section \ref{subsec:columndensity}).

\subsection{TSAS and Turbulence} \label{subsec:turbulence}

\cite{2000ApJ...543..227D} showed that $\Delta \tau /\tau \propto \Delta n /n$.
This assumes that the optical depth fluctuations are small and that the relationship between temperature fluctuations and density fluctuations does not change with physical scale. However, if the relationship between these fluctuations changes with scale (e.g., small scale perturbations are isobaric while larger scale perturbations are isothermal), then this straightforward relationship no longer holds. We find that the fractional change in \hi{} optical depth depends strongly on the optical depth (Figure \ref{fig:maxdtauratio_v_ew_cnm}), which indicates that $\Delta \tau /\tau$---and, under Deshpande's assumptions, $\Delta n /n$---has a strong environmental dependence.  This suggests that the amplitude of density fluctuations is not universal, as would be the case for the universal interstellar turbulence. Instead, density fluctuations appear stronger at low Galactic latitudes and high \hi{} column densities, consistent with the idea that excess energy injection happens in such environments (\citetalias{doi:10.1146/annurev-astro-081817-051810}).

The turbulent pressure dominates the thermal pressure for most features (Figure \ref{fig:r_hist_stacked_log}). This ratio, $r$, does not depend on assumptions about the environment (such as the pressure and density, which are still controversial for TSAS). Nevertheless, the magnitude of \hi{} optical depth variation does not appear to be strongly correlated with turbulence (Figure \ref{fig:dew_v_vturb}). Moreover, Figure \ref{fig:vth_v_vturb} indicates that TSAS structures are not pressure confined (following the argument outlined in \citetalias{doi:10.1146/annurev-astro-081817-051810} that the turbulent velocity in pressure confined structures should exceed the rms velocity by a factor of $>10$). Instead, these structures must be transient, be confined by some other mechanism, or be near pressure equilibrium with their surroundings. It is possible, though, that turbulent motions contribute to the the optical depth variation by altering the linewidths of \hi{} structures (see Section \ref{subsec:source_of_variability}).

\subsection{Future work} \label{subsec:future_work}

We have found strong evidence that TSAS is correlated with the total quantity of cold \hi{} gas, investigated the role of turbulence in producing TSAS, and estimated TSAS densities for a handful of sightlines, but a number of important questions about the formation and role of TSAS remain. 

For example, a geometric understanding of the variable \hi{} structures is still unclear, i.e., are they sheets or filaments as suggested by \cite{1997ApJ...481..193H}, large scale structures as suggested by \cite{2000MNRAS.317..199D}, true tiny dense structures, or none of the above?  Obtaining \hi{} optical depth maps, as in \cite{2000ApJ...543..227D} and \cite{2012ApJ...749..144R}, can constrain the slope of the optical depth power spectrum in different environments, testing the \cite{2000MNRAS.317..199D} hypothesis. It is particularly important to probe the intermediate scales not covered by either of those experiments. If the hypothesis is correct, then TSAS densities (e.g., Table \ref{tab:knownsources}) may be overestimated by over an order of magnitude.

\cite{1985A&A...146..223C} showed that the change in central velocity of the Gaussian features between adjacent lines of sight could be used as a proxy for the velocity dispersion of the absorbing \hi{} structures ($\Delta v_0 = |v_{0,1}-v_{0,2}|\sim\sigma_v$). If the distances to the \hi{} structures are known, this dispersion can be measured as a function of linear separation, $L$, and thereby indicate the source(s) of turbulence (e.g.,  \citealt{1979MNRAS.186..479L}) associated with tiny scale structures. Here, we only have 9 features for which we know $\Delta v_0$ and $L$. The scatter in $\Delta v_0$ versus $L$ is far too large to make a strong statement about the source of turbulence.

The role of temperature in explaining
the observed optical depth variation remains unconstrained.
\cite{2007A&A...465..431H}
and \cite{2012ApJ...759...35I} showed that steep temperature gradients can form on the outskirts of sub-parsec scale CNM \hi{} clumps that collapse in thermally unstable regions of the WNM. Dramatic temperature variation could alter the linewidth of a given feature, leading to variation in the observed \hi{} optical depth spectrum. Although we have more precise estimates of \ts{} than previous TSAS searches, which have generally assumed some (constant) value of \ts{} of $\sim50$ K, we still lack independent measurements of \ts{} across each component pair, and so cannot reliably measure the temperature variation. Higher angular resolution \hi{} emission observations are needed to investigate the question of whether \ts{} changes significantly between multiple components (Section \ref{subsec:dts_g}).

The above considerations also demonstrate the need for high resolution, non-isothermal numerical models of the formation of tiny scale \hi{} structure. \cite{2002ApJ...564L..97K}, \cite{2007A&A...465..431H}, and
\cite{2012ApJ...759...35I}, for example, have shown that the fragmentation of warm \hi{} into cold, dense structures with TSAS-like properties can occur as the result of shock compression and thermal instability. But a detailed description of these structures' turbulent properties, thermal properties, and overall abundance is needed to compare to TSAS observed in \hi{} absorption experiments. It is especially important to probe a range of scales, including the $\lesssim100$ AU scale, which is smaller than what was probed by any of the above experiments. \cite{2012ApJ...759...35I}, the only of those experiments that modeled the flow of \hi{} in three dimensions, reached a spatial resolution of $\sim0.02$ pc.

Finally, complementary observations of TSAS in molecular lines at millimeter wavelengths could be used to better understand the environments associated with TSAS by further constraining TSAS densities and assessing the importance of TSAS for interstellar chemistry. Such observations can also test for shocks in the direction of TSAS using tracers such as SiO (e.g., \citealt{1997ApJ...482L..45M,2016A&A...595A.122L}) or HNCO (e.g., \citealt{2000A&A...361.1079Z,2010A&A...516A..98R,2017A&A...597A..11K}).

\section{Conclusions} \label{sec:conclusions}
We have used \hi{} absorption spectra toward multiple-component background sources from the 21-SPONGE survey (\citealt{2018ApJS..238...14M}) and the Millennium Survey (\citealt{2003ApJS..145..329H,2003ApJ...586.1067H}) to probe AU-scale atomic structure in the Milky Way. We measured changes in both the integrated properties of the \hi{} optical depth spectra and the Gaussian components fit to the optical depth spectra between component pairs. 
Our main results can be summarized as follows.
\begin{itemize}
    \setlength\itemsep{2em}
     \item In 13/14 pairs of HI absorption spectra, the 
     \hi{} optical depth varies by at least $\sim0.05$ on angular scales non-continuously spanning from $5\arcsec{}$ to $4\arcmin{}$. This suggests that this level of variation may be common in the ISM.
    \item The \hi{} optical depth variations are strongest for lines of sight probing a higher \hi{} column density and higher CNM fraction.
    \item The highest fractional variations in \hi{} optical depth occur for lines of sight with the lowest optical depth, as seen in studies of narrow quasar absorption lines in other galaxies (\citealt{1993PhDT........56B,2007ApJ...656...73L,2013MNRAS.434..163H}).
    \item From the variations in the properties of the fitted Gaussian components, we find evidence that both changes in the peak \hi{} optical depth and the linewidth of \hi{} features contribute to the observed optical depth variation, while the central velocity of Gaussian features does not change much.
    \item Most of the optical depth variation is associated with CNM structures, but there is not a correlation between the level of \hi{} optical depth variation and spin temperature. 
    \item The apparent AU-scale structures do not appear to be turbulently confined, although
    turbulent broadening is an important component of measured velocity linewidths.
    \item A majority of the structures we probed
    have sufficiently high column densities that they can be in thermal equilibrium with the rest of the ISM despite being overpressured.
    \item Under a simple geometric assumption, structures for which we can estimate the density appear to be overdense by at least an order of magnitude
(e.g., \citetalias{doi:10.1146/annurev-astro-081817-051810}).
\end{itemize}
Future analogous studies at molecular wavelengths can determine if the observed AU-scale structures are associated with shocks. Higher resolution emission measurements can both assess the role of temperature variation in producing dramatic changes in \hi{} optical depth and also quantify the relative importance of thermal motions and turbulent motions in changing the linewidths of \hi{} absorption features. An expanded sample size can be used to better constrain the statistical properties of TSAS in the Milky Way. Finally, we stress the need for high resolution models to investigate the formation of \hi{} structures on tiny scales and compare to observations.

\acknowledgements
{S.S. acknowledges the support by the Vilas funding provided by the University of Wisconsin, and the John Simon Guggenheim fellowship. This research was partially supported by the Munich Institute for Astro- and Particle Physics (MIAPP) which is funded by the Deutsche Forschungsgemeinschaft (DFG, German Research Foundation) under Germany's Excellence Strategy - EXC-2094-390783311. D.R. thanks Trey V. Wenger for helpful conversations regarding the kinematic distance estimates to the sources in this study. We thank the anonymous referee for constructive comments.
21-SPONGE data were obtained by the Arecibo Observatory and the Karl G. Jansky Very Large Array (VLA). The Millennium survey data were obtained by the Arecibo Observatory.
The Arecibo Observatory is operated by SRI International under a cooperative agreement with the National Science Foundation (AST-1100968), and in alliance with Ana G. Méndez-Universidad Metropolitana, and the Universities Space Research Association.
The National Radio Astronomy Observatory is a facility of the National Science Foundation operated under cooperative agreement by Associated Universities, Inc. 
This research made use of TMBIDL (\citealt{2016ascl.soft05005B}).
This research made use of Astropy,\footnote{http://www.astropy.org} a community-developed core Python package for Astronomy \citep{astropy:2013, astropy:2018}.}

\bibliography{refs}{}
\bibliographystyle{aasjournal}

\startlongtable
\begin{deluxetable*}{l|c|c|c|c|c|c|c}
\tablecaption{Summary of Gaussian features fit to the \hi{} absorption spectra from \cite{2018ApJS..238...14M} and \cite{2003ApJS..145..329H}. \textbf{Column 1:} name of the multiple-component background source; \textbf{Column 2:} label of the component in each row; \textbf{Column 3:} central velocity of the Gaussian feature; \textbf{Column 4:} peak optical depth of the Gaussian feature; \textbf{Column 5:} FWHM of the Gaussian feature; \textbf{Column 6:} estimated spin temperature of the \hi{} structure; \textbf{Column 7:} estimated column density of the \hi{} structure ($N(\hi{})=1.064 \times 1.823\times10^{18}$ \persc{}/K\kms{} $\times \tau_0 \times \fwhm{} \times T_S$); \textbf{Column 8:} $\delta$ parameter (Equation \ref{eq:delta}) for matched components. Rows for matched features are merged. For Columns 1 and 2, we use the naming conventions from \cite{2018ApJS..238...14M} and \cite{2003ApJS..145..329H} (e.g., A/B versus 1/2 for the labeling of components). \label{tab:allgausstable}}
\tablehead{\colhead{Source} & \colhead{Component} & \colhead{$v_{0}$} & \colhead{$\tau_{0}$} & \colhead{\fwhm{}} & \colhead{\ts{}} & \colhead{$N(\hi{})$} &  \colhead{$\langle\delta\rangle$}  \\ \colhead{} & \colhead{} & \colhead{\kms{}} & \colhead{} & \colhead{\kms{}} & \colhead{K} & \colhead{$10^{20}$ \persc{}} & \colhead{}}
\startdata
\input{end_table.tex}
\enddata
\end{deluxetable*}

.

\end{document}

%% file: summarytable.tex
                                                                          $\langle \tau_{0} \rangle$  &  $ 0.348\pm 0.369$  &  $ 0.174\pm 0.299$  &  $  0.30\pm  0.18$  &  $  0.41\pm  0.21/  0.42\pm  0.28$  \\
                                                                                  $\Delta \tau_{0} $  &  $ 0.167\pm 0.280$  &  &  $  0.05\pm  0.04$  &  $  0.08\pm  0.08/  0.06\pm  0.06$  \\
                                                  $ \frac{\Delta \tau_{0}}{\langle\tau_{0}\rangle} $  &  $ 0.463\pm 0.500$  &  &  $  0.19\pm  0.20$  &  $  0.27\pm  0.30/  0.18\pm  0.18$  \\
\hline
                                                                                      $\Delta v_{0}$  &  $  0.31\pm  0.47$  &  &  $  0.21\pm  0.20$  &  $  0.52\pm  0.34/  1.60\pm  1.47$  \\
                                                 $ \frac{\Delta v_{0}}{\langle\fwhm{}\rangle} $  &  $  0.07\pm  0.06$  &  &  $  0.06\pm  0.05$  &  $  0.09\pm  0.08/  0.14\pm  0.11$  \\
\hline
                                                                      $\langle \fwhm{} \rangle$  &  $  4.10\pm  2.83$  &  $  2.57\pm  1.88$  &  $  2.35\pm  1.81$  &  $  7.47\pm  3.81/  11.59\pm  3.93$  \\
                                                                               $\Delta \fwhm{}$  &  $  1.46\pm  2.46$  &  &  $  0.49\pm  0.57$  &  $  0.89\pm  0.57/  2.19\pm  1.24$  \\
                                          $ \frac{\Delta \fwhm{}}{\langle\fwhm{}\rangle} $  &  $  0.30\pm  0.30$  &  &  $  0.31\pm  0.24$  &  $  0.33\pm  0.24/  0.54\pm  0.38$  \\
\hline
                                                                      $\langle EW_0 \rangle$  &  $  1.17\pm  1.05$  &  $  0.58 \pm  1.13$  &  $  1.03\pm  0.80$  &  $  3.35\pm  2.57/  4.48\pm  2.76$  \\
                                                                               $\Delta EW_0$  &  $ 0.74 \pm  1.19$  &  &  $  0.15\pm  0.15$  &  $  0.53\pm  0.39/  1.37\pm  1.20$  \\
                                          $ \frac{\Delta EW_0}{\langle EW_0 \rangle} $  &  $  0.52 \pm  0.50$  &  &  $  0.17\pm  0.17$  &  $  0.29\pm  0.34/  0.32\pm  0.28$  \\

%% file: densities.tex
3C111A  & 3C111B  & $ -1.78 $ & $ -1.74 $ &  $ 145$  & $ 0.085$  & $  0.68\pm  0.52$  & $    259  \pm     199 $\\
3C111A  & 3C111B  & $  1.77 $ & $  2.09 $ &  $ 145$  & $ 0.085$  & $  2.78\pm  0.48$  & $   1059  \pm     210 $\\
3C111A  &    3C111B   & $  6.05 $ & $   \cdot\cdot\cdot $ &  $ 145$  & $ 0.085$  & $  3.02\pm  0.10$  & $   1152  \pm     117 $\\
3C111A  &    3C111B   & $ -4.72 $ & $   \cdot\cdot\cdot $ &  $ 145$  & $ 0.085$  & $  0.38\pm  0.10$  & $    144  \pm      39 $\\
 3C111A & 3C111B  & $   \cdot\cdot\cdot $ & $  3.69 $ &  $ 145$  & $ 0.085$  & $  4.51\pm  0.70$  & $   1719  \pm     313 $\\
3C111A  & 3C111C  & $ -1.78 $ & $ -2.15 $ &  $ 145$  & $ 0.059$  & $  3.82\pm  0.53$  & $    2108  \pm     295 $\\
3C111A  & 3C111C  & $  1.77 $ & $  1.63 $ &  $ 145$  & $ 0.059$  & $  3.15\pm  1.43$  & $    1740  \pm     790 $\\
3C111A  & 3C111C  & $ -4.72 $ & $   \cdot\cdot\cdot $ &  $ 145$  & $ 0.059$  & $  0.38\pm  0.10$  & $    209  \pm      53 $\\
3C111B  & 3C111C  & $  2.09 $ & $  1.63 $ &  $ 145$  & $ 0.143$  & $  5.93\pm  1.41$  & $   1341  \pm     319 $\\
3C111B  & 3C111C & $ -1.74 $ & $   \cdot\cdot\cdot $ &  $ 145$  & $ 0.143$  & $  3.88\pm  0.42$  & $   878  \pm     94 $\\
3C111B  & 3C111C & $  3.69 $ & $   \cdot\cdot\cdot $ &  $ 145$  & $ 0.143$  & $  4.51\pm  0.70$  & $   1019  \pm     157 $\\
3C123A  & 3C123B  & $  5.51 $ & $  5.50 $ &  $ 145$  & $ 0.015$  & $  0.19\pm  0.40$  & $    401  \pm     848 $\\
3C123A  & 3C123B  & $  5.28 $ & $  4.94 $ &  $ 145$  & $ 0.015$  & $  1.31\pm  1.95$  & $   2787  \pm    4166 $\\
3C123A  & 3C123B  & $  3.75 $ & $  4.37 $ &  $ 145$  & $ 0.015$  & $  0.28\pm  2.13$  & $    599  \pm    4543 $\\
3C123A  & 3C123B  & $  1.58 $ & $   \cdot\cdot\cdot $ &  $ 145$  & $ 0.015$  & $  0.32\pm  0.23$  & $    677  \pm     487 $\\
3C225A  & 3C225B  & $  3.97 $ & $  3.99 $ &  $  17$  & $ 4\times10^{-4}$  &  $ 0.05\pm  0.04$  &    $4248  \pm    3798$ \\

%% file: end_table.tex
  3C018  &   A  &  $ -9.1  \pm  0.0  $  &  $ 0.565  \pm 0.007  $  &  $  2.46  \pm  0.02  $  &  $   17  \pm    1  $  &  $  0.48  \pm  0.04  $  &  0.16  \\
         &   B  &  $ -9.0  \pm  0.0  $  &  $ 0.524  \pm 0.003  $  &  $  2.41  \pm  0.01  $  &  $   16  \pm    2  $  &  $  0.41  \pm  0.06  $  &      \\
\hline
         &   A  &  $ -6.2  \pm  0.1  $  &  $ 0.134  \pm 0.003  $  &  $  5.49  \pm  0.17  $  &  $  196  \pm    5  $  &  $  2.81  \pm  0.13  $  &  0.24  \\
         &   B  &  $ -6.8  \pm  0.0  $  &  $ 0.149  \pm 0.002  $  &  $  6.16  \pm  0.06  $  &  $  162  \pm    4  $  &  $  2.89  \pm  0.10  $  &      \\
\hline
         &   A  &  $ -5.0  \pm  0.0  $  &  $ 0.084  \pm 0.003  $  &  $  1.49  \pm  0.06  $  &  $\cdot\cdot\cdot  $  &  $   \cdot\cdot\cdot  $  &      \\
\hline
         &   A  &  $ 24.4  \pm  0.1  $  &  $ 0.007  \pm 0.003  $  &  $  0.75  \pm  0.31  $  &  $\cdot\cdot\cdot  $  &  $   \cdot\cdot\cdot  $  &      \\
\hline
\hline
  3C041  &   A  &  $ -1.4  \pm  0.1  $  &  $ 0.033  \pm 0.001  $  &  $  8.87  \pm  0.16  $  &  $  351  \pm    7  $  &  $  1.99  \pm  0.07  $  &  0.02  \\
         &   B  &  $ -1.3  \pm  0.1  $  &  $ 0.042  \pm 0.001  $  &  $  7.14  \pm  0.22  $  &  $  150  \pm    5  $  &  $  0.87  \pm  0.05  $  &      \\
\hline
         &   A  &  $  1.8  \pm  0.1  $  &  $ 0.011  \pm 0.002  $  &  $  1.21  \pm  0.22  $  &  $   98  \pm   16  $  &  $  0.02  \pm  0.01  $  &      \\
\hline
         &   A  &  $-30.5  \pm  0.0  $  &  $ 0.029  \pm 0.001  $  &  $  1.75  \pm  0.06  $  &  $\cdot\cdot\cdot  $  &  $   \cdot\cdot\cdot  $  &      \\
\hline
         &   B  &  $-10.5  \pm  0.1  $  &  $ 0.022  \pm 0.002  $  &  $  1.65  \pm  0.19  $  &  $   14  \pm    3  $  &  $  0.01  \pm  0.00  $  &      \\
\hline
         &   B  &  $ -8.0  \pm  0.1  $  &  $ 0.023  \pm 0.003  $  &  $  0.79  \pm  0.13  $  &  $\cdot\cdot\cdot  $  &  $   \cdot\cdot\cdot  $  &      \\
\hline
\hline
  3C111  &   A  &  $ -1.8  \pm  0.0  $  &  $ 0.790  \pm 0.009  $  &  $  3.86  \pm  0.13  $  &  $   77  \pm    4  $  &  $  4.57  \pm  0.31  $  &  0.02  \\
         &   B  &  $ -1.7  \pm  0.0  $  &  $ 0.967  \pm 0.020  $  &  $  3.69  \pm  0.05  $  &  $   56  \pm    5  $  &  $  3.88  \pm  0.42  $  &      \\
\hline
         &   A  &  $  1.8  \pm  0.0  $  &  $ 0.643  \pm 0.020  $  &  $  3.31  \pm  0.10  $  &  $   93  \pm    8  $  &  $  3.85  \pm  0.38  $  &  0.25  \\
         &   B  &  $  2.1  \pm  0.0  $  &  $ 0.347  \pm 0.054  $  &  $  2.72  \pm  0.15  $  &  $   58  \pm   13  $  &  $  1.07  \pm  0.30  $  &      \\
\hline
         &   A  &  $-16.8  \pm  0.0  $  &  $ 0.322  \pm 0.003  $  &  $  2.60  \pm  0.03  $  &  $   34  \pm    5  $  &  $  0.56  \pm  0.10  $  &  0.05  \\
         &   B  &  $-16.8  \pm  0.0  $  &  $ 0.111  \pm 0.002  $  &  $  2.69  \pm  0.06  $  &  $   30  \pm    5  $  &  $  0.17  \pm  0.03  $  &      \\
\hline
         &   A  &  $  8.4  \pm  0.0  $  &  $ 0.069  \pm 0.004  $  &  $  1.27  \pm  0.08  $  &  $\cdot\cdot\cdot  $  &  $   \cdot\cdot\cdot  $  &  0.57  \\
         &   B  &  $  8.0  \pm  0.0  $  &  $ 0.212  \pm 0.005  $  &  $  2.07  \pm  0.04  $  &  $\cdot\cdot\cdot  $  &  $   \cdot\cdot\cdot  $  &      \\
\hline
         &   A  &  $ -8.3  \pm  0.0  $  &  $ 0.114  \pm 0.006  $  &  $  2.09  \pm  0.13  $  &  $\cdot\cdot\cdot  $  &  $   \cdot\cdot\cdot  $  &  0.03  \\
         &   B  &  $ -8.3  \pm  0.0  $  &  $ 0.400  \pm 0.005  $  &  $  2.81  \pm  0.03  $  &  $   14  \pm    8  $  &  $  0.32  \pm  0.20  $  &      \\
\hline
         &   A  &  $ -9.1  \pm  0.2  $  &  $ 0.172  \pm 0.005  $  &  $  8.65  \pm  0.31  $  &  $  248  \pm   10  $  &  $  7.18  \pm  0.46  $  &  0.41  \\
         &   B  &  $-10.8  \pm  0.3  $  &  $ 0.092  \pm 0.002  $  &  $ 12.14  \pm  0.34  $  &  $  409  \pm   10  $  &  $  8.86  \pm  0.41  $  &      \\
\hline
         &   A  &  $  6.1  \pm  0.1  $  &  $ 0.232  \pm 0.003  $  &  $  5.20  \pm  0.13  $  &  $  129  \pm    1  $  &  $  3.02  \pm  0.10  $  &      \\
\hline
         &   A  &  $ -4.7  \pm  0.0  $  &  $ 0.201  \pm 0.011  $  &  $  1.48  \pm  0.07  $  &  $   65  \pm   15  $  &  $  0.38  \pm  0.10  $  &      \\
\hline
         &   A  &  $-21.9  \pm  0.1  $  &  $ 0.041  \pm 0.002  $  &  $  2.57  \pm  0.13  $  &  $  131  \pm    6  $  &  $  0.27  \pm  0.02  $  &      \\
\hline
         &   A  &  $-28.9  \pm  0.4  $  &  $ 0.013  \pm 0.001  $  &  $  5.66  \pm  0.97  $  &  $  487  \pm   44  $  &  $  0.69  \pm  0.15  $  &      \\
\hline
         &   A  &  $-32.4  \pm  0.0  $  &  $ 0.037  \pm 0.003  $  &  $  1.68  \pm  0.14  $  &  $\cdot\cdot\cdot  $  &  $   \cdot\cdot\cdot  $  &      \\
\hline
         &   A  &  $  6.6  \pm 99    $  &  $ 0.025  \pm 99     $  &  $   0.06  \pm 99    $  &  $\cdot\cdot\cdot  $  &  $   \cdot\cdot\cdot  $  &      \\
\hline
         &   B  &  $  3.7  \pm  0.4  $  &  $ 0.342  \pm 0.032  $  &  $  5.63  \pm  0.36  $  &  $  120  \pm   12  $  &  $  4.51  \pm  0.70  $  &      \\
\hline
         &   B  &  $-54.8  \pm  0.1  $  &  $ 0.019  \pm 0.001  $  &  $  1.72  \pm  0.15  $  &  $   26  \pm    1  $  &  $  0.02  \pm  0.00  $  &      \\
\hline
         &   B  &  $ -2.3  \pm  0.0  $  &  $ 0.181  \pm 0.006  $  &  $  0.78  \pm  0.03  $  &  $\cdot\cdot\cdot  $  &  $   \cdot\cdot\cdot  $  &      \\
\hline
         &   B  &  $-51.3  \pm  0.1  $  &  $ 0.015  \pm 0.001  $  &  $  2.16  \pm  0.20  $  &  $\cdot\cdot\cdot  $  &  $   \cdot\cdot\cdot  $  &      \\
\hline
\hline
  3C111  &   A  &  $ -1.8  \pm  0.0  $  &  $ 0.790  \pm 0.009  $  &  $  3.86  \pm  0.13  $  &  $   77  \pm    4  $  &  $  4.57  \pm  0.31  $  &  0.30  \\
         &   C   &  $ -2.1  \pm  0.0  $  &  $ 0.816  \pm 0.050  $  &  $  2.30  \pm  0.07  $  &  $   20  \pm   11  $  &  $  0.75  \pm  0.43  $  &      \\
\hline
         &   A  &  $  6.1  \pm  0.1  $  &  $ 0.232  \pm 0.003  $  &  $  5.20  \pm  0.13  $  &  $  129  \pm    1  $  &  $  3.02  \pm  0.10  $  &  0.18  \\
         &   C   &  $  6.3  \pm  0.1  $  &  $ 0.058  \pm 0.005  $  &  $  1.79  \pm  0.16  $  &  $\cdot\cdot\cdot  $  &  $   \cdot\cdot\cdot  $  &      \\
\hline
         &   A  &  $-16.8  \pm  0.0  $  &  $ 0.322  \pm 0.003  $  &  $  2.60  \pm  0.03  $  &  $   34  \pm    5  $  &  $  0.56  \pm  0.10  $  &  0.13  \\
         &   C   &  $-16.9  \pm  0.1  $  &  $ 0.282  \pm 0.045  $  &  $  2.06  \pm  0.08  $  &  $   20  \pm    4  $  &  $  0.23  \pm  0.07  $  &      \\
\hline
         &   A  &  $  1.8  \pm  0.0  $  &  $ 0.643  \pm 0.020  $  &  $  3.31  \pm  0.10  $  &  $   93  \pm    8  $  &  $  3.85  \pm  0.38  $  &  0.07  \\
         &   C   &  $  1.6  \pm  0.4  $  &  $ 0.456  \pm 0.044  $  &  $  7.32  \pm  0.29  $  &  $  108  \pm   18  $  &  $  7.00  \pm  1.38  $  &      \\
\hline
         &   A  &  $ -8.3  \pm  0.0  $  &  $ 0.114  \pm 0.006  $  &  $  2.09  \pm  0.13  $  &  $\cdot\cdot\cdot  $  &  $   \cdot\cdot\cdot  $  &  0.08  \\
         &   C   &  $ -8.4  \pm  0.0  $  &  $ 0.416  \pm 0.008  $  &  $  1.60  \pm  0.03  $  &  $\cdot\cdot\cdot  $  &  $   \cdot\cdot\cdot  $  &      \\
\hline
         &   A  &  $ -9.1  \pm  0.2  $  &  $ 0.172  \pm 0.005  $  &  $  8.65  \pm  0.31  $  &  $  248  \pm   10  $  &  $  7.18  \pm  0.46  $  &  0.23  \\
         &   C   &  $ -8.1  \pm  0.6  $  &  $ 0.182  \pm 0.011  $  &  $ 11.25  \pm  0.59  $  &  $  118  \pm   28  $  &  $  4.69  \pm  1.18  $  &      \\
\hline
         &   A  &  $ -4.7  \pm  0.0  $  &  $ 0.201  \pm 0.011  $  &  $  1.48  \pm  0.07  $  &  $   65  \pm   15  $  &  $  0.38  \pm  0.10  $  &      \\
\hline
         &   A  &  $-21.9  \pm  0.1  $  &  $ 0.041  \pm 0.002  $  &  $  2.57  \pm  0.13  $  &  $  131  \pm    6  $  &  $  0.27  \pm  0.02  $  &      \\
\hline
         &   A  &  $-28.9  \pm  0.4  $  &  $ 0.013  \pm 0.001  $  &  $  5.66  \pm  0.97  $  &  $  487  \pm   44  $  &  $  0.69  \pm  0.15  $  &      \\
\hline
         &   A  &  $-32.4  \pm  0.0  $  &  $ 0.037  \pm 0.003  $  &  $  1.68  \pm  0.14  $  &  $\cdot\cdot\cdot  $  &  $   \cdot\cdot\cdot  $  &      \\
\hline
         &   A  &  $  6.6  \pm  99  $  &  $ 0.025  \pm 99      $  &  $   0.06  \pm   99  $  &  $\cdot\cdot\cdot  $  &  $   \cdot\cdot\cdot  $  &      \\
\hline
         &   A  &  $  8.4  \pm  0.0  $  &  $ 0.069  \pm 0.004  $  &  $  1.27  \pm  0.08  $  &  $\cdot\cdot\cdot  $  &  $   \cdot\cdot\cdot  $  &      \\
\hline
         &   C   &  $-15.3  \pm  0.5  $  &  $ 0.088  \pm 0.027  $  &  $  2.54  \pm  0.59  $  &  $   90  \pm   34  $  &  $  0.39  \pm  0.21  $  &      \\
\hline
         &   C   &  $-27.8  \pm  0.1  $  &  $ 0.029  \pm 0.001  $  &  $  2.61  \pm  0.20  $  &  $   39  \pm    2  $  &  $  0.06  \pm  0.01  $  &      \\
\hline
         &   C   &  $-31.4  \pm  0.1  $  &  $ 0.027  \pm 0.001  $  &  $  2.38  \pm  0.19  $  &  $   35  \pm    3  $  &  $  0.04  \pm  0.01  $  &      \\
\hline
         &   C   &  $  1.1  \pm  0.0  $  &  $ 0.584  \pm 0.043  $  &  $  2.47  \pm  0.07  $  &  $\cdot\cdot\cdot  $  &  $   \cdot\cdot\cdot  $  &      \\
\hline
         &   C   &  $ -5.5  \pm  0.0  $  &  $ 0.211  \pm 0.014  $  &  $  1.49  \pm  0.08  $  &  $\cdot\cdot\cdot  $  &  $   \cdot\cdot\cdot  $  &      \\
\hline
\hline
  3C111  &   B  &  $-16.8  \pm  0.0  $  &  $ 0.111  \pm 0.002  $  &  $  2.69  \pm  0.06  $  &  $   30  \pm    5  $  &  $  0.17  \pm  0.03  $  &  0.18  \\
         &  C    &  $-16.9  \pm  0.1  $  &  $ 0.282  \pm 0.045  $  &  $  2.06  \pm  0.08  $  &  $   20  \pm    4  $  &  $  0.23  \pm  0.07  $  &      \\
\hline
         &   B  &  $ -8.3  \pm  0.0  $  &  $ 0.400  \pm 0.005  $  &  $  2.81  \pm  0.03  $  &  $   14  \pm    8  $  &  $  0.32  \pm  0.20  $  &  0.11  \\
         &  C    &  $ -8.4  \pm  0.0  $  &  $ 0.416  \pm 0.008  $  &  $  1.60  \pm  0.03  $  &  $\cdot\cdot\cdot  $  &  $   \cdot\cdot\cdot  $  &      \\
\hline
         &   B  &  $  2.1  \pm  0.0  $  &  $ 0.347  \pm 0.054  $  &  $  2.72  \pm  0.15  $  &  $   58  \pm   13  $  &  $  1.07  \pm  0.30  $  &  0.27  \\
         &   C   &  $  1.6  \pm  0.4  $  &  $ 0.456  \pm 0.044  $  &  $  7.32  \pm  0.29  $  &  $  108  \pm   18  $  &  $  7.00  \pm  1.38  $  &      \\
\hline
         &   B  &  $ -2.3  \pm  0.0  $  &  $ 0.181  \pm 0.006  $  &  $  0.78  \pm  0.03  $  &  $\cdot\cdot\cdot  $  &  $   \cdot\cdot\cdot  $  &  0.31  \\
         &   C   &  $ -2.1  \pm  0.0  $  &  $ 0.816  \pm 0.050  $  &  $  2.30  \pm  0.07  $  &  $   20  \pm   11  $  &  $  0.75  \pm  0.43  $  &      \\
\hline
         &   B  &  $-10.8  \pm  0.3  $  &  $ 0.092  \pm 0.002  $  &  $ 12.14  \pm  0.34  $  &  $  409  \pm   10  $  &  $  8.86  \pm  0.41  $  &  0.55  \\
         &   C   &  $ -8.1  \pm  0.6  $  &  $ 0.182  \pm 0.011  $  &  $ 11.25  \pm  0.59  $  &  $  118  \pm   28  $  &  $  4.69  \pm  1.18  $  &      \\
\hline
         &   B  &  $ -1.7  \pm  0.0  $  &  $ 0.967  \pm 0.020  $  &  $  3.69  \pm  0.05  $  &  $   56  \pm    5  $  &  $  3.88  \pm  0.42  $  &      \\
\hline
         &   B  &  $  3.7  \pm  0.4  $  &  $ 0.342  \pm 0.032  $  &  $  5.63  \pm  0.36  $  &  $  120  \pm   12  $  &  $  4.51  \pm  0.70  $  &      \\
\hline
         &   B  &  $-54.8  \pm  0.1  $  &  $ 0.019  \pm 0.001  $  &  $  1.72  \pm  0.15  $  &  $   26  \pm    1  $  &  $  0.02  \pm  0.00  $  &      \\
\hline
         &   B  &  $-51.3  \pm  0.1  $  &  $ 0.015  \pm 0.001  $  &  $  2.16  \pm  0.20  $  &  $\cdot\cdot\cdot  $  &  $   \cdot\cdot\cdot  $  &      \\
\hline
         &   B  &  $  8.0  \pm  0.0  $  &  $ 0.212  \pm 0.005  $  &  $  2.07  \pm  0.04  $  &  $\cdot\cdot\cdot  $  &  $   \cdot\cdot\cdot  $  &      \\
\hline
         &   C   &  $-15.3  \pm  0.5  $  &  $ 0.088  \pm 0.027  $  &  $  2.54  \pm  0.59  $  &  $   90  \pm   34  $  &  $  0.39  \pm  0.21  $  &      \\
\hline
         &   C   &  $-27.8  \pm  0.1  $  &  $ 0.029  \pm 0.001  $  &  $  2.61  \pm  0.20  $  &  $   39  \pm    2  $  &  $  0.06  \pm  0.01  $  &      \\
\hline
         &   C   &  $-31.4  \pm  0.1  $  &  $ 0.027  \pm 0.001  $  &  $  2.38  \pm  0.19  $  &  $   35  \pm    3  $  &  $  0.04  \pm  0.01  $  &      \\
\hline
         &  C    &  $  1.1  \pm  0.0  $  &  $ 0.584  \pm 0.043  $  &  $  2.47  \pm  0.07  $  &  $\cdot\cdot\cdot  $  &  $   \cdot\cdot\cdot  $  &      \\
\hline
         &  C    &  $ -5.5  \pm  0.0  $  &  $ 0.211  \pm 0.014  $  &  $  1.49  \pm  0.08  $  &  $\cdot\cdot\cdot  $  &  $   \cdot\cdot\cdot  $  &      \\
\hline
         &  C    &  $  6.3  \pm  0.1  $  &  $ 0.058  \pm 0.005  $  &  $  1.79  \pm  0.16  $  &  $\cdot\cdot\cdot  $  &  $   \cdot\cdot\cdot  $  &      \\
\hline
\hline
  3C123  &   A  &  $-19.9  \pm  0.0  $  &  $ 0.064  \pm 0.001  $  &  $  3.13  \pm  0.03  $  &  $   26  \pm    2  $  &  $  0.10  \pm  0.01  $  &  0.43  \\
         &   B  &  $-19.3  \pm  0.0  $  &  $ 0.064  \pm 0.001  $  &  $  3.23  \pm  0.04  $  &  $   16  \pm    2  $  &  $  0.07  \pm  0.01  $  &      \\
\hline
         &   A  &  $ 20.2  \pm  0.1  $  &  $ 0.008  \pm 0.000  $  &  $  4.05  \pm  0.27  $  &  $  124  \pm    9  $  &  $  0.08  \pm  0.01  $  &  0.09  \\
         &   B  &  $ 20.1  \pm  0.2  $  &  $ 0.004  \pm 0.001  $  &  $  3.21  \pm  0.54  $  &  $  334  \pm   52  $  &  $  0.09  \pm  0.02  $  &      \\
\hline
         &   A  &  $  5.5  \pm  0.0  $  &  $ 0.628  \pm 0.037  $  &  $  1.75  \pm  0.03  $  &  $   18  \pm   11  $  &  $  0.40  \pm  0.24  $  &  0.02  \\
         &   B  &  $  5.5  \pm  0.0  $  &  $ 0.253  \pm 0.007  $  &  $  1.47  \pm  0.04  $  &  $   81  \pm   43  $  &  $  0.59  \pm  0.31  $  &      \\
\hline
         &   A  &  $  5.3  \pm  0.2  $  &  $ 0.810  \pm 0.071  $  &  $  4.40  \pm  0.15  $  &  $   67  \pm   25  $  &  $  4.67  \pm  1.82  $  &  0.12  \\
         &   B  &  $  4.9  \pm  0.1  $  &  $ 0.042  \pm 0.003  $  &  $ 10.95  \pm  0.33  $  &  $  664  \pm   52  $  &  $  5.98  \pm  0.69  $  &      \\
\hline
         &   A  &  $  3.7  \pm  0.1  $  &  $ 0.044  \pm 0.004  $  &  $ 10.13  \pm  0.28  $  &  $  619  \pm   79  $  &  $  5.36  \pm  0.89  $  &  0.24  \\
         &   B  &  $  4.4  \pm  0.0  $  &  $ 1.648  \pm 0.005  $  &  $  4.31  \pm  0.01  $  &  $   36  \pm   14  $  &  $  5.08  \pm  1.94  $  &      \\
\hline
         &   A  &  $  1.6  \pm  0.0  $  &  $ 0.379  \pm 0.035  $  &  $  2.18  \pm  0.05  $  &  $   19  \pm   14  $  &  $  0.32  \pm  0.23  $  &      \\
\hline
         &   A  &  $  3.7  \pm  0.0  $  &  $ 1.020  \pm 0.096  $  &  $  2.28  \pm  0.04  $  &  $\cdot\cdot\cdot  $  &  $   \cdot\cdot\cdot  $  &      \\
\hline
         &   B  &  $  8.5  \pm  0.0  $  &  $ 0.064  \pm 0.003  $  &  $  1.73  \pm  0.09  $  &  $\cdot\cdot\cdot  $  &  $   \cdot\cdot\cdot  $  &      \\
\hline
\hline
  3C225  &   A  &  $  4.0  \pm  0.0  $  &  $ 0.805  \pm 0.001  $  &  $  1.26  \pm  0.00  $  &  $   11  \pm    2  $  &  $  0.23  \pm  0.04  $  &  0.03  \\
         &   B  &  $  4.0  \pm  0.0  $  &  $ 0.774  \pm 0.001  $  &  $  1.28  \pm  0.00  $  &  $   14  \pm    0  $  &  $  0.28  \pm  0.01  $  &      \\
\hline
         &   A  &  $-40.2  \pm  0.0  $  &  $ 0.043  \pm 0.002  $  &  $  1.82  \pm  0.07  $  &  $   22  \pm    0  $  &  $  0.03  \pm  0.00  $  &  0.13  \\
         &   B  &  $-40.3  \pm  0.0  $  &  $ 0.044  \pm 0.002  $  &  $  2.00  \pm  0.09  $  &  $   18  \pm    0  $  &  $  0.03  \pm  0.00  $  &      \\
\hline
         &   A  &  $-27.2  \pm  0.0  $  &  $ 0.048  \pm 0.001  $  &  $  2.54  \pm  0.04  $  &  $   22  \pm    0  $  &  $  0.05  \pm  0.00  $  &  0.09  \\
         &   B  &  $-27.3  \pm  0.0  $  &  $ 0.053  \pm 0.001  $  &  $  2.36  \pm  0.05  $  &  $   20  \pm    0  $  &  $  0.05  \pm  0.00  $  &      \\
\hline
         &   A  &  $-37.4  \pm  0.2  $  &  $ 0.020  \pm 0.001  $  &  $  4.76  \pm  0.30  $  &  $   60  \pm    1  $  &  $  0.11  \pm  0.01  $  &  0.12  \\
         &   B  &  $-37.2  \pm  0.1  $  &  $ 0.023  \pm 0.001  $  &  $  4.02  \pm  0.31  $  &  $  145  \pm    4  $  &  $  0.26  \pm  0.02  $  &      \\
\hline
         &   A  &  $ -5.2  \pm  0.1  $  &  $ 0.013  \pm 0.000  $  &  $  7.71  \pm  0.29  $  &  $  327  \pm   11  $  &  $  0.64  \pm  0.04  $  &  0.10  \\
         &   B  &  $ -5.6  \pm  0.2  $  &  $ 0.013  \pm 0.001  $  &  $  8.32  \pm  0.38  $  &  $  458  \pm   17  $  &  $  1.00  \pm  0.07  $  &      \\
\hline
\hline
  3C245  &   A  &  $ -9.1  \pm  0.1  $  &  $ 0.010  \pm 0.001  $  &  $  5.30  \pm  0.26  $  &  $  385  \pm   30  $  &  $  0.40  \pm  0.05  $  &      \\
\hline
         &   A  &  $ -9.8  \pm  0.1  $  &  $ 0.006  \pm 0.001  $  &  $  1.55  \pm  0.28  $  &  $\cdot\cdot\cdot  $  &  $   \cdot\cdot\cdot  $  &      \\
\hline
\hline
3C327.1  &   A  &  $ -0.0  \pm  0.0  $  &  $ 0.425  \pm 0.010  $  &  $  1.91  \pm  0.05  $  &  $   63  \pm    4  $  &  $  1.00  \pm  0.08  $  &  0.00  \\
         &   B  &  $ -0.0  \pm  0.0  $  &  $ 0.359  \pm 0.011  $  &  $  1.94  \pm  0.07  $  &  $   74  \pm    4  $  &  $  1.01  \pm  0.08  $  &      \\
\hline
         &   A  &  $  2.0  \pm  0.0  $  &  $ 0.401  \pm 0.006  $  &  $  2.17  \pm  0.04  $  &  $   69  \pm    3  $  &  $  1.18  \pm  0.06  $  &  0.07  \\
         &   B  &  $  1.9  \pm  0.0  $  &  $ 0.419  \pm 0.008  $  &  $  2.16  \pm  0.04  $  &  $   68  \pm    3  $  &  $  1.20  \pm  0.07  $  &      \\
\hline
         &   A  &  $ -2.7  \pm  0.1  $  &  $ 0.126  \pm 0.002  $  &  $  3.36  \pm  0.12  $  &  $  140  \pm    3  $  &  $  1.16  \pm  0.06  $  &  0.04  \\
         &   B  &  $ -2.7  \pm  0.1  $  &  $ 0.118  \pm 0.002  $  &  $  3.22  \pm  0.13  $  &  $  142  \pm    3  $  &  $  1.05  \pm  0.05  $  &      \\
\hline
         &   B  &  $ 46.1  \pm  0.1  $  &  $ 0.013  \pm 0.002  $  &  $  0.67  \pm  0.13  $  &  $\cdot\cdot\cdot  $  &  $   \cdot\cdot\cdot  $  &      \\
\hline
\hline
  3C409  &   A  &  $  4.2  \pm  0.0  $  &  $ 0.443  \pm 0.001  $  &  $  3.19  \pm  0.02  $  &  $   64  \pm    9  $  &  $  1.77  \pm  0.26  $  &  0.19  \\
         &   B  &  $  4.0  \pm  0.0  $  &  $ 0.429  \pm 0.003  $  &  $  2.86  \pm  0.02  $  &  $   49  \pm    9  $  &  $  1.18  \pm  0.23  $  &      \\
\hline
         &   A  &  $  7.9  \pm  0.0  $  &  $ 0.332  \pm 0.002  $  &  $  3.00  \pm  0.03  $  &  $  122  \pm   11  $  &  $  2.36  \pm  0.22  $  &  0.03  \\
         &   B  &  $  8.0  \pm  0.0  $  &  $ 0.280  \pm 0.007  $  &  $  3.01  \pm  0.09  $  &  $  120  \pm   13  $  &  $  1.97  \pm  0.23  $  &      \\
\hline
         &   A  &  $ 14.6  \pm  0.0  $  &  $ 0.440  \pm 0.008  $  &  $  6.53  \pm  0.07  $  &  $  145  \pm   20  $  &  $  8.11  \pm  1.15  $  &  0.03  \\
         &   B  &  $ 14.5  \pm  0.2  $  &  $ 0.106  \pm 0.004  $  &  $ 13.09  \pm  0.19  $  &  $  285  \pm   12  $  &  $  7.67  \pm  0.45  $  &      \\
\hline
         &   A  &  $ 13.8  \pm  0.0  $  &  $ 0.732  \pm 0.007  $  &  $  2.15  \pm  0.02  $  &  $   13  \pm    5  $  &  $  0.41  \pm  0.16  $  &  0.29  \\
         &   B  &  $ 13.6  \pm  0.0  $  &  $ 0.631  \pm 0.081  $  &  $  1.85  \pm  0.06  $  &  $   66  \pm   22  $  &  $  1.51  \pm  0.54  $  &      \\
\hline
         &   A  &  $ 15.9  \pm  0.0  $  &  $ 0.735  \pm 0.006  $  &  $  1.70  \pm  0.01  $  &  $   12  \pm   18  $  &  $  0.29  \pm  0.44  $  &  0.67  \\
         &   B  &  $ 15.3  \pm  0.1  $  &  $ 0.890  \pm 0.065  $  &  $  3.01  \pm  0.05  $  &  $   47  \pm   13  $  &  $  2.45  \pm  0.71  $  &      \\
\hline
         &   A  &  $ 24.3  \pm  0.1  $  &  $ 0.020  \pm 0.001  $  &  $  4.33  \pm  0.20  $  &  $  366  \pm   39  $  &  $  0.63  \pm  0.08  $  &      \\
\hline
         &   A  &  $-53.8  \pm  0.1  $  &  $ 0.004  \pm 0.001  $  &  $  0.64  \pm  0.22  $  &  $\cdot\cdot\cdot  $  &  $   \cdot\cdot\cdot  $  &      \\
\hline
         &   B  &  $ 12.3  \pm  0.9  $  &  $ 0.193  \pm 0.053  $  &  $  3.63  \pm  1.24  $  &  $\cdot\cdot\cdot  $  &  $   \cdot\cdot\cdot  $  &      \\
\hline
\hline
  3C410  &   A  &  $  2.7  \pm  0.1  $  &  $ 0.648  \pm 0.023  $  &  $  3.24  \pm  0.11  $  &  $   49  \pm   14  $  &  $  2.02  \pm  0.59  $  &  0.21  \\
         &   B  &  $  2.5  \pm  0.0  $  &  $ 0.426  \pm 0.010  $  &  $  2.28  \pm  0.06  $  &  $\cdot\cdot\cdot  $  &  $   \cdot\cdot\cdot  $  &      \\
\hline
         &   A  &  $ -0.2  \pm  0.1  $  &  $ 0.613  \pm 0.018  $  &  $  3.58  \pm  0.11  $  &  $   48  \pm   12  $  &  $  2.07  \pm  0.54  $  &  0.01  \\
         &   B  &  $ -0.2  \pm  0.0  $  &  $ 0.476  \pm 0.005  $  &  $  2.65  \pm  0.04  $  &  $   28  \pm    2  $  &  $  0.71  \pm  0.07  $  &      \\
\hline
         &   A  &  $ 17.9  \pm  0.0  $  &  $ 0.186  \pm 0.001  $  &  $  3.33  \pm  0.05  $  &  $   71  \pm    8  $  &  $  0.86  \pm  0.11  $  &  0.37  \\
         &   B  &  $ 18.4  \pm  0.0  $  &  $ 0.119  \pm 0.003  $  &  $  2.53  \pm  0.08  $  &  $\cdot\cdot\cdot  $  &  $   \cdot\cdot\cdot  $  &      \\
\hline
         &   A  &  $-30.2  \pm  0.1  $  &  $ 0.014  \pm 0.001  $  &  $  1.46  \pm  0.14  $  &  $   15  \pm    3  $  &  $  0.01  \pm  0.00  $  &  0.01  \\
         &   B  &  $-30.2  \pm  0.1  $  &  $ 0.019  \pm 0.002  $  &  $  1.79  \pm  0.27  $  &  $   18  \pm    3  $  &  $  0.01  \pm  0.00  $  &      \\
\hline
         &   A  &  $ 25.3  \pm  0.1  $  &  $ 0.060  \pm 0.003  $  &  $  5.25  \pm  0.12  $  &  $   69  \pm   14  $  &  $  0.43  \pm  0.09  $  &  0.10  \\
         &   B  &  $ 25.2  \pm  0.0  $  &  $ 0.112  \pm 0.002  $  &  $  2.54  \pm  0.07  $  &  $   32  \pm    3  $  &  $  0.18  \pm  0.02  $  &      \\
\hline
         &   A  &  $  8.1  \pm  0.0  $  &  $ 1.864  \pm 0.125  $  &  $  1.54  \pm  0.03  $  &  $  104  \pm   60  $  &  $  5.79  \pm  3.39  $  &  0.17  \\
         &   B  &  $  8.0  \pm  0.0  $  &  $ 2.798  \pm 0.014  $  &  $  2.19  \pm  0.01  $  &  $   18  \pm    1  $  &  $  2.15  \pm  0.23  $  &      \\
\hline
         &   A  &  $ 11.1  \pm  0.1  $  &  $ 0.149  \pm 0.081  $  &  $  1.63  \pm  0.32  $  &  $  106  \pm   45  $  &  $  0.50  \pm  0.36  $  &  0.23  \\
         &   B  &  $ 10.9  \pm  0.1  $  &  $ 0.430  \pm 0.137  $  &  $  1.92  \pm  0.12  $  &  $   65  \pm   21  $  &  $  1.06  \pm  0.48  $  &      \\
\hline
         &   A  &  $ 11.3  \pm  0.4  $  &  $ 0.575  \pm 0.096  $  &  $  4.39  \pm  0.37  $  &  $  111  \pm   21  $  &  $  5.46  \pm  1.45  $  &  0.46  \\
         &   B  &  $ 12.7  \pm  3.1  $  &  $ 0.041  \pm 0.017  $  &  $ 19.16  \pm  1.99  $  &  $ 1006  \pm  403  $  &  $ 15.52  \pm  8.97  $  &      \\
\hline
         &   A  &  $  7.4  \pm  0.1  $  &  $ 1.693  \pm 0.150  $  &  $  3.16  \pm  0.14  $  &  $   70  \pm   13  $  &  $  7.33  \pm  1.60  $  &      \\
\hline
         &   A  &  $-22.7  \pm  0.1  $  &  $ 0.020  \pm 0.001  $  &  $  3.99  \pm  0.20  $  &  $  647  \pm   31  $  &  $  0.99  \pm  0.08  $  &      \\
\hline
         &   A  &  $-46.5  \pm  0.0  $  &  $ 0.055  \pm 0.001  $  &  $  1.91  \pm  0.04  $  &  $\cdot\cdot\cdot  $  &  $   \cdot\cdot\cdot  $  &      \\
\hline
         &   A  &  $ -4.8  \pm  0.1  $  &  $ 0.048  \pm 0.002  $  &  $  2.97  \pm  0.17  $  &  $\cdot\cdot\cdot  $  &  $   \cdot\cdot\cdot  $  &      \\
\hline
         &   A  &  $ 24.6  \pm  0.0  $  &  $ 0.076  \pm 0.003  $  &  $  1.79  \pm  0.07  $  &  $\cdot\cdot\cdot  $  &  $   \cdot\cdot\cdot  $  &      \\
\hline
         &   B  &  $  5.9  \pm  0.0  $  &  $ 0.292  \pm 0.011  $  &  $  1.45  \pm  0.04  $  &  $   31  \pm    3  $  &  $  0.26  \pm  0.03  $  &      \\
\hline
         &   B  &  $  5.1  \pm  0.3  $  &  $ 0.344  \pm 0.023  $  &  $ 11.43  \pm  0.50  $  &  $  153  \pm   12  $  &  $ 11.71  \pm  1.33  $  &      \\
\hline
         &   B  &  $-47.6  \pm  0.2  $  &  $ 0.021  \pm 0.001  $  &  $  5.27  \pm  0.42  $  &  $   39  \pm    3  $  &  $  0.08  \pm  0.01  $  &      \\
\hline
         &   B  &  $ 12.2  \pm  0.4  $  &  $ 0.347  \pm 0.087  $  &  $  2.57  \pm  0.33  $  &  $\cdot\cdot\cdot  $  &  $   \cdot\cdot\cdot  $  &      \\
\hline
\hline
   3C33  &   1  &  $ -4.6  \pm  0.1  $  &  $ 0.034  \pm 0.001  $  &  $  9.44  \pm  0.30  $  &  $  309  \pm    4  $  &  $  1.95  \pm  0.00  $  &  0.15  \\
         &   2  &  $ -4.2  \pm  0.2  $  &  $ 0.059  \pm 0.002  $  &  $  9.31  \pm  0.42  $  &  $  177  \pm    4  $  &  $  1.90  \pm  0.00  $  &      \\
\hline
\hline
   3C75  &   1  &  $-10.4  \pm  0.0  $  &  $ 0.729  \pm 0.014  $  &  $  2.06  \pm  0.03  $  &  $   34  \pm    4  $  &  $  1.02  \pm  0.00  $  &  0.00  \\
         &   2  &  $-10.4  \pm  0.0  $  &  $ 0.647  \pm 0.015  $  &  $  2.33  \pm  0.05  $  &  $   36  \pm    4  $  &  $  1.06  \pm  0.00  $  &      \\
\hline
         &   1  &  $ -6.1  \pm  0.1  $  &  $ 0.082  \pm 0.007  $  &  $  3.01  \pm  0.30  $  &  $   16  \pm    5  $  &  $  0.08  \pm  0.00  $  &  0.14  \\
         &   2  &  $ -6.0  \pm  0.1  $  &  $ 0.094  \pm 0.009  $  &  $  2.34  \pm  0.26  $  &  $   34  \pm    4  $  &  $  0.15  \pm  0.00  $  &      \\
\hline
         &   1  &  $  5.0  \pm  0.1  $  &  $ 0.127  \pm 0.006  $  &  $  4.58  \pm  0.22  $  &  $   84  \pm    7  $  &  $  0.95  \pm  0.00  $  &  0.08  \\
         &   2  &  $  4.9  \pm  0.1  $  &  $ 0.139  \pm 0.007  $  &  $  4.41  \pm  0.23  $  &  $   78  \pm    8  $  &  $  0.93  \pm  0.00  $  &      \\
\hline
\hline
   3C98  &   1  &  $ -1.2  \pm  0.1  $  &  $ 0.081  \pm 0.004  $  &  $  3.23  \pm  0.18  $  &  $   21  \pm    6  $  &  $  0.11  \pm  0.00  $  &  0.33  \\
         &   2  &  $ -1.5  \pm  0.1  $  &  $ 0.090  \pm 0.004  $  &  $  3.16  \pm  0.14  $  &  $   58  \pm    5  $  &  $  0.33  \pm  0.00  $  &      \\
\hline
         &   1  &  $  9.4  \pm  0.0  $  &  $ 0.209  \pm 0.011  $  &  $  1.47  \pm  0.09  $  &  $   41  \pm   16  $  &  $  0.25  \pm  0.00  $  &  0.25  \\
         &   2  &  $  9.5  \pm  0.0  $  &  $ 0.203  \pm 0.013  $  &  $  1.39  \pm  0.10  $  &  $   24  \pm   15  $  &  $  0.13  \pm  0.00  $  &      \\
\hline
         &   1  &  $  9.7  \pm  0.0  $  &  $ 0.368  \pm 0.008  $  &  $  6.07  \pm  0.08  $  &  $  114  \pm    5  $  &  $  5.00  \pm  0.00  $  &  0.07  \\
         &   2  &  $  9.6  \pm  0.0  $  &  $ 0.452  \pm 0.011  $  &  $  4.62  \pm  0.06  $  &  $  100  \pm    8  $  &  $  4.08  \pm  0.00  $  &      \\
\hline
         &   1  &  $ 22.8  \pm  0.3  $  &  $ 0.028  \pm 0.003  $  &  $  5.40  \pm  0.65  $  &  $  215  \pm    6  $  &  $  0.63  \pm  0.00  $  &  0.22  \\
         &   2  &  $ 22.5  \pm  0.2  $  &  $ 0.035  \pm 0.003  $  &  $  4.56  \pm  0.42  $  &  $  165  \pm    6  $  &  $  0.52  \pm  0.00  $  &      \\
\hline